\documentclass[12pt]{iopart}

\usepackage{graphicx}
\usepackage{dcolumn}
\usepackage{bm}
\usepackage[mathlines]{lineno}
\usepackage{caption}
\usepackage{subcaption}
\usepackage{float}
\usepackage{mathrsfs}
\usepackage{amsthm}
\usepackage{bbm}
\usepackage{amsfonts}
\usepackage{cite}
\usepackage[hidelinks]{hyperref}

\makeatletter
\expandafter\let\csname equation*\endcsname\relax
\expandafter\let\csname endequation*\endcsname\relax
\makeatother

\usepackage{physics}
\usepackage{mathtools}

\begin{document}

\title[]{Exploring the near-kernel sector of quantum reduced loop gravity}

\author{Ilkka Mäkinen\footnote{Current address: Department of Physics, Nanoscience Center, University of Jyv\"{a}skyl\"{a}, 40014 Jyv\"{a}skyl\"{a}, Finland}}

\address{National Centre for Nuclear Research, Pasteura 7, 02-093 Warsaw, Poland}

\ead{ilkka.makinen@ncbj.gov.pl}

\author{Hanno Sahlmann $^{1}$, Waleed Sherif $^{2}$\footnote{Author to whom any correspondence should be addressed}}

\address{Institute for Quantum Gravity, Department of Physics, Friedrich-Alexander-Universit\"{a}t Erlangen-N\"{u}rnberg (FAU), Staudtstraße 7, 91058 Erlangen, Germany}

\ead{$^1$ hanno.sahlmann@fau.de, $^2$ waleed.sherif@fau.de}

%
%

\begin{abstract}
We study the near-kernel sector of the Hamiltonian constraint operator in the one-vertex model of quantum reduced loop gravity using exact diagonalisation methods and variational Monte Carlo methods with neural quantum states. 
The analysis is based on the symmetric Hamiltonian containing both Euclidean and Lorentzian contributions, and on the variational minimization of the positive quadratic operator $\hat{Q}=\hat{C}^2$, where $\hat{C}$ is the symmetric Hamilton constraint, in truncated Hilbert spaces with spin cutoff up to $j_\text{max}=1001$. 
In the cases where exact diagonalization is possible, the eigenspace of lowest eigenvalue is one- or two-dimensional, and the eigenvalue, as well as the structure 
of the eigenstates depends on the cutoff. These ground states are entangled across the three edges.  
The variational method does not find the lowest eigenspace, but near kernel states, i.e., states with components in a number of low lying eigenspaces. For low cutoffs, they also do not factorize across the three edge degrees of freedom. At larger cutoffs, it increasingly finds near kernel states that factorize over the edges, matched with near-unit fidelity by reduced Thiemann coherent states. We discuss some indication that these states are the optimal near-kernel states if factorization is assumed.  
\end{abstract}

%
%

\section{\label{sec:introduction}Introduction}

Loop quantum gravity \cite{Rovelli:1997yv,Thiemann:2001gmi,Thiemann:2007pyv,Ashtekar:2004eh,Ashtekar:2021kfp} is a canonical and background-independent approach to quantum gravity in which quantum states of spatial geometry are described by spin networks and geometric operators acquire discrete spectra already at the kinematical level \cite{Ashtekar:1996eg,Rovelli:1994ge,Ashtekar:1997fb}. The central dynamical problem is then to construct and interpret physical states selected by the quantum constraints. Even if one simplifies the problem, by considering it on fixed graphs, and with a cutoff for the spins, this is highly nontrivial, since the constraint operators are complicated and the size of truncated Hilbert spaces grows rapidly with the allowed spin range and the size of the graph.

Quantum reduced loop gravity provides a framework in which this problem can be studied in a simpler setting, while maintaining a direct connection the full theory \cite{Alesci:2013xd,Alesci:2013xya,Alesci:2016gub,Makinen:2020rda,Makinen:2023shj}. 
In particular, QRLG is a sector of loop quantum gravity in which one can ask whether states selected by the quantum dynamics exhibit recognizable semiclassical properties. This question becomes especially concrete in the one-vertex model \cite{Makinen:2024rbg,Makinen:2026rof}, where the  graph is formed by a single six-valent vertex. In this model, the Hamiltonian constraint operator is still rather complicated, but tractable with numerical methods. In particular, it was found that the quantum dynamics of semiclassical states in the one-vertex model, when deparametrized by a scalar field, can reproduce a behavior closely resembling the semiclassical effective dynamics of a homogeneous and isotropic universe in loop quantum cosmology \cite{Makinen:2026rof}.

Recent work has shown that neural quantum states \cite{Carleo:2016svm} combined with variational Monte Carlo can address nontrivial fixed-graph constraint problems of loop quantum gravity type \cite{Sahlmann:2024pba,Sahlmann:2024kat,Sahlmann:2026qvs} allowing one to obtain candidate approximate near-kernel states by minimising the residual of quadratic constraints constructed from the Hamilton constraint. These developments open the possibility of using scalable numerical methods not only to approximate kernels of constraint operators, but also to extract physical information from the resulting states. In the present paper, the question is whether states in the near-kernel sector of the one-vertex model of quantum reduced loop gravity exhibit organization that can be interpreted semiclassically.

To address this question, we start from a symmetric Hamiltonian constraint $\hat C$ including both the Euclidean and Lorentzian parts, and study the positive operator 
\begin{equation}
    \hat{\mathcal{Q}} = \hat C  \hat C^\dagger = \hat C^\dagger  \hat C = \hat C^2,
\end{equation}
whose expectation value is the squared norm of the constraint action. Since the full one-vertex Hilbert space is infinite dimensional, the numerical analysis is performed in finite Hilbert spaces obtained by imposing a cutoff $j_\mathrm{max}$ on the representation labels. We study the near-kernel sector of the resulting operator for different cutoffs and the quality of the obtained states in terms of being candidate solutions for the full model under removal of such a cutoff.

We then examine the structure of the obtained states, in particular from the perspective of the so-called complexifier coherent states, which were introduced to loop quantum gravity by Thiemann, and which have long played a central role in semiclassical analyses of the theory \cite{Thiemann:2000bw,Thiemann:2000ca, Thiemann:2000bx,Thiemann:2002vj,Bahr:2007xa}. 
They are closely related to the mathematical notion of coherent states on compact Lie groups, first introduced and studied by Hall \cite{HALL1994103}. In this work, we examine whether such coherent states arise in the near-kernel sector without them being built into the variational ansätze. Namely, we ask whether they arise dynamically from the constraint minimisation and, if so, in what mathematically qualified sense they below to its near-kernel sector.

We investigate this problem in two complementary regimes. At small odd cutoffs from $j_\mathrm{max} = 5$ to 49, the low-lying spectrum of the finite quadratic constraint $\hat{\mathcal{Q}}$ can be determined by dense and sparse exact methods. Its ground space is found to be either a singlet or a doublet, as permitted by the permutation symmetry of the three edges, and is generally correlated. We show that the spectral gap, the ground space multiplicity and the one-edge spin content change non-trivially with the cutoff and hence such finite-matrix ground states, while they provide a precise reference at each cutoff, they do not identify a controlled sequence converging to a state in the kernel after cutoff removal.

We then use two neural quantum state ansätze and variational Monte Carlo to search for low-residual states at cutoffs up to $j_\mathrm{max} = 1001$. At the larger cutoffs, both ansätze yield states that factorize over the three edge Hilbert space to high accuracy, although they occupy different spin regimes. Further, one ansatz produces one-edge factors with high fidelity to reduced Thiemann coherent states, while the factors produced by another dense multilayer perceptron (MLP) remain concentrated near the lower spin boundary and are not equally well described by the same coherent state family. At low cutoff, correlated MLP solutions also occur, showing that factorization is not imposed by the variational machinery. Using such correlated states, we discuss how one can qualitatively build a notion of a quantum clock using such variational techniques.

The exact eigensystems allow the obtained variational states to be resolved spectrally, which shows that most of their norm, as well as the coherent product organisation, can be reconstructed from a small low-lying fraction of the finite-matrix spectrum. In the remainder of this work, therefore, we denote by \emph{near-kernel} such states which are supported on a small amount of low-lying eigenstates in the fixed cutoff setting. 

To test whether the coherent profile is induced by the neural parametrisation, and motivated by the fact that all obtained variational states are factorized states, we minimize the same quadratic constraint directly over the unrestricted product manifold $\ket{a} \otimes \ket{b} \otimes \ket{c}$ where we do not impose any coherent state shape nor equality between the independent normalized factors $a, b$ and $c$. The minima found from independent states reduce to permutation invariant products $\ket{\phi}^{\otimes 3}$ whose common factor has near-unit fidelity with a reduced Thiemann coherent state, supplying an independent variational sense in which the coherent profile is dynamically selected by the constraint within the product state manifold. Lastly, we also qualitatively compare whether the states can be candidate near-kernel states in the full model by canonically embedding them in larger cutoff spaces and  examining their sensitivity under such enlargement.

%
%

\section{\label{sec:qrlg}Quantum reduced loop gravity}

Quantum reduced loop gravity (QRLG) is a model of loop quantum gravity adapted to cuboidal graphs and diagonal triads aligned with three Cartesian fiducial directions \cite{Alesci:2013xd,Alesci:2013xya,Alesci:2016gub}. The motivation is twofold. On the one hand, one retains a direct link with the kinematics and operator framework of the full $\operatorname{SU}(2)$ theory. On the other hand, the resulting operators acquire a much simpler action on the reduced states, which makes explicit dynamical calculations possible. The reduced states are usually understood as quantum states satisfying the diagonal-triad gauge conditions approximately in the large-spin regime. This large-spin sector is also the regime in which one expects a semiclassical geometric interpretation to emerge. For that reason QRLG provides a natural setting in which one can ask whether solutions of the quantum constraints admit a recognizable semiclassical structure.

The Hilbert space of QRLG is not introduced as a construction unrelated to full loop quantum gravity, but rather as a sector selected from the kinematical Hilbert space of the full $\operatorname{SU}(2)$ theory by a suitable quantum implementation of the diagonal-triad gauge conditions \cite{Alesci:2013xya, Makinen:2023shj}. Moreover, for a wide class of loop quantum gravity operators, the action of the full $\operatorname{SU}(2)$ operator on a reduced state contains a dominant contribution that remains inside the reduced Hilbert space, together with subleading terms outside it. The reduced operator corresponding to a given full theory operator is then obtained by retaining only this leading contribution \cite{Makinen:2020rda}. Consequently, in this sense the reduced theory keeps the operator content of loop quantum gravity in a form that is tailored to the large-spin sector relevant for semiclassical questions and for numerical work.

In the present work we restrict our attention to the one-vertex model \cite{Makinen:2024rbg,Makinen:2026rof}, whose Hilbert space is formed by states defined on a graph consisting of three mutually orthogonal closed edges $e_x$, $e_y$, and $e_z$ meeting at a single six-valent vertex. The reduced spin network states on this graph take the form
\begin{equation}
    \ket{j_xj_yj_z}
    =
    \mathscr{D}^{(j_x)}_{j_xj_x}(h_{e_x})_x
    \mathscr{D}^{(j_y)}_{j_yj_y}(h_{e_y})_y
    \mathscr{D}^{(j_z)}_{j_zj_z}(h_{e_z})_z
    \label{eq:1v-state}
\end{equation}
where
\begin{equation}
    \mathscr{D}^{(j)}_{mn}(h)_a
    = \sqrt{2j+1} \, \prescript{}{a}{\bra{jm}}D^{(j)}(h)\ket{jn}_a
    \label{}
\end{equation}
denotes the normalized matrix elements of the $SU(2)$ Wigner matrices with respect to the eigenbasis of the operators $\hat J^2$ and $\hat J_a$ (with $a = x, y, z$).

In general, the reduced spin network states spanning the Hilbert space of QRLG are characterized by the property that the magnetic indices of each holonomy take the maximal or the minimal value with respect to the basis corresponding to the direction of the given edge. However, here we will consider exclusively the sector in which the quantum numbers $j_a$ in \eqref{eq:1v-state} take positive integer values; this sector forms a closed subspace under the action of the Hamiltonian constraint operator introduced below.

The elementary operators of the one-vertex model are reduced fluxes associated with the three coordinate directions and reduced holonomies associated with the three edges of the one-vertex graph. The action of the reduced flux operators $\hat p_a$ on the basis \eqref{eq:1v-state} is diagonal,
\begin{equation}
    \hat p_a\ket{j_xj_yj_z} = j_a\ket{j_xj_yj_z}.
    \label{eq:pa-action}
\end{equation}
We will also use the inverse flux operators $1/\hat p_a$, which are defined according to the Tikhonov regularization, i.e.
\begin{equation}
    \frac{1}{\hat p_a}\ket{j_xj_yj_z} = \begin{cases}
        j_a^{-1}\ket{j_xj_yj_z} & \text{if} \; j_a \neq 0 \\
        0 & \text{if} \; j_a = 0
    \end{cases}
    \label{eq:Tikhonov_reg}
\end{equation}
The reduced volume operator associated with the single vertex will be denoted by $\hat v$, and it acts as
\begin{equation}
    \hat v\ket{j_xj_yj_z} = \sqrt{j_xj_yj_z}\ket{j_xj_yj_z}.
    \label{eq:v}
\end{equation}

Lastly, the reduced holonomy operators act by shifting the quantum number carried by the corresponding edge. If the holonomy operator is expressed in the basis matching the direction of the edge, only the diagonal matrix elements of the operator have a non-vanishing action on the basis \eqref{eq:1v-state}. For example,
\begin{equation}
    {}^R\hat{D}^{(s)}_{mn}(h_{e_x})_x\ket{j_xj_yj_z}
    =
    \delta_{mn}\ket{j_x+m,j_y,j_z},
    \label{eq:DR}
\end{equation}
where $s$ is a label for the SU(2) irrep $D$. Analogous expressions hold for the $y$ and $z$ edges. The action of the reduced holonomies is formally identical to a $\mathrm{U}(1)$ multiplication law, with the shift determined by the magnetic number $m$. For the Hamiltonian it is convenient to introduce the symmetric combinations
\begin{equation}
    \hat c_a^{(m)}
    =
    \frac{1}{2}
    \Bigl(
    {}^R\hat{D}^{(s)}_{mm}(h_{e_a})_a
    +
    {}^R\hat{D}^{(s)}_{-m,-m}(h_{e_a})_a
    \Bigr)
    \label{eq:c_a}
\end{equation}
and
\begin{equation}
    \hat s_a^{(m)}
    =
    \frac{1}{2i}
    \Bigl(
    {}^R\hat{D}^{(s)}_{mm}(h_{e_a})_a
    -
    {}^R\hat{D}^{(s)}_{-m,-m}(h_{e_a})_a
    \Bigr).
    \label{eq:s_a}
\end{equation}
Their action is determined entirely by $m$. For instance,
\begin{align}
    \hat c_x^{(m)}\ket{j_xj_yj_z}
    &=
    \frac{1}{2}
    \Bigl(
    \ket{j_x+m,j_y,j_z}
    +
    \ket{j_x-m,j_y,j_z}
    \Bigr), \nonumber \\[1ex]
    \hat s_x^{(m)}\ket{j_xj_yj_z}
    &=
    \frac{1}{2i}
    \Bigl(
    \ket{j_x+m,j_y,j_z}
    -
    \ket{j_x-m,j_y,j_z}
    \Bigr),
    \label{eq:cs-action}
\end{align}
with corresponding formulas for $a=y$ and $a=z$.

The dynamics is based on the Hamiltonian constraint operator, which arises from the quantization of the Hamiltonian constraint on a fixed cubic graph. The resulting operator, including both the Euclidean and Lorentzian parts, is written as
\begin{equation}
    {}^R\hat C(N)
    =
    \frac{1}{\beta^2}{}^R\hat C_E(N)
    +
    \frac{1+\beta^2}{\beta^2}{}^R\hat C_L(N)
    \label{eq:C^R}
\end{equation}
where $\beta$ is the Barbero--Immirzi parameter. Throughout the numerical analysis, we set $\beta = 1$. For the Hamiltonian used here, the Lorentzian part corresponds to an operator representing the scalar curvature of the spatial manifold \cite{Lewandowski:2021iun,Lewandowski:2022xox}. On the one-vertex Hilbert space, the Euclidean and Lorentzian parts of the Hamiltonian constraint in a specific non-symmetric factor ordering take the form \cite{Makinen:2024rbg}
\begin{align}
    {}^R\hat C_E(N)
    &=
    -N(v)\Biggl[
    \hat s_x^{(1)}\hat s_y^{(1)}\sqrt{\frac{\hat p_x\hat p_y}{\hat p_z}}
    + \hat s_x^{(1)}\hat s_z^{(1)}\sqrt{\frac{\hat p_x\hat p_z}{\hat p_y}}
    + \hat s_y^{(1)}\hat s_z^{(1)}\sqrt{\frac{\hat p_y\hat p_z}{\hat p_x}}
    \Biggr]
    \label{eq:C_E^R}
\end{align}
and
\begin{align}
    {}^R\hat C_L(N)
    &=
    -16N(v)\Biggl[
    \bigl(\hat s_x^{(1/2)}\bigr)^4\frac{\hat p_x^{3/2}}{\sqrt{\hat p_y\hat p_z}}
    + \bigl(\hat s_y^{(1/2)}\bigr)^4\frac{\hat p_y^{3/2}}{\sqrt{\hat p_x\hat p_z}}
    + \bigl(\hat s_z^{(1/2)}\bigr)^4\frac{\hat p_z^{3/2}}{\sqrt{\hat p_x\hat p_y}}
    \Biggr].
    \label{eq:C_L^R}
\end{align}

For the variational approach adopted in this work we use the symmetric version of the Hamiltonian constraint,
\begin{equation}
    \hat{C}
    =
    \frac{1}{\beta^2}\hat{C}_E^{\rm sym}
    +
    \frac{1+\beta^2}{\beta^2}\hat{C}_L^{\rm sym}.
    \label{eq:C_sym}
\end{equation}
Following the presentation in \cite{Makinen:2026rof}, we introduce the auxiliary operator
\begin{equation}
    \hat X_a
    =
    \sqrt{\frac{\hat p_b\hat p_c}{\hat p_a}}
    =
    \frac{\hat v}{\hat p_a},
    \label{eq:X_a}
\end{equation}
with $(a,b,c)$ any permutation of $(x,y,z)$, in terms of which the Euclidean part is written as
\begin{align}
    \hat{C}_E^{\rm sym}
    &=
    -\hat X_z^{1/2}\hat s_x^{(1)}\hat s_y^{(1)}\hat X_z^{1/2}
    -\hat X_y^{1/2}\hat s_x^{(1)}\hat s_z^{(1)}\hat X_y^{1/2}
    -\hat X_x^{1/2}\hat s_y^{(1)}\hat s_z^{(1)}\hat X_x^{1/2}.
    \label{eq:C_E^sym}
\end{align}
For the Lorentzian part, introducing
\begin{equation}
    \hat Y_a
    =
    \frac{\hat p_a^{3/2}}{\sqrt{\hat p_b\hat p_c}}
    =
    \frac{\hat p_a^2}{\hat v},
    \label{eq:Y_a}
\end{equation}
we specify the symmetric ordering
\begin{equation}
    \hat{C}_L^{\rm sym}
    =
    -16\sum_{a=x,y,z}
    \hat Y_a^{1/4}
    \bigl(\hat s_a^{(1/2)}\bigr)^2
    \hat Y_a^{1/2}
    \bigl(\hat s_a^{(1/2)}\bigr)^2
    \hat Y_a^{1/4}.
    \label{eq:C_L^sym}
\end{equation}
The identity $\bigl(\hat s_a^{(1/2)}\bigr)^2 = \tfrac{1}{2}\bigl(\mathbb{I} - \hat c_a^{(1)}\bigr)$ shows that $\hat{C}_L^{\rm sym}$ shifts the spins in integer steps. With the specific factor ordering chosen in \eqref{eq:C_E^sym} and \eqref{eq:C_L^sym}, the operator \eqref{eq:C_sym} preserves the subspace spanned by the states \eqref{eq:1v-state} where all $j_a$ are positive integers, and therefore this subspace forms a closed sector for the numerical problem.

The numerical spin domain has two distinct boundaries. At its lower end, we work in the positive integer sector $j_a \geq 1$. The Tikhonov prescription in \eqref{eq:Tikhonov_reg} assigns zero to the inverse flux at $j_a = 0$ and together with the factor ordering above makes the positive integer sector closed under the constraint action. At its upper end we impose a hard truncation $j_a \leq j_\mathrm{max}$. Each elementary shift whose target lies above this boundary is set to zero and no periodic or cyclic identifications of the spin labels are used, as usually done in other work \cite{Sahlmann:2026qvs}. The symmetric expressions in \eqref{eq:C_E^sym} and \eqref{eq:C_L^sym} preserve Hermiticity with these boundary prescriptions and in particularly, numerically $\hat{C} = \hat{C}^\dagger$ for the operator used throughout this work.

Following the variational approach of \cite{Sahlmann:2026qvs}, we therefore define the positive quadratic constraint as
\begin{equation}
    \hat{\mathcal{Q}} = \hat{C}^\dagger \hat{C} = \hat{C} \hat{C}^\dagger = \hat{C}^2,
\end{equation}
and search variationally for states with small expectation value of $\hat{\mathcal{Q}}$ which give access to states close to the kernel of $\hat{C}$. In what follows, we investigate the exact spectrum of the constraint in low cutoffs and further obtain low-residual states for cutoffs up to $j_\mathrm{max} = 1001$ using two neural parameterisations in Section \ref{sec:results}. In Section \ref{sec:ed_vs_vmc} we then show that such variational states are near-kernel in the qualified sense of them being supported on a small low-lying spectral band and lastly, we investigate whether these near-kernel states admit any semiclassical organisation, and in particular such an organisation, if present, is compatible with a factorised product of reduced Thiemann coherent states.

%
%

\section{\label{sec:exact}Exact diagonalisation of the constraint at small cutoffs}

Before turning to the variational calculations at large cutoffs, we first examine the finite-dimensional constraint problem directly. At sufficiently small cutoffs, the dimension of the Hilbert space remains modest enough that its low constraint residual spectrum and eigenspaces can be determined by direct numerical spectral methods, without introducing a variational ansatz. This analysis serves several purposes. It reveals how the low-lying spectrum, its degeneracies and the structure of the lowest eigenspace depend on the cutoff. Further, it provides a reference against which the variational states can be compared, allowing one to distinguish properties intrinsic to the (cutoff-dependent) constraint from the structures selected by the variational parametrization. Lastly, it sheds light on whether the sequence of finite-cutoff problems admits a consistent relation to the theory obtained upon removal of the spin cutoff. 

For a cutoff $j_\mathrm{max}$, we denote by $\mathscr{H}_{j_\mathrm{max}}$ the space spanned by $\ket{j_xj_yj_z}$ with $j_a \in \{1, \ldots, j_\mathrm{max}\}$, which has $\dim\mathscr{H}_{j_\mathrm{max}} = j_\mathrm{max}^3$. The finite symmetric constraint matrix constructed on this space will be denoted by $\hat{C}_{j_\mathrm{max}}$ and its quadratic constraint by $\hat{\mathcal{Q}}_{j_\mathrm{max}} = \hat{C}_{j_\mathrm{max}}^2$ respectively. In this section, we will use the term \emph{ground state} to always refer to the eigenvector(s) associated to the lowest eigenspace (which we refer to as the ground space) of the finite matrix $\hat{\mathcal{Q}}_{j_\mathrm{max}}$.

%
%

\subsection{\label{subsec:exact_diag}Exact diagonalisation results}

The exact analysis below is conducted on a range of odd-valued cutoffs $j_\mathrm{max} \in \{5, 7, 9, \ldots, 49\}$ \footnote{The restriction to odd-valued cutoffs $j_\mathrm{max}$ is solely a numerical choice and does not follow from any physical requirement of the model. The analysis of the finite-matrix problem done throughout this work is equally well defined at even-valued cutoffs.}. At each cutoff, we determine the low-lying spectrum of the finite matrix constraint and retain the respective eigenspaces. For small enough cutoffs of $j_\mathrm{max} = 5, 7, 9$ and 11, exact diagonalisation is performed. Sparse iterative calculations were performed throughout the complete range of selected cutoffs. At the four cutoffs accessible to both methods, the iteratively obtained lowest eigenvalues of $\hat{\mathcal{Q}}_{j_\mathrm{max}}$ agree with the dense spectra to a maximum absolute difference of the order of $10^{-12}$, providing a direct check for the sparse method before applying it to larger cutoffs.

For the sparse calculations, an iterative eigensolver is applied to $(\hat{C}_{j_\mathrm{max}} - \sigma \mathbb{I})^{-1}$ rather than directly to $\hat{C}_{j_\mathrm{max}}$. For eigenvalues $\lambda$, transformation $\lambda \mapsto (\lambda - \sigma)^{-1}$ turns eigenvalues nearest the chosen shift $\sigma$ into extremal eigenvalues, allowing the corresponding modes to be extracted efficiently without forming the inverse explicitly, but only evaluating its action through sparse linear solves. To set the shift relative to the scale of each matrix, the maximum absolute row sum
\begin{equation}
    M_{j_\mathrm{max}} = \max_r \sum_s |(C_{j_\mathrm{max}})_{rs}|
\end{equation}
is used to determine the shifts $\sigma_\pm = \pm \eta_{j_\mathrm{max}} M_{j_\mathrm{max}}$, with $\eta_{j_\mathrm{max}} = 10^{-8}$. We use a smaller value of $\eta_{j_\mathrm{max}}$, when required, to recover the complete and mutually consistent set of eigenvalues nearest to zero. Since $\hat{C}_{j_\mathrm{max}}$ is Hermitian and $\hat{\mathcal{Q}}_{j_\mathrm{max}} = \hat{C}^2_{j_\mathrm{max}}$, the low-lying spectrum of $\hat{\mathcal{Q}}_{j_\mathrm{max}}$ can be obtained without explicitly forming the square. We therefore compute the eigenvalues of $\hat{C}_{j_\mathrm{max}}$ nearest zero, combine shift-invert solves on the two sides of the origin, perform a Rayleigh-Ritz diagonalisation in the combined candidate space, and finally denote the squared eigenvalues, in order of increasing magnitude, by $E_n$. We note that using both signs of the shift is required because the low-residual spectrum of $\hat{\mathcal{Q}}_{j_\mathrm{max}}$ receives contributions from both the positive and negative parts of the spectrum of $\hat{C}_{j_\mathrm{max}}$.

%
%

\subsection{\label{subsec:perm_symm}Low-lying spectrum and permutation symmetry}

The symmetric constraint is invariant under the permutations of the three edges. Namely, if $U_\pi$ denotes the unitary action of $\pi \in S_3$ obtained by permuting $(j_x, j_y, j_z)$, then $[\hat{C}_{j_\mathrm{max}}, U_\pi] = [\hat{\mathcal{Q}}_{j_\mathrm{max}}, U_\pi] = 0$. Each eigenspace of $\hat{\mathcal{Q}}_{j_\mathrm{max}}$ therefore carries a representation of $S_3$. Since the irreducible representations of $S_3$ have dimensions one (symmetric), one (anti-symmetric) and two (mixed), we refer to the one-dimensional eigenspaces as singlets and to the two-dimensional standard representation as a doublet. The observed doublets discussed below are therefore symmetry multiplets rather than unresolved pairs of different numerical eigenvectors. For every numerically identified eigenspace, the six permutations of $S_3$ map its basis vectors back into the same eigenspace, therefore reflecting the analytically expected symmetry. 

Across the cutoff ranges considered, the exact ground space $\mathcal{G}_{j_\mathrm{max}}$ is always either a singlet or a doublet and thus the ground dimension is $m_{j_\mathrm{max}} = \dim \mathcal{G}_{j_\mathrm{max}} \in \{1, 2\}$, and no higher ground space degeneracy is found. When $m_{j_\mathrm{max}} = 2$, the individual eigenvectors returned by the eigensolver are not unique, since any orthonormal basis of the doublet spans the same eigenspace. Therefore, when needed the projector onto $\mathcal{G}_{j_\mathrm{max}}$ is used when necessary. For example, the excitation gap (i.e. the gap between the lowest and first eigenspace) must consequently be defined above the complete ground multiplet. Ordering the eigenvalues with their multiplicities included, we denote such a quantity by
\begin{equation}
    \Delta_{j_\mathrm{max}} = E_{m_{j_\mathrm{max}}} - E_0,
\end{equation}
where $E_0$ is the eigenvalue of the lowest eigenspace. Hence, $\Delta_{j_{\mathrm{max}}} = E_1 - E_0$ for a singlet ground state, whereas $\Delta_{j_{\mathrm{max}}} = E_2 - E_0$ for a ground state doublet.

\begin{figure}
    \centering
    \includegraphics[width=0.85\linewidth]{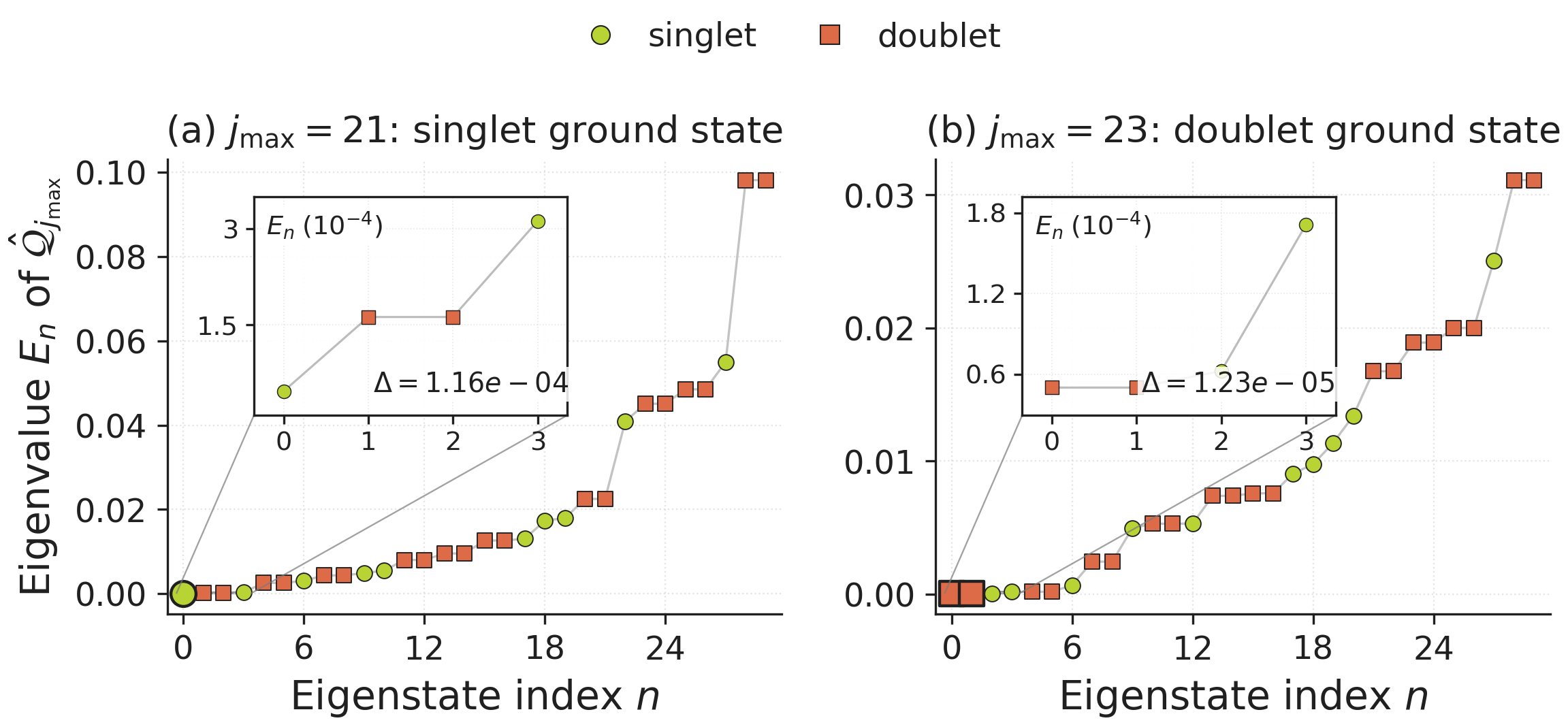}
    \caption{The lowest 30 eigenvalues of $\hat{\mathcal{Q}}_{j_\mathrm{max}}$ at two representative cutoffs $j_{\mathrm{max}} = 21$ and 23. Colour and marker shape distinguish singlet from doublet eigenspaces. At $j_{\mathrm{max}} = 21$ the ground state is a singlet whereas at $j_{\mathrm{max}} = 23$ it is a doublet. The insets resolve the spectral gap $\Delta_{j_{\mathrm{max}}}$ above the ground space.}
    \label{fig:exact_spectrum}
\end{figure}

Figure \ref{fig:exact_spectrum} illustrates the two possibilities. At $j_\mathrm{max} = 21$, the ground state is a singlet with $E_0 = 4.5808 \times 10^{-5}$ and $\Delta_{j_\mathrm{max}} = 1.1626\times 10^{-4}$. At $j_\mathrm{max} = 23$ the ground space is a doublet with $E_0 = 5.0027 \times 10^{-5}$ and $\Delta_{j_{\mathrm{max}}} = 1.2335 \times 10^{-5}$. Resolving the spectra of the finite constraints of each cutoff shows that both the ground eigenvalue and the gap reduce as the cutoff is increased.

\begin{figure}
    \centering
    \includegraphics[width=0.65\linewidth]{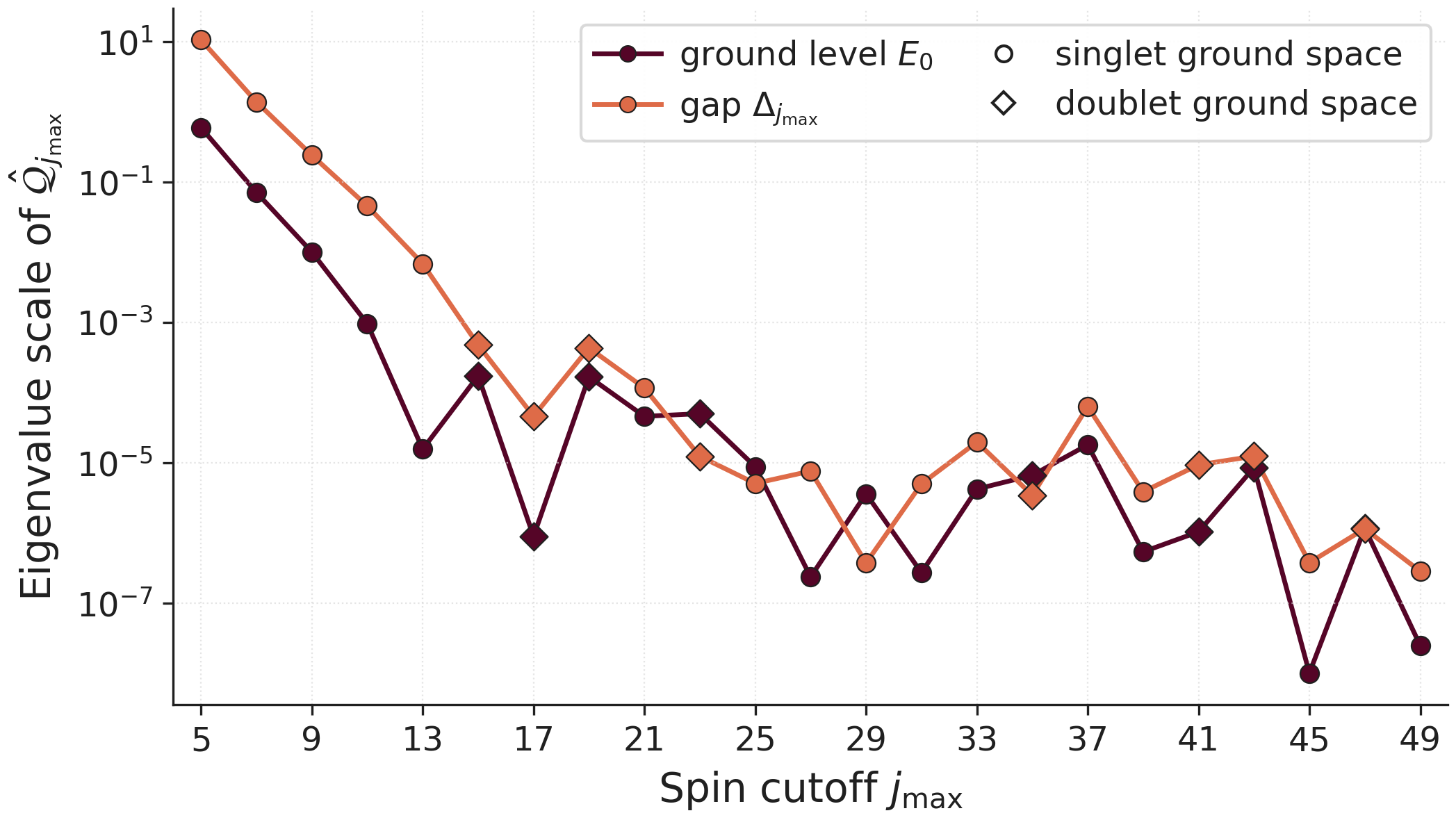}
    \caption{Lowest eigenvalue $E_0$ of $\hat{\mathcal{Q}}_{j_\mathrm{max}}$ and the multiplet-resolved spectral gap $\Delta_{j_{\mathrm{max}}}$ for every odd spin cutoff $j_\mathrm{max} \in \{5, \ldots, 49\}$. The markers distinguish singlet and doublet ground spaces respectively. The numerical uncertainty of each point is smaller than the marker size and is hence omitted.}
    \label{fig:exact_ground_and_gap}
\end{figure}

As shown in Figure \ref{fig:exact_ground_and_gap}, both $E_0$ and $\Delta_{j_{\mathrm{max}}}$ decrease overall by several orders of magnitude, but neither varies monotonically with the cutoff. The identity of the lowest symmetry sector also changes with the cutoff, and both singlet and doublet ground spaces occur over the range considered. The exact finite-matrix spectrum therefore has degeneracies fully explained by symmetry. No accidental degeneracies that would produce higher dimensional ground spaces are observed. The cutoff-dependence of the dimension of the ground space as well as the ground level and its gap from the remainder of the spectrum is complicated.

%
%

\subsection{\label{subsec:fact_ed}Correlations in the exact ground space}

Beyond the eigenvalues and their multiplicities, the finite matrix problem allows one to ask what correlation structure is selected by the constraint in this setting. To quantify how closely the exact finite-matrix ground space approaches the set of three-edge product states, and since the ground eigenspace can be degenerate, we therefore introduce the basis-independent fidelity
\begin{equation}
    F_\mathrm{prod}(\mathcal{G}_{j_\mathrm{max}}) = \max_{\|\phi_x\| = \|\phi_y\| = \|\phi_z\| = 1} \bra{\phi_x \otimes \phi_y \otimes \phi_z} P_{\mathcal{G}_{j_{\mathrm{max}}}} \ket{\phi_x \otimes \phi_y \otimes \phi_z},
    \label{eq:ed_fprod}
\end{equation}
where $P_{\mathcal{G}_{j_{\mathrm{max}}}}$ is the orthogonal projector onto the complete ground space. Here, $\ket{\phi_a} \in \mathscr{H}_{j_\mathrm{max}}^{(a)}$, with $a \in \{x, y, z\}$, are arbitrary normalized states of the corresponding one-edge Hilbert space. Thus $\ket{\phi_x \otimes \phi_y \otimes \phi_z}$ ranges over all normalized three-edge product states. The three factors are neither required to be identical nor assigned any particular functional form. For a non-degenerate ground state, \eqref{eq:ed_fprod} is its maximal squared overlap with a product state. For a degenerate ground space, it is equivalently the largest product fidelity attainable by any normalized vector in that space and is invariant under changes of basis within the multiplet. Considering some representative cutoffs, we find
\begin{center}
    \begin{tabular}{c|cccc}
         $j_\mathrm{max}$ & 11  & 21 & 31 & 41 \\ \hline
         $\dim \mathcal{G}_{j_\mathrm{max}}$ & 1 & 1 & 1 &2 \\ \hline
         $F_\mathrm{prod}(\mathcal{G}_{j_\mathrm{max}})$ & 0.9485 & 0.6468 & 0.3725 & 0.1939
    \end{tabular}
\end{center}
The decrease of $F_\mathrm{prod}$ from unity shows that the exact ground spaces considered generally do not exhibit approximate factorization across the three edges. For these representative cutoffs, the departure from product structure is particularly pronounced at $j_\mathrm{max} = 41$ where no normalized state in the complete ground doublet has more than 19.39\% fidelity with any three-edge product state. Correspondingly, the one-edge reduced density matrices are mixed rather than rank one. The exact finite-matrix problem therefore selects correlated, and in some cases strongly entangled, ground states rather than product states. We note, lastly, that these observations concern the spectral and correlation structure of the finite matrix problem. Their relation to certain variational low-residual states, as well as to the full one-vertex model, are examined below.

%
%

\section{\label{sec:results}Low-residual variational states}

Having investigated the finite-matrix constraint problem at small cutoffs, we now use variational Monte Carlo to search for low-residual states at substantially larger cutoffs where exact diagonalisation is no longer feasible. To fix the notation, we will denote by $\mathscr{H}$ the Hilbert space defined over the single-vertex graph whose basis elements are given by \eqref{eq:1v-state}. A reduced spin network function on $\mathscr{H}$ is then denoted by 
\begin{equation}
    \ket{\Psi} = \sum_{j_x, j_y, j_z} \psi_{j_x, j_y, j_z} \ket{j_xj_yj_z}.
    \label{eq:svsnf}
\end{equation}
To render the problem computationally tractable, we work with a truncated Hilbert space obtained by imposing an upper bound $j_{\mathrm{max}}$ on the allowed spins such that $j_x, j_y, j_z \in \{1, 2, \cdots, j_{\mathrm{max}}\}$ and search for low-residual variational states of the quadratic constraint $\hat{\mathcal{Q}}$ for increasing values of $j_{\mathrm{max}}$ up to 1001. The Hilbert space dimension $j_\text{max}^3$ thus ranges from $\sim 10^3$ for $j_\text{max} \sim 20$ to 
$\sim 10^9$ for $j_\text{max} \sim 1000$.

A priori, there is no reason to expect that low-residual states of $\hat{C}$ should exhibit any particular factorization or low-order structure in the spin variables. For this reason, we consider two qualitatively different variational ansätze with distinct inductive biases. The first ansatz is a dense multilayer perceptron (MLP), which represents the logarithmic wavefunction as a generic nonlinear function of $(j_x, j_y, j_z)$ and does not impose any explicit structure. In particular, it allows for entangled states in which all degrees of freedom are coupled. The second ansatz is a structured model in which the logarithmic wavefunction is decomposed into unary, pairwise, and three-body contributions, supplemented by a residual nonlinear term (see \ref{app:networks} and \ref{app:vmc_params} for details on the architecture and the simulation parameters). This architecture is biased towards states that admit a low-order interaction structure and, in particular, towards approximately separable solutions.

Despite their very different inductive biases, both ansätze converge to low residual states across the range of cutoffs considered, specifically $j_\mathrm{max} \in \{51, 101, 201, \cdots, 901, 1001\}$ for the structured ansatz and up to 801 for the MLP \footnote{Once again, the choice of odd-valued cutoffs in the variational analysis is simply a numerical choice and the analysis conducted throughout, as well as the results, are equally valid in even-valued cutoffs.}.

\begin{figure}[h]
    \centering
    \includegraphics[width=0.65\textwidth]{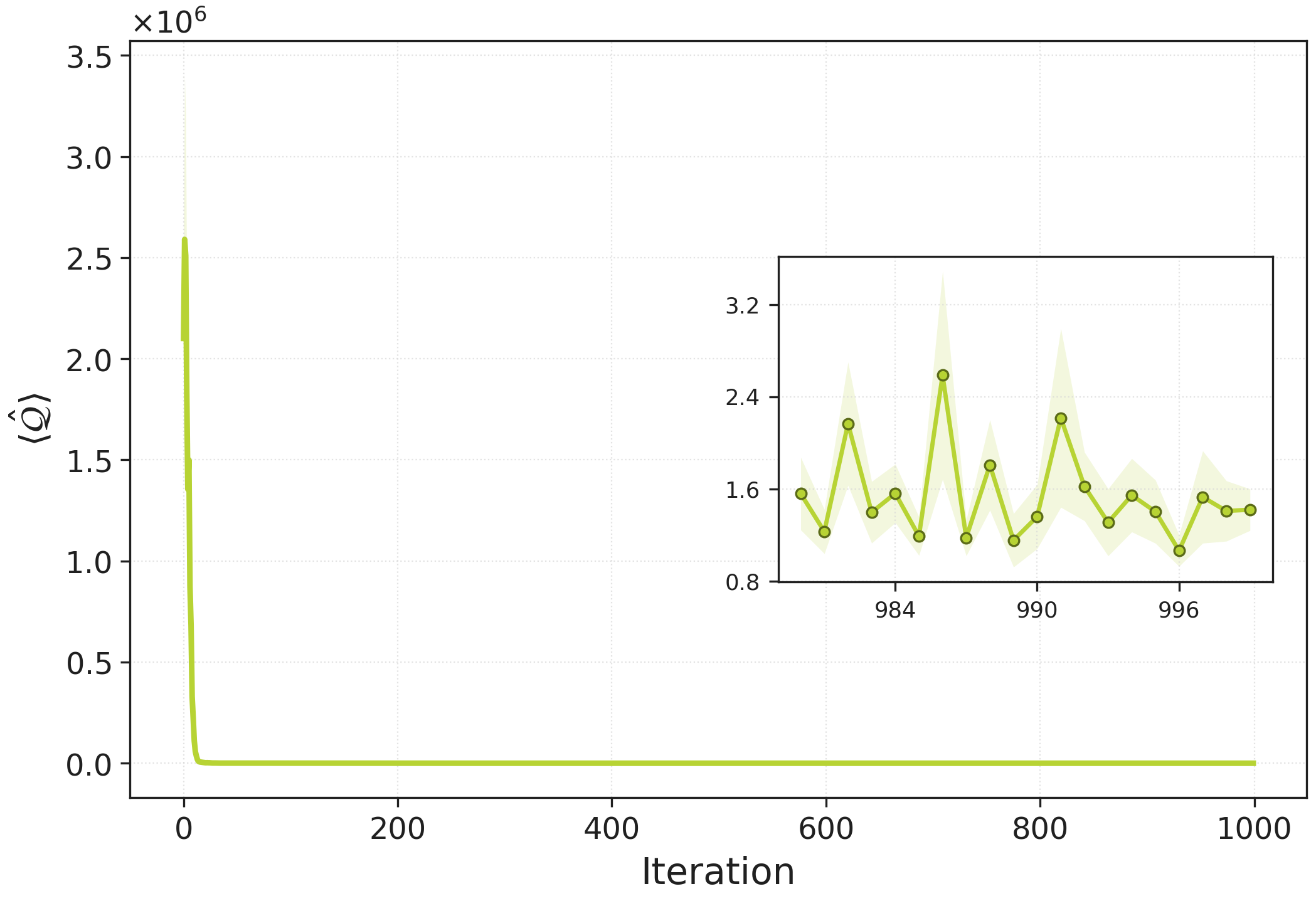}
    \caption{Representative convergence for the structured variational ansatz at $j_{\mathrm{max}}=1001$. The expectation value $\langle \hat{\mathcal{Q}}\rangle $ drops rapidly during optimization and stabilizes at a low-residual value. The inset magnifies the final iterations. The shaded region denotes the Monte Carlo error.}
    \label{fig:conv-1001}
\end{figure}

Figure~\ref{fig:conv-1001} shows a representative optimization curve for the structured ansatz at the largest cutoff considered, $j_{\mathrm{max}}=1001$. The expectation value $\expval{\hat{\mathcal{Q}}}$ drops rapidly by several orders of magnitude during the early stages of training and then settles into a stable late-time plateau of order unity. The inset resolves the final iterations and shows that the residual fluctuations are consistent with the Monte Carlo uncertainty.

\begin{figure}[h]
    \centering
    \includegraphics[width=0.65\textwidth]{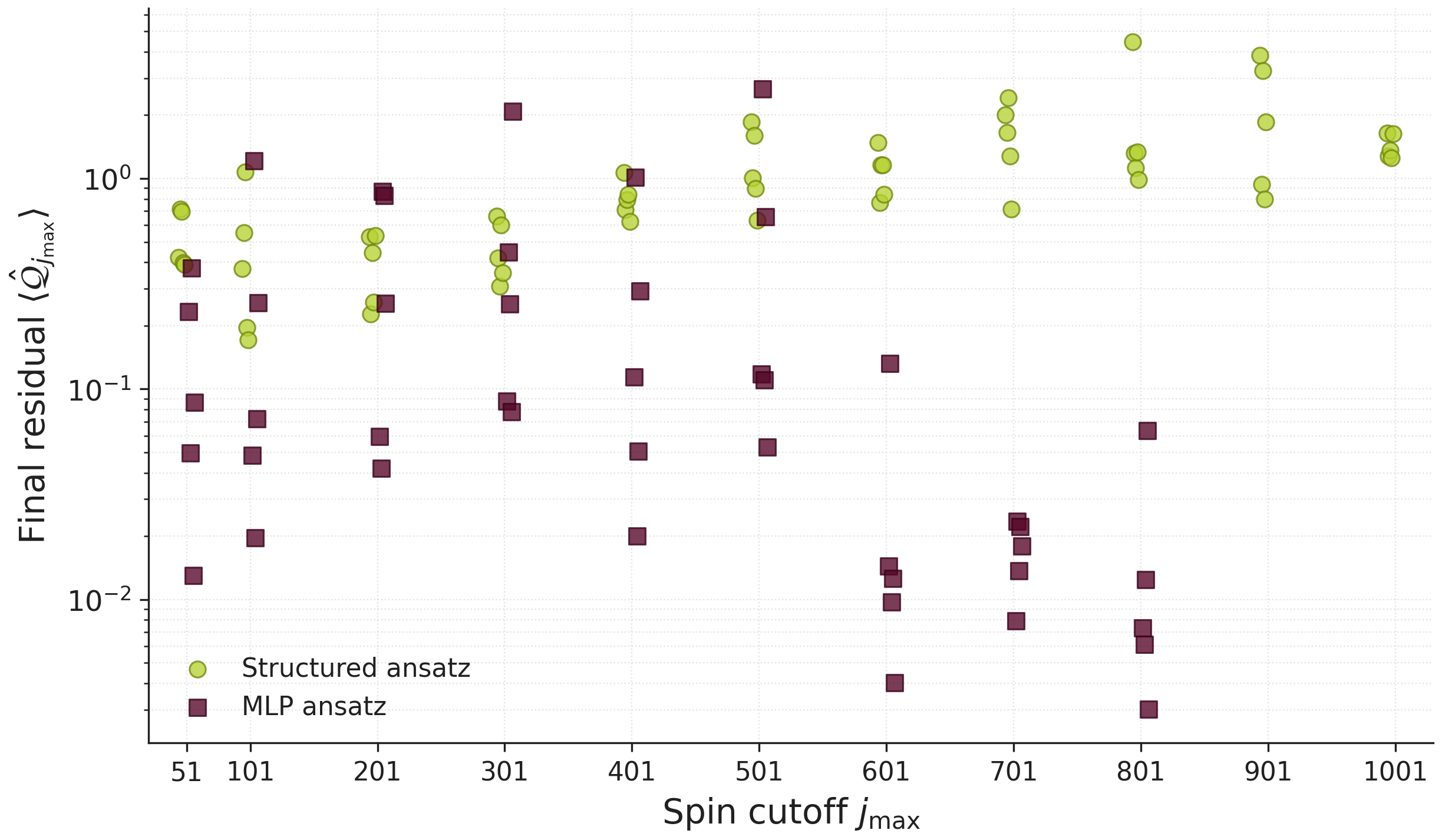}
    \caption{Distribution of the final quadratic constraint residuals over different seed simulations for different spin cutoffs. Each point represents one independently initialised simulation. Data points along one cutoff being shifted slightly along the x-axis is purely for illustration purposes.}
    \label{fig:vmc_residuals_scatter}
\end{figure}

Figure~\ref{fig:vmc_residuals_scatter} on the other hand shows a scatter plot of the final constraint residual for five simulations per cutoff for both the structured and MLP states for different cutoffs. Each simulation is initialised with a different random seed and thus a different random initialisation for both the weights and the sampling process. For a given cutoff, data points are shifted slightly along the x-axis for illustration purposes to allow for distinguishing different, possibly overlapped, data points. As shown in the figure, the MLP state regularly obtains a smaller final residual, while the structured ansatz maintains an $O(1)$ residual throughout with the worst residual being $4.463 \pm 0.452$ at $j_\mathrm{max} = 801$. As evident from the figure, the distributions exhibit a marked dependence on the initialisation and, at several cutoffs, span multiple orders of magnitude.

%
%

\subsection{\label{subsec:statefactorisationprobes}State factorization probes}

Having obtained converged low-residual variational states, their internal structure has to be examined. 

For a quick overview, we will be plotting some probability distributions that arise from the states. On the one hand, this will be the one-edge marginals for the flux, 
\begin{equation}
    P_x(j_x) = \langle\, \ket{j_x}\bra{j_x} \,\rangle_\Psi = \sum_{j_y,j_z} P(j_x,j_y,j_z),
    \label{eq:marginal}
\end{equation}
and similarly for $P_y(j_y)$ and $P_z(j_z)$,  
with $P(j_x,j_y,j_z)=|\Psi_{j_x,j_y,j_z}|^2$ the probability distribution of the state.
On the other hand we will show the probability distributions that are induced by the states on the spectrum of the operators $\hat{s}$ and $\hat{c}$, 
\begin{equation}
\label{eq:probabilities_sc}
    P^{(\hat{s})}_x (\lambda) =  \langle \hat{P}_{S,x}\rangle_\Psi = |\langle \lambda |  \Psi \rangle |^2, \quad \text{ with } \quad\hat{s}_x \ket{\lambda} = \lambda \ket{\lambda}
\end{equation}
and similar for $y,z$ and $\hat{c}$. 

An  important deeper question is whether the wavefunction approximately factorizes across the three edge degrees of freedom. In terms of the coefficients introduced in \eqref{eq:svsnf}, this amounts to asking whether they admit an approximation of the form
\begin{equation}
    \psi_{j_x,j_y,j_z} \approx \phi_x(j_x)\phi_y(j_y)\phi_z(j_z),
    \label{eq:factorized-coeffs}
\end{equation}
or, equivalently, whether the state itself can be written approximately as
\begin{equation}
    \ket{\Psi} \approx \ket{\phi_x}\otimes\ket{\phi_y}\otimes\ket{\phi_z},
    \qquad
    \ket{\phi_a}
    =
    \sum_{j_a=1}^{j_{\mathrm{max}}}\phi_a(j_a)\ket{j_a},
    \label{eq:factorized-state}
\end{equation}
where $a \in \{x,y,z\}$. To assess this possibility, we apply a sequence of complementary diagnostics of increasing sensitivity.

We begin with the probability distribution induced by the state, $P(j_x,j_y,j_z)=|\Psi_{j_x,j_y,j_z}|^2$, and its one-edge marginals \eqref{eq:marginal}.
For an exactly factorized state, the corresponding joint distribution must satisfy
\begin{equation}
    P(j_x,j_y,j_z)=P_x(j_x)P_y(j_y)P_z(j_z) =: P_{\mathrm{prod}} .
    \label{eq:prob-factorization}
\end{equation}
We therefore compare the full distribution with the product of its marginals by means of standard information-theoretic diagnostics. The first is the total variation distance,
\begin{equation}
\begin{aligned}
    D_{\mathrm{TV}}(P,P_{\mathrm{prod}})
    &=
    \frac{1}{2}
    \sum_{j_x,j_y,j_z}
    \Bigl|
        P(j_x,j_y,j_z)
        -
        P_{\mathrm{prod}}
    \Bigr| .
\end{aligned}
\label{eq:tv}
\end{equation}
The second considered probe is the total correlation,
\begin{equation}
\begin{aligned}
    T(P)
    = D_{\mathrm{KL}}(P\|P_{\mathrm{prod}})
    = \sum_{j_x,j_y,j_z}
    P(j_x,j_y,j_z)\,
    \log\!\left(
        \frac{P(j_x,j_y,j_z)}{P_{\mathrm{prod}}}
    \right) .
\end{aligned}
\label{eq:total-correlation}
\end{equation}
In addition, we evaluate the pairwise mutual informations
\begin{equation}
    I(a : b)
    = \sum_{j_a, j_b} P_{ab}(j_a, j_b)
    \log\frac{P_{ab}(j_a, j_b)}{P_a(j_a) P_b(j_b)}
    \label{eq:mutual-information}
\end{equation}
derived from the corresponding two-edge marginals $P_{xy}(j_x, j_y) = \sum_{j_z} P(j_x, j_y, j_z)$, etc.

While these quantities test statistical independence at the level of probabilities for the $\hat{p}_a$, they do not constrain correlations carried purely by the phases of the amplitudes. To probe entanglement at the level of the full quantum state, we next consider the one-edge reduced density matrices, i.e.
\begin{equation}
    \rho_x = \Tr_{yz}\ket{\Psi}\bra{\Psi}
    \label{eq:density-matrix}
\end{equation}
and similarly for $\rho_y$ and $\rho_z$. If $\ket{\Psi}$ is a product state, then each reduced density matrix is itself pure and thus its von Neumann entropy must vanish,
\begin{equation}
    S(\rho_a) = -\Tr\!\bigl(\rho_a \log \rho_a\bigr)=0,
    \qquad a \in \{x,y,z\}.
    \label{eq:entropy}
\end{equation}
Nonzero entropies therefore quantify the  entanglement between a given edge and the complementary two-edge subsystem.

Finally, we characterize factorization directly through the best rank-one product approximation to $\ket{\Psi}$. More precisely, we determine the normalized product state
\begin{equation}
    \ket{\Phi_{\mathrm{prod}}}
    =
    \ket{\phi_x}\otimes\ket{\phi_y}\otimes\ket{\phi_z}
    \label{eq:phiprod}
\end{equation}
that maximizes its overlap with $\ket{\Psi}$. This defines the product fidelity
\begin{equation}
    F_{\mathrm{prod}}
    =
    \max_{\substack{
        \ket{\phi_x},\ket{\phi_y},\ket{\phi_z} \\
        \|\phi_x\|=\|\phi_y\|=\|\phi_z\|=1
    }}
    \left|
        \braket{\phi_x\otimes\phi_y\otimes\phi_z}{\Psi}
    \right|^2,
    \label{eq:fprod}
\end{equation}
and the geometric measure of entanglement $E_G=1-F_{\mathrm{prod}}$ \cite{Wei:2003qfk,Weinbrenner:2025uwb} which provide a geometric measure of proximity of the variational state to a separable one, and a measure of multipartite entanglement, respectively. 

We end the discussion of probes of correlations with consideration of a quantity that is perhaps a bit closer to a physical interpretation. We use the momentum (or area) observable $j_a$ as a quantity relative to which another, $j_b$ can be measured. 
To this end, we introduce the projector onto the subspace with $j_a=j$,
\begin{equation}
    \hat{\Pi}_a^{(j)}\ket{j_xj_yj_z}
    =
    \delta_{j_a,j}\ket{j_xj_yj_z}.
    \label{eq:charge-projector}
\end{equation}
For a second edge $b \neq a$, we then consider the diagonal operator $\hat{\mathcal{O}}_{b|a}^{(j)} = \hat p_b \hat{\Pi}_a^{(j)}$ whose action is
\begin{equation}
    \hat{\mathcal{O}}_{b|a}^{(j)}\ket{j_xj_yj_z}
    =
    j_b\,\delta_{j_a,j}\ket{j_xj_yj_z}.
    \label{eq:conditional-operator-action}
\end{equation}
Its expectation value is therefore simply
\begin{equation}
    \expval{\hat{\mathcal{O}}_{b|a}^{(j)}}
    =
    \sum_{j_x,j_y,j_z}
    j_b\,\delta_{j_a,j}\,P(j_x,j_y,j_z),
    \label{eq:conditional-operator-exp}
\end{equation}
while $\expval{\hat{\Pi}_a^{(j)}} = P_a(j)$. Consequently, the ratio
\begin{equation}
    \mu_{b|a}(j)
    :=
    \frac{\expval{\hat{\mathcal{O}}_{b|a}^{(j)}}}{\expval{\hat{\Pi}_a^{(j)}}}
    =
    \frac{1}{P_a(j)}
    \sum_{j_x,j_y,j_z}
    j_b\,\delta_{j_a,j}\,P(j_x,j_y,j_z)
    \label{eq:conditional-mean}
\end{equation}
is simply the conditional mean of the $b$-edge spin at fixed $j_a=j$,
\begin{equation}
    \mu_{b|a}(j)
    =
    \sum_{j_b} j_b\, P(j_b \mid j_a=j).
    \label{eq:conditional-mean-prob}
\end{equation}
This observable provides a direct probe of inter-edge dependence. For a product state, or more generally if the probability distribution satisfies $P(j_x,j_y,j_z)=P_x(j_x)P_y(j_y)P_z(j_z)$, conditioning on $j_a=j$ does not affect the statistics of $j_b$. In that case one finds
\begin{equation}
    \mu_{b|a}(j)
    =
    \sum_{j_b} j_b\,P_b(j_b)
    =
    \expval{\hat p_b},
    \label{eq:conditional-mean-factorized}
\end{equation}
which is independent of $j$.

\subsection{\label{subsec:statefactorisation}Structure of low-residual variational states}

We now apply the diagnostics introduced in Sec.~\ref{subsec:statefactorisationprobes} to representative converged states produced by the two variational ansätze at cutoff $j_{\mathrm{max}}=501$. We note that the results are exclusive neither to this representative state nor this representative cutoff. 

\begin{figure}[htbp]
    \centering
    \begin{subfigure}{0.49\textwidth}
        \centering
        \includegraphics[width=\textwidth]{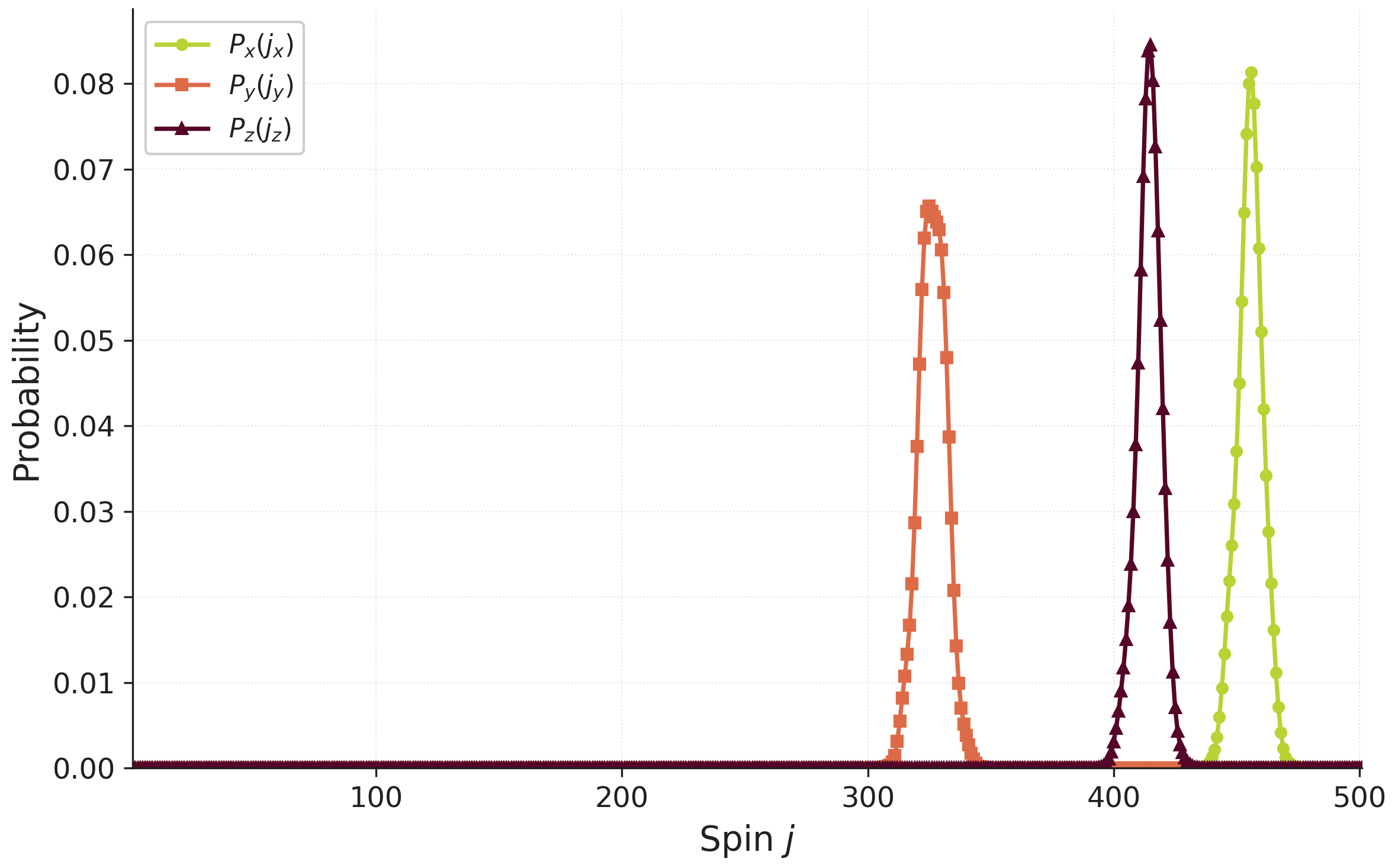}
        \caption{Structured ansatz.}
    \end{subfigure}
    \hfill
    \begin{subfigure}{0.49\textwidth}
        \centering
        \includegraphics[width=\textwidth]{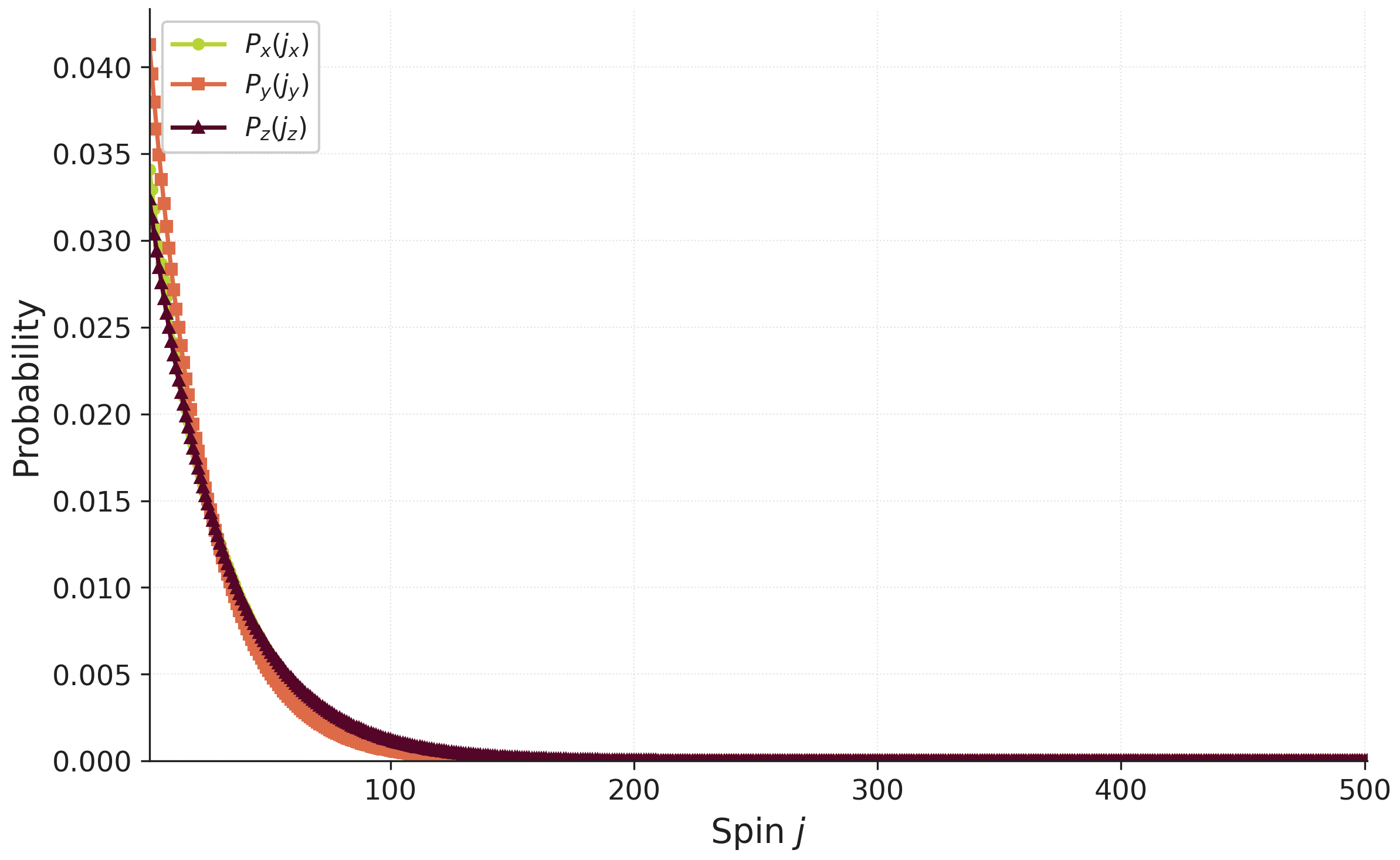}
        \caption{MLP ansatz.}
    \end{subfigure}
    \caption{Exact one-edge spin marginals at cutoff $j_{\mathrm{max}}=501$ for the two converged variational states. In both cases the three one-edge marginals are sharply localized, but the two ansätze select different spin sectors. The structured ansatz yields narrow peaks at comparatively large and edge-dependent spins, whereas the MLP solution is concentrated near the lowest admissible spin.}
    \label{fig:charge-marginals-comparison}
\end{figure}

As shown in Fig.~\ref{fig:charge-marginals-comparison}, the two converged states have markedly different one-edge spin marginals. The MLP solution is concentrated near the lower boundary of the positive-spin sector and decays rapidly with increasing spin. By contrast, the structured ansatz produces narrow peaks at substantially larger spins, with visibly different peak locations on the three edges. The two variational families therefore do not converge to the same local spin profile, even though both produce highly factorized low-residual states as shown below.

\begin{figure}[htbp]
    \centering
    \begin{subfigure}{0.49\textwidth}
        \centering
        \includegraphics[width=\textwidth]{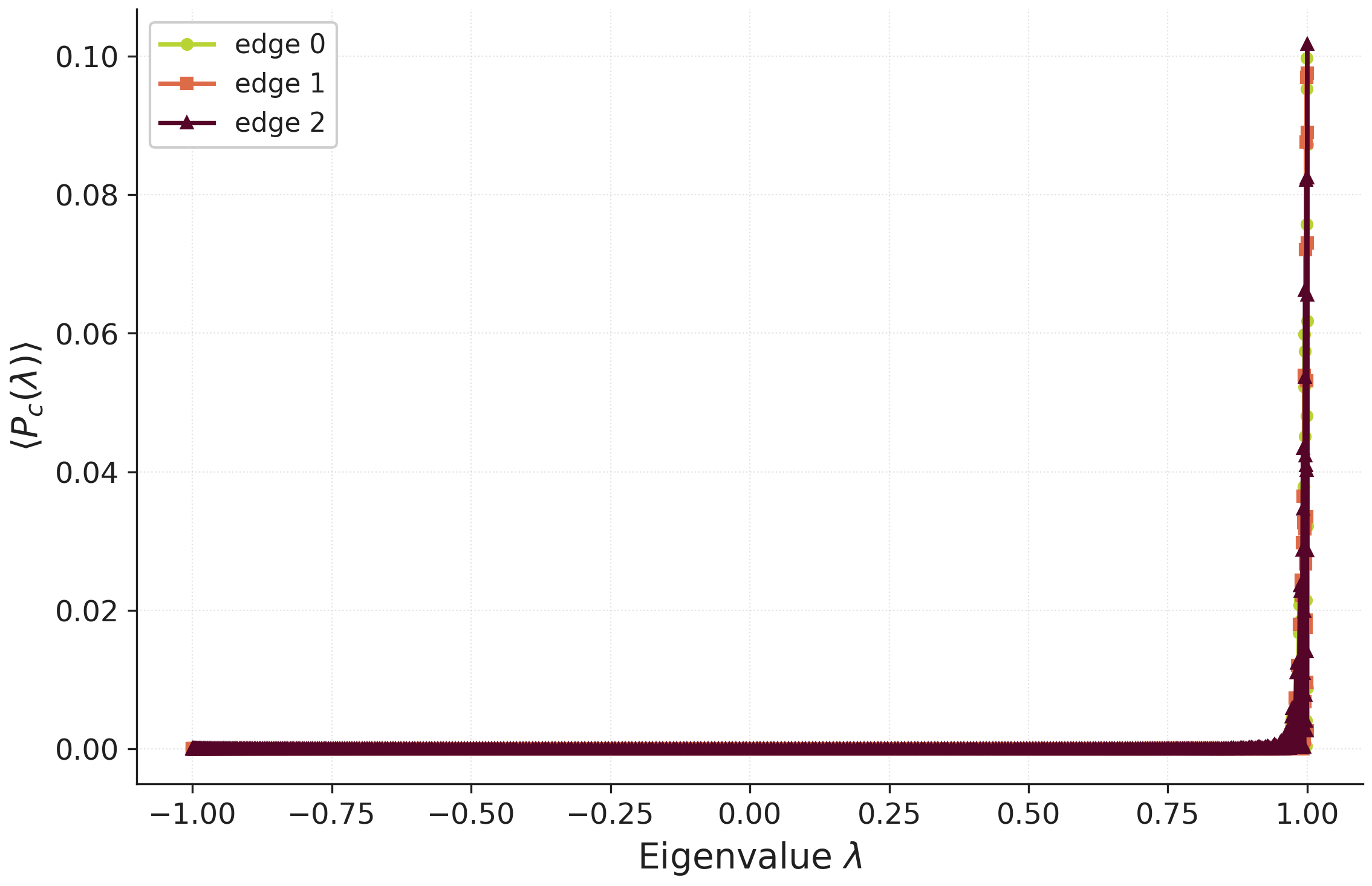}
        \caption{Structured ansatz, $\hat{c}$ profile.}
    \end{subfigure}
    \hfill
    \begin{subfigure}{0.49\textwidth}
        \centering
        \includegraphics[width=\textwidth]{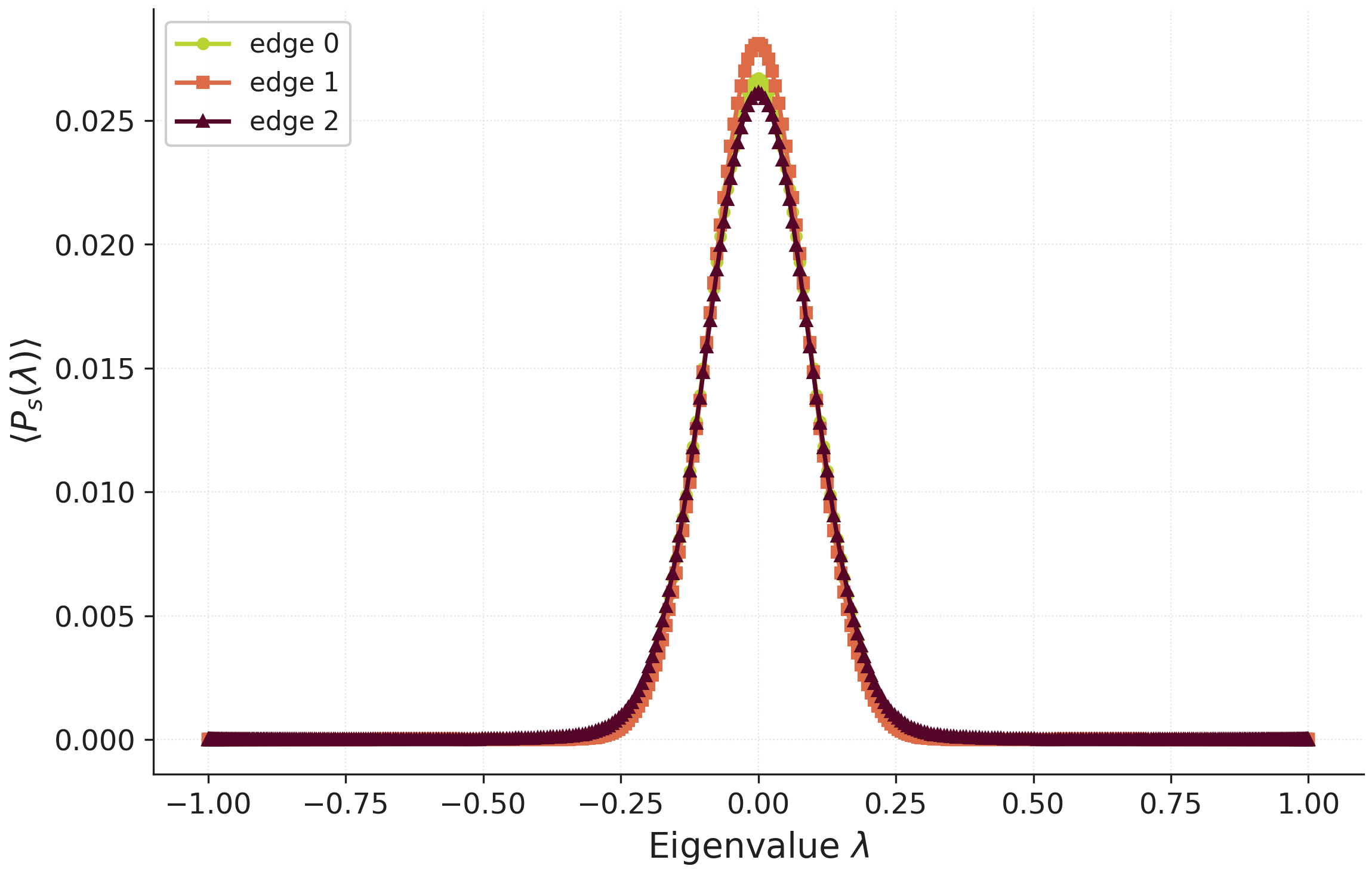}
        \caption{Structured ansatz, $\hat{s}$ profile.}
    \end{subfigure}

    \vspace{0.5em}

    \begin{subfigure}{0.49\textwidth}
        \centering
        \includegraphics[width=\textwidth]{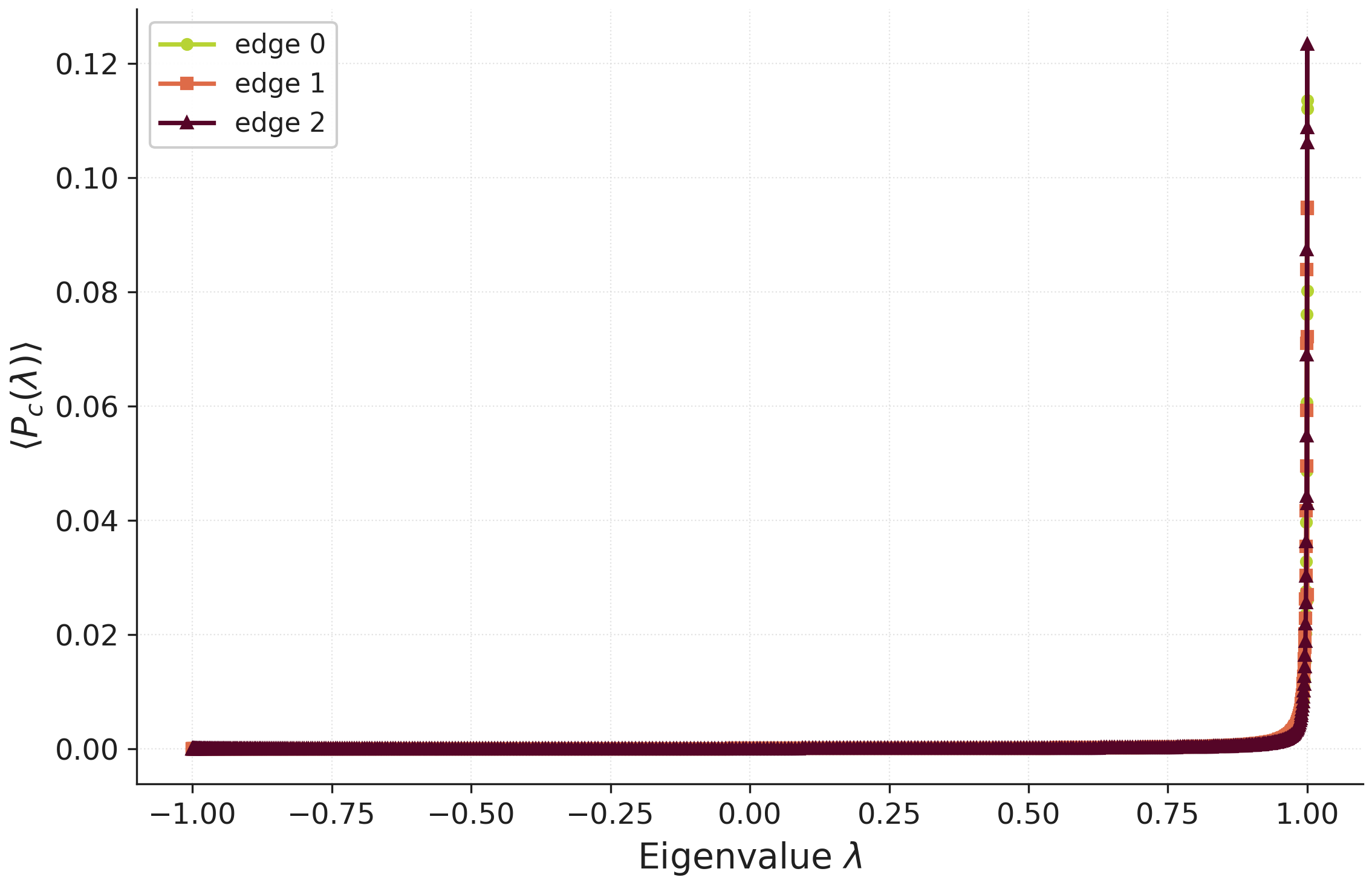}
        \caption{MLP ansatz, $\hat{c}$ profile.}
    \end{subfigure}
    \hfill
    \begin{subfigure}{0.49\textwidth}
        \centering
        \includegraphics[width=\textwidth]{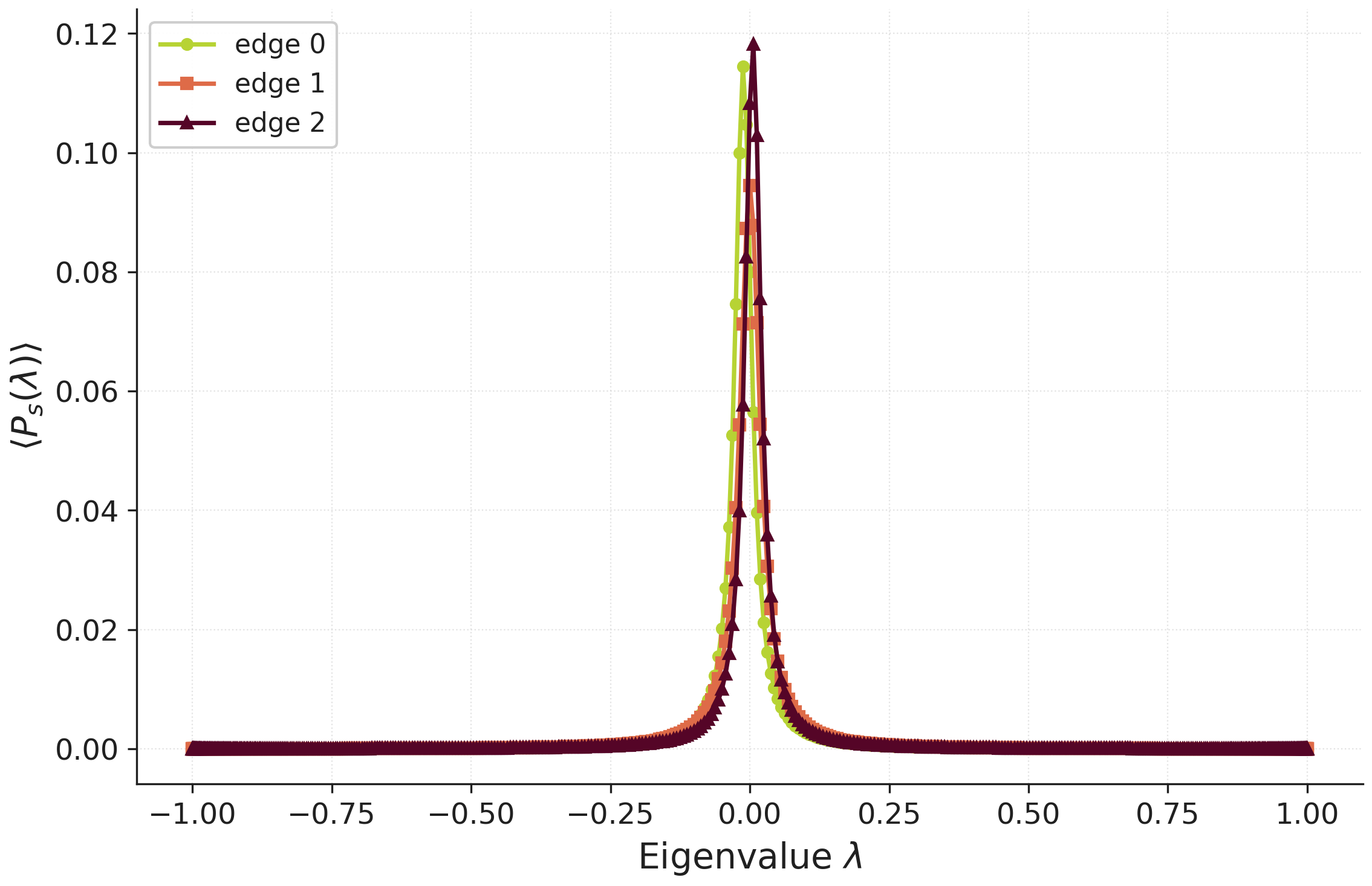}
        \caption{MLP ansatz, $\hat{s}$ profile.}
    \end{subfigure}
    \caption{One-edge spectral projector profiles for the two representative factorized low-residual states at cutoff $j_\mathrm{max} = 501$. The top row shows the structured ansatz and the bottom row shows the MLP ansatz. The left column displays the spectral weights associated with $\hat{c}_a^{(1)}$ while the right column shows the corresponding spectral weights for $\hat{s}_{a}^{(1)}$}
    \label{fig:holonomy_projectors_profiles}
\end{figure}

Figure \ref{fig:holonomy_projectors_profiles} complements the spin-space picture by resolving the same states in spectral bases adapted to the reduced holonomy. The MLP state is sharply concentrated near $\lambda=1$ in the $\hat{c}$ profile and near $\lambda=0$ in the $\hat{s}$ profile. The structured ansatz shows the same qualitative pattern although with broader support. Thus, despite the clear difference between the two states in the spin representation, both exhibit similar local holonomy content which corresponds to a small effective holonomy phase.

For the MLP solution, the factorization diagnostics gives a total variation distance from the product distribution 
$
D_{\mathrm{TV}} = 2.346 \times 10^{-3}
$,
total correlation of
$
T(P) = 4.328 \times 10^{-5}
$, as well as pairwise mutual informations of
$
I(x:y)=1.796\times10^{-5}
$,
$
I(x:z)=4.110\times10^{-6}
$,
and
$
I(y:z)=1.341\times10^{-5}
$.
The phase sensitive diagnostics lead to the same conclusion, namely the one-edge von Neumann entropies are computed to be
$
S(\rho_x)=3.168\times10^{-4}
$,
$
S(\rho_y)=3.521\times10^{-4}
$,
and
$
S(\rho_z)=3.197\times10^{-4}
$
and with a best rank-one product fidelity $F_{\mathrm{prod}}$ being almost at unity all together therefore show that the MLP state is a product state to numerical precision.

The structured state is likewise very close to a product, although its residual correlations are larger than those of the MLP state. In this case the total variation distance is
$
D_{\mathrm{TV}} = 1.575 \times 10^{-2}
$,
the total correlation is
$
T(P) = 1.240 \times 10^{-3}
$,
and the pairwise mutual informations are
$
I(x:y)=9.796\times10^{-4}
$,
$
I(x:z)=1.412\times10^{-4}
$,
and
$
I(y:z)=1.241\times10^{-4}
$.
The reduced-density-matrix entropies remain very small,
$
S(\rho_x)=4.766\times10^{-3}
$,
$
S(\rho_y)=4.724\times10^{-3}
$,
and
$
S(\rho_z)=1.354\times10^{-3}
$,
while the optimal product fidelity is
$
F_{\mathrm{prod}} = 0.999382
$.
Thus more than $99.93\%$ of the state norm is captured by a single product state even in this second, visibly different spin sector.

The overall conclusion is that both variational ansätze produce low-residual states that are almost exact product states across the three edges for higher $j_\mathrm{max}$. This feature is robust despite the substantial change of inductive bias as well as different cutoffs $j_\mathrm{max}$. At the same time, the resulting factorized states belong to different spin regimes. The MLP solution is localized near the smallest admissible spins, whereas the structured solution is supported at much larger and edge-dependent spins which differ both within cutoff and across different cutoffs.

\begin{figure}[h]
    \centering
    \begin{minipage}[t]{0.48\textwidth}
        \centering
        \includegraphics[width=\textwidth]{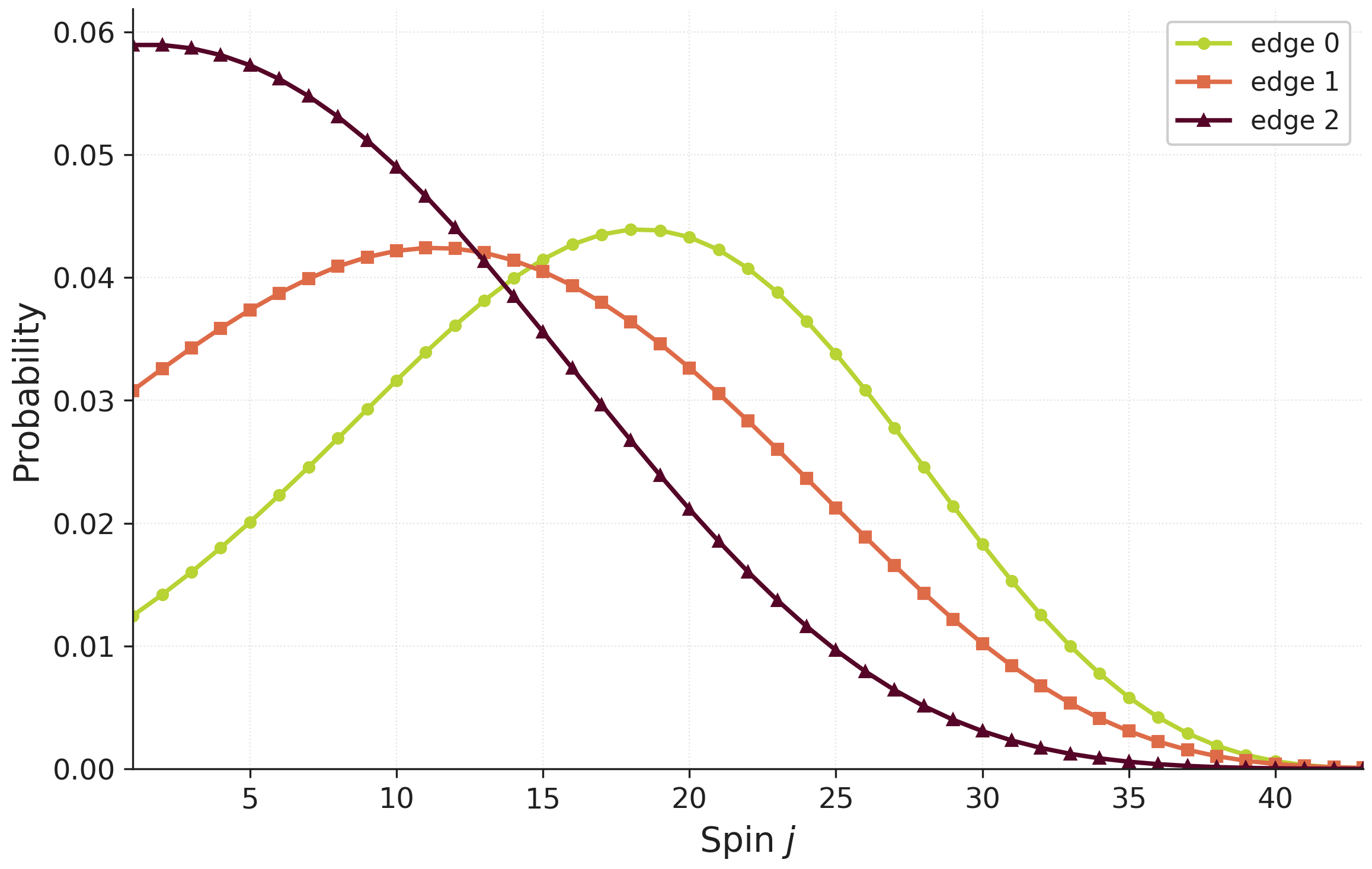}
    \end{minipage}
    \hfill
    \begin{minipage}[t]{0.48\textwidth}
        \centering
        \includegraphics[width=\textwidth]{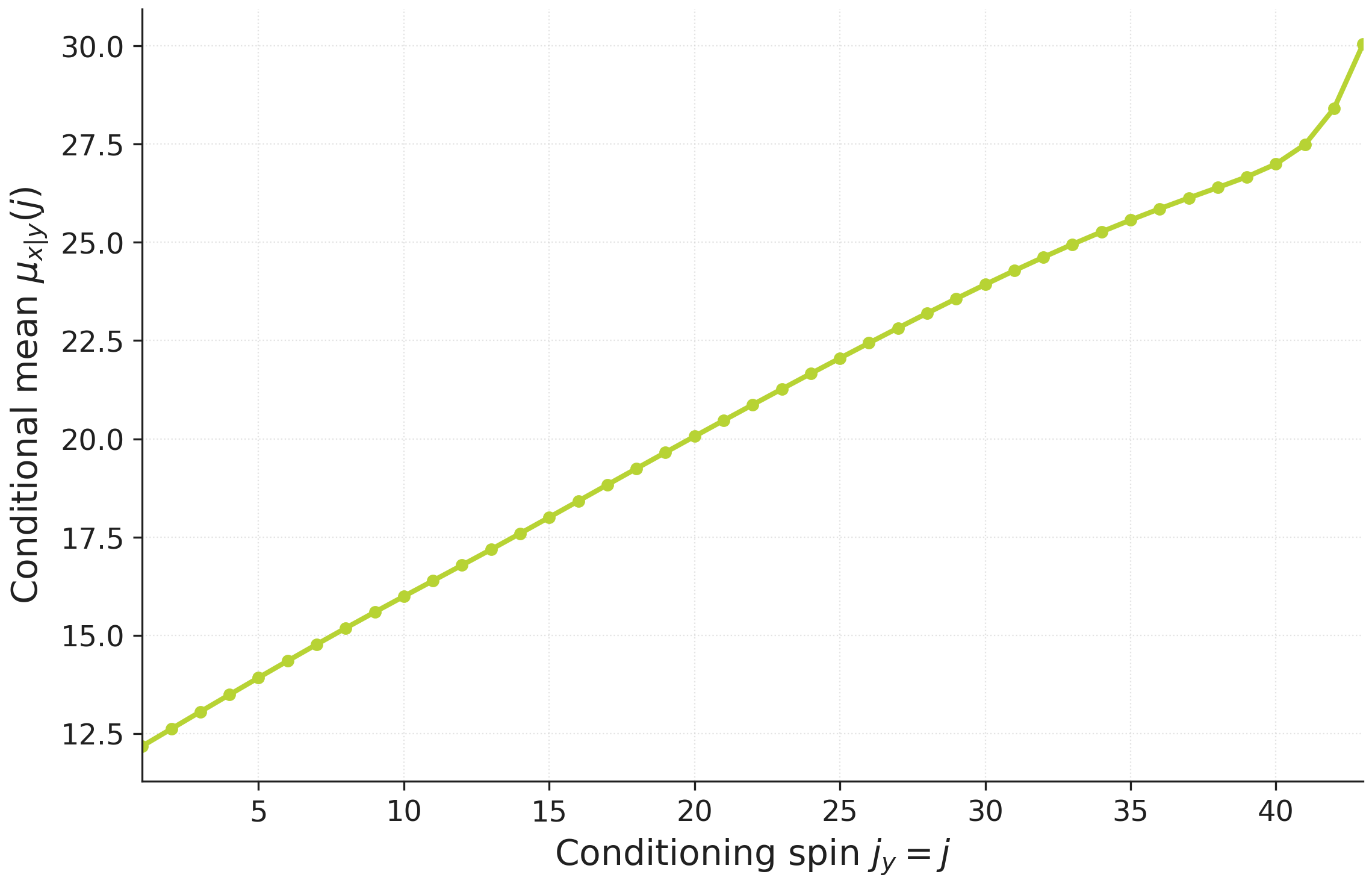}
    \end{minipage}
    \caption{One-edge marginals (left) and the conditional mean $\mu_{x|y}(j)$ (right) for a representative correlated MLP solution at cutoff $j_{\mathrm{max}}=43$, both computed using a full-state sum.}
    \label{fig:correlated-low-cutoff}
\end{figure}

Before ending this section, we should note however, that the MLP states for low cutoff have a slightly larger entangled component. We will consider the MLP state $j_{\mathrm{max}}=43$ which provides a representative example. It has  $F_{\mathrm{prod}}\simeq 0.9301$ and hence $E_G=1-F_{\mathrm{prod}}\simeq 6.99\times10^{-2}$, where $E_G$ is the geometric measure of entanglement as described in Section \ref{subsec:statefactorisationprobes}. This is markedly larger than that of the non-correlated (high-cutoff states), where $F_{\mathrm{prod}}$ can exceed $0.999$ (and therefore $E_G<10^{-3}$), as described above. 
The entanglement in the state can have interesting consequences. As an example, we consider the conditional expectation $\mu_{b|a}(j)$ defined in \eqref{eq:conditional-mean-prob} for $b=x$ and $a=y$ in the MLP state $j_{\mathrm{max}}=43$. 
The $j \mapsto \mu_{x|y}(j)$ curve shown for that state is shown on the right panel in Fig.~\ref{fig:correlated-low-cutoff}.

It clearly shows a non-zero, almost constant slope. In particular, for spins $j$ less than approximately 40, instead of being constant, it increases linearly across the supported range of spins and develops a higher slope above $j=40$. This behavior shows that the low-cutoff solution retains genuine inter-edge dependence. We note that in the present plot, the conditional mean is computed from a full-state sum and therefore no Monte Carlo estimates have been used.

It is tempting to interpret the  observed correlations between $p_x$ and $p_y$ as the system  expanding proportionately in $x$ and $y$ directions for the MLP solution we consider. We point out, however, that while $p_x$ and $p_y$ can be regarded as partial observables, $\mu_{x|y}(j)$ is not a complete observable, as it does not commute with the constraint. Physically speaking, $p_x$ can not be written as a function of $p_y$ alone under the present dynamics. There are situations, however, in which the gravitational momentum can be used as a gravitational clock, see for example  \cite{Bakhoda:2024mth}. 

For completeness, we also show the the one-edge marginals of the probability distribution in momentum space in \eqref{eq:marginal}, and in the $\hat{c}, \hat{s}$ representation.  The  left panel in Figure \ref{fig:correlated-low-cutoff} displays the former.  
It shows probability concentrated on lower spin configurations, thus limiting the influence of cutoff-effects. 
\begin{figure}[htbp]
    \centering
    \begin{subfigure}{0.49\textwidth}
        \centering
        \includegraphics[width=\textwidth]{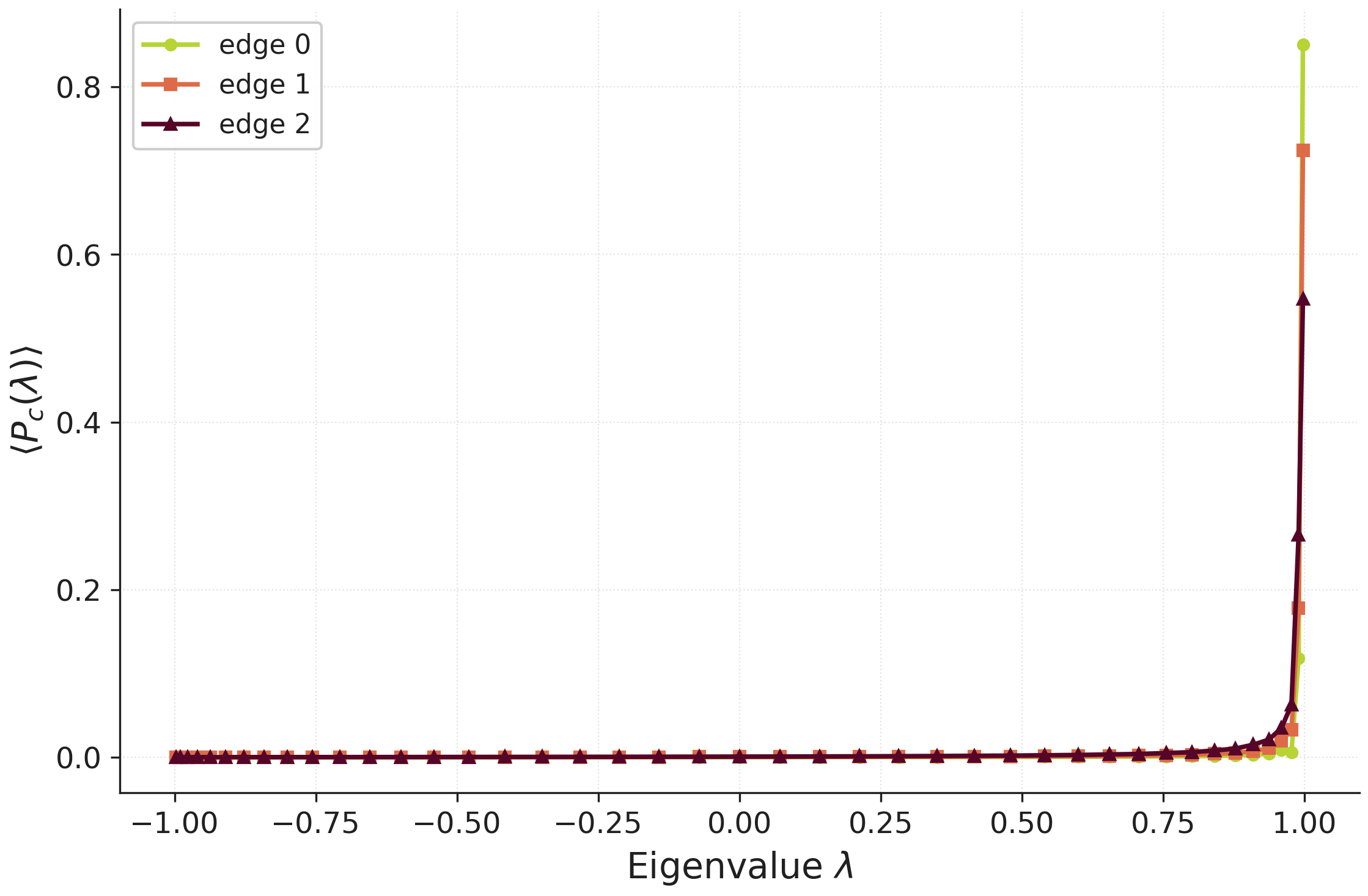}
        \caption{$\hat{c}$ profile.}
    \end{subfigure}
    \hfill
    \begin{subfigure}{0.49\textwidth}
        \centering
        \includegraphics[width=\textwidth]{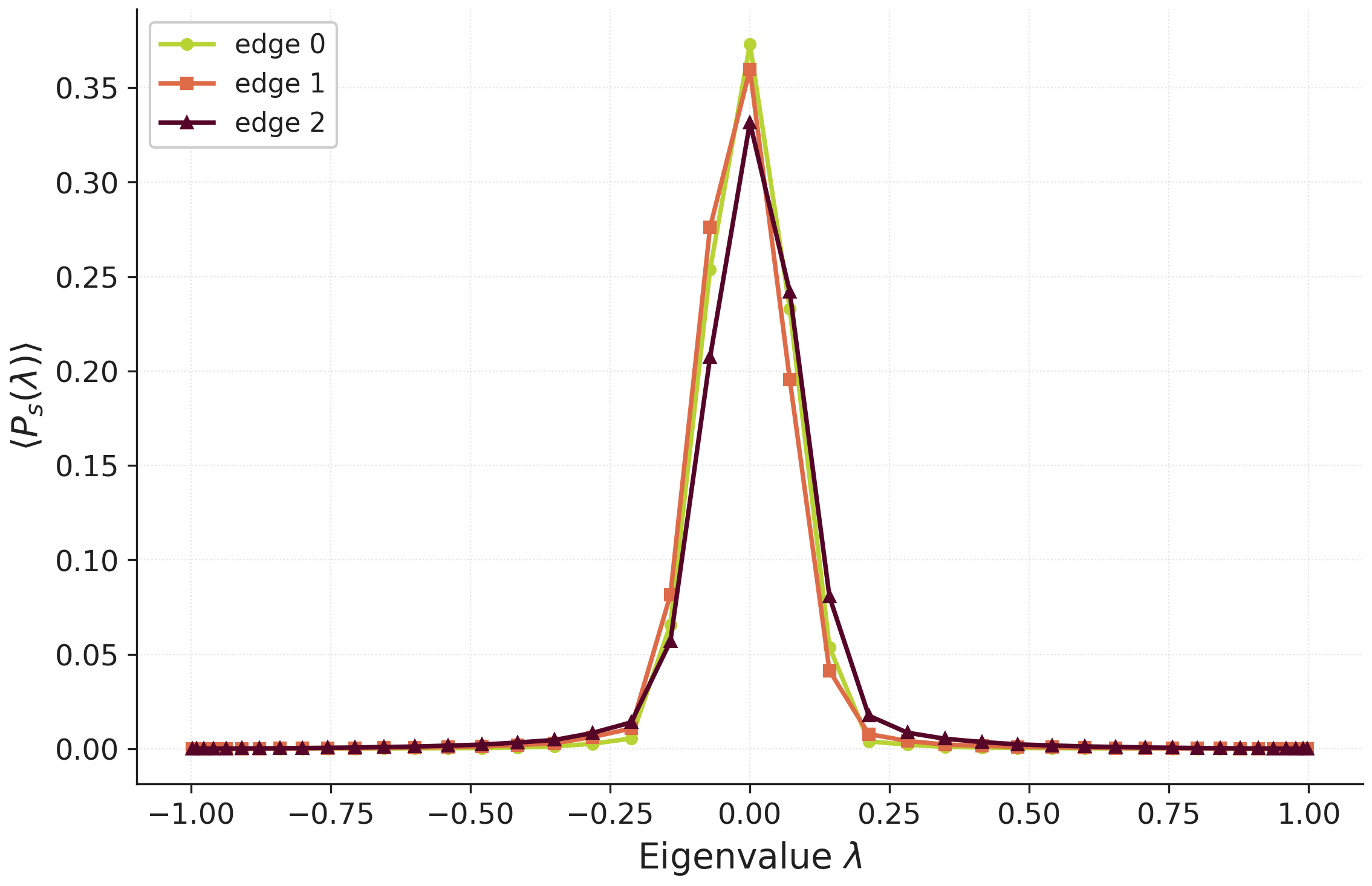}
        \caption{$\hat{s}$ profile.}
    \end{subfigure}
    \caption{One-edge spectral projector profiles for the same correlated MLP state at cutoff $j_\mathrm{max}= 43$ shown in Figure \ref{fig:correlated-low-cutoff}. The left panel shows the $\hat{c}$ spectral weights and the right panel shows the corresponding spectral weights for $\hat{s}$}
    \label{fig:holonomy_profile_corr}

\end{figure}
The probability distributions on $\hat{c}, \hat{s}$ displayed in Figure \ref{fig:holonomy_profile_corr} also show peakedness.

Altogether, the correlated low cutoff state still has relatively simple one-edge spectral distributions, but these local diagnostics no longer determine the structure of the full three edge state. The non-flat conditional mean in Figure \ref{fig:correlated-low-cutoff} shows that despite the localized one-edge profiles, the edges remain genuinely coupled.

%
%

\subsection{\label{subsec:tcs}One-edge factors as Thiemann coherent states}

Having established that the variational low-residual states are very close to product states, we now ask whether the resulting one-edge factors admit an interpretation in terms of reduced coherent states of the heat-kernel type. Thiemann and collaborators have developed the coherent states on compact Lie groups, in particular SU(2), introduced by Hall \cite{HALL1994103} as semiclassical states for loop quantum gravity \cite{Thiemann:2000bw,Thiemann:2000ca, Thiemann:2000bx,Thiemann:2002vj,Bahr:2007xa}. 
Such Thiemann coherent states are tensor products of Hall coherent states over edge Hilbert spaces, with the SL(2,$\mathbb{C}$) parameter of the states interpreted in terms of classical holonomy-flux data \cite{Thiemann:2000ca,Thiemann:2002vj}. 

Under projection onto the QRLG-sector, the Thiemann coherent states take the form of tensor products of heat kernel states over the group U(1) \cite{Alesci:2014uha}. The coefficient in the eigenbasis of momentum can be written 
\begin{equation}
    \chi_{t,p,\theta}(j)
    =
    \mathcal{N}(t,p,\theta)\,
    \sqrt{2j+1}\,
    \exp\!\left(
        -\frac{t}{2}j^2 + pj
    \right)
    e^{i\theta j},
    \label{eq:u1-thiemann-state}
\end{equation}
where, with the cutoff $j \in \{1,\dots,j_{\mathrm{max}}\}$, $\mathcal{N}(t,p,\theta)$ denotes the appropriate normalization. The parameter $t>0$ controls the width of the heat-kernel packet between the momentum and holonomy representation, $p$ together with $t$ fixes its location in spin space, and $\theta$ together with $t$ in holonomy space. We will refer to tensor products of such states over the edges $e_x,e_y,e_z$ as reduced Thiemann coherent states.

For each extracted factor $\phi_a$, we determine the best-fit coherent state by maximizing the one-edge fidelity
\begin{equation}
    F_a(t,p,\theta)
    =
    \left|
        \braket{\chi_{t,p,\theta}}{\phi_a}
    \right|^2
    =
    \left|
        \sum_{q=1}^{j_{\mathrm{max}}}
        \chi_{t,p,\theta}(j)^*\,\phi_a(j)
    \right|^2,
    \label{eq:one-edge-fidelity}
\end{equation}
with both states normalized on the same finite spin grid. The fidelity is evaluated as a deterministic sum over the complete one-edge grid, rather than estimated by Monte Carlo sampling. For the representative states below, we minimize $1 - F_a$ over $t \in [10^{-10}, 5], p \in [-50, 50]$ and $\theta \in [-\pi, \pi]$ using bounded L-BFGS-B with analytic derivatives. To guard against local optima, every representative factor is tested form 72 initial points, including moment-informed and Latin-hypercube starts and the result is independently checked by three differential evolution searches before the best fidelity is retained. Repeated best solutions agree with $10^{-9}$ in fidelity. Stability with respect to the VMC initialisation and cutoff is tested separately on five independently optimised structured states at each of the cutoffs. For this ensemble calculation, we use the wider ranges $t \in [10^{-10}, 10]$ and $p \in [-100, 100]$ with the same range for $\theta$ and perform five moment-informed and 16 Latin hypercube starts per factor.

Since the fidelity alone can obscure systematic shape mismatches, we also compare the corresponding spin distributions $|\phi_a(j)|^2$ and $|\chi_{t,p,\theta}(j)|^2$, together with their first two moments and effective linear phase slopes.

\begin{figure}[htbp]
    \centering
    \begin{minipage}[t]{0.48\textwidth}
        \centering
        \includegraphics[width=\textwidth]{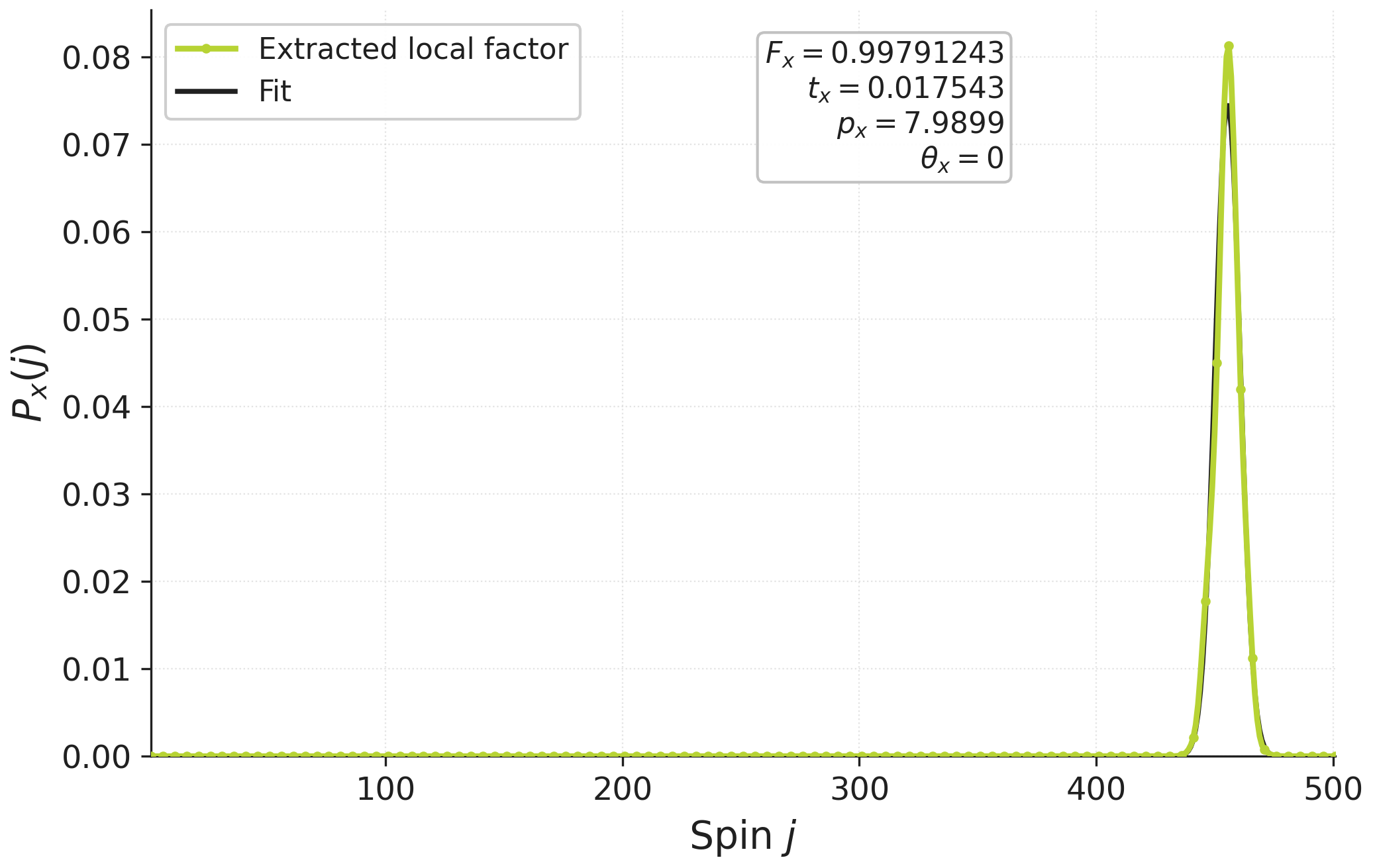}
    \end{minipage}
    \hfill
    \begin{minipage}[t]{0.48\textwidth}
        \centering
        \includegraphics[width=\textwidth]{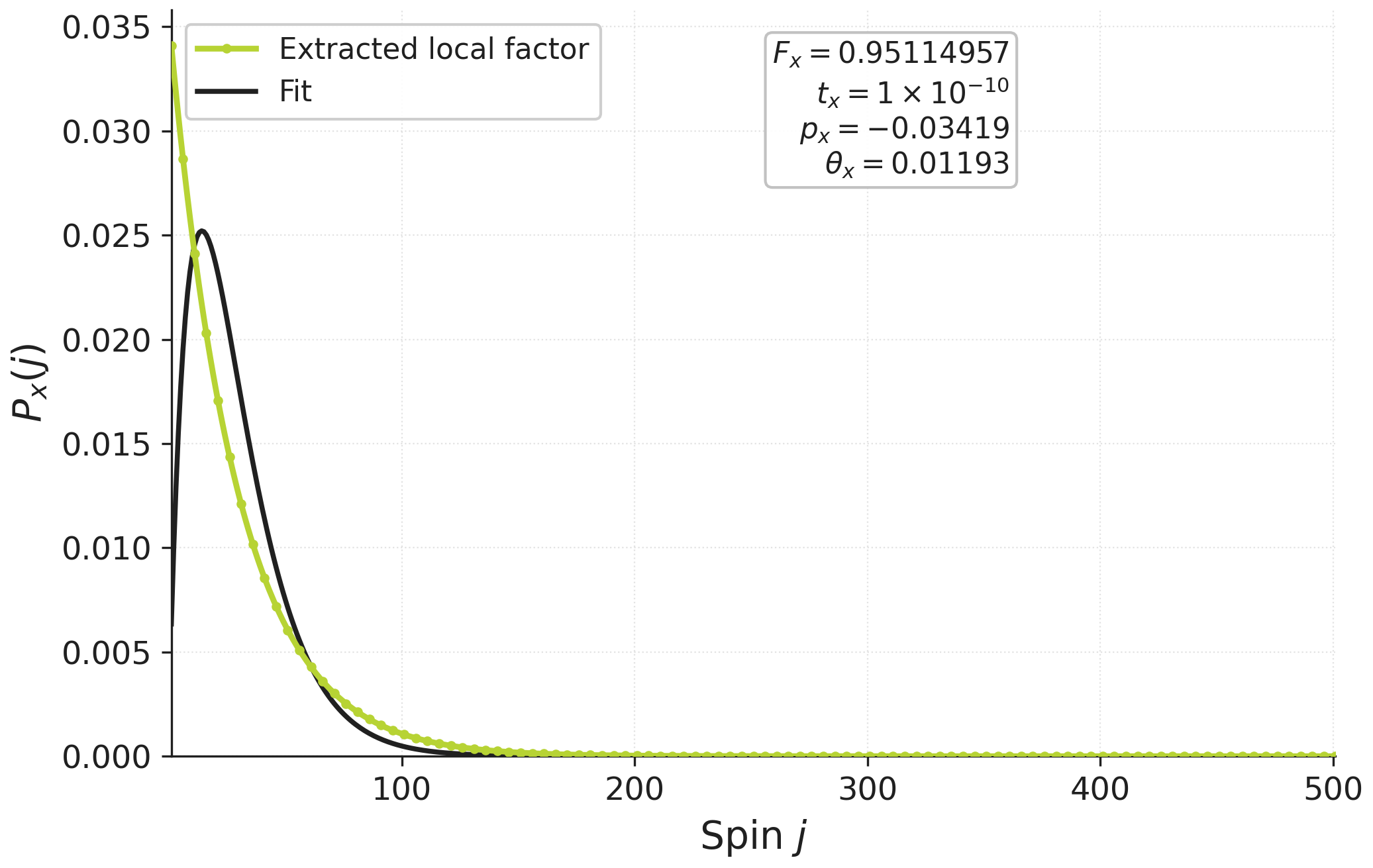}
    \end{minipage}
    \caption{Representative one-edge fits of the extracted local factors to the reduced Thiemann coherent-state family at cutoff $j_{\mathrm{max}}=501$. Left: structured ansatz, shown here for edge $x$. Right: MLP ansatz, again for edge $x$. In each case, the plot compares the one-edge probability distribution with the best-fit coherent-state profile.}
    \label{fig:tcs-fit-comparison}
\end{figure}

For the MLP state, the comparison reveals only a partial agreement with the coherent-state family. The fitted fidelities are $F_x = 0.951150, F_y = 0.953872$ and $F_z = 0.951536$ and the corresponding overlap moduli are $0.975269$, $0.976664$, and $0.975467$. At first sight these numbers might appear reasonably high. However, as shown in Fig.~\ref{fig:tcs-fit-comparison}, the fit exhibits a clear and systematic mismatch in the probability profile. The extracted MLP factors are boundary-dominated and decay monotonically from the smallest admissible spins, whereas the best-fit coherent states develop interior peaks and are visibly narrower. This is also reflected in the fitted moments, which reproduce the mean spins only approximately while underestimating the widths on every edge.

The fitted phase slopes do reproduce the extracted ones accurately, namely $\theta_x = 1.19 \times 10^{-2}, \theta_y = -1.39 \times 10^{-4}$ and $\theta_z = -5.35 \times 10^{-3}$. However, this agreement should not be over-interpreted. The amplitude profile is the more constraining part of the comparison, and there the fit is visibly imperfect. A further indication of this is that the fitted width parameter reaches the lower search boundary on all three edges,
\begin{equation}
    t_x = t_y = t_z = 10^{-10},
    \label{eq:mlp-tcs-t}
\end{equation}
which shows that the optimization is pushing the reference family towards the broadest state allowed by the fitting domain without achieving a genuinely good match. Repeating the fit while successfully lowering the bound on $t$ produces the same limiting behaviour. Lastly, the fitted values of $p$ are $p_x = -0.03419, p_y = -0.04144$ and $p_z = -0.03270$. We therefore conclude that the MLP factors are not genuinely compatible with the reduced Thiemann family. The fit captures their approximate location and phase gradient, but not their detailed boundary-dominated amplitude profile. We note that since the MLP factors are maximal at the smallest admissible spin and their shape is controlled by the hard lower boundary, they are not to be interpreted as semiclassical states.

The situation is markedly different for the structured ansatz. This is already visible in Fig.~\ref{fig:tcs-fit-comparison}, where the fitted coherent state tracks both the amplitude profile and the phase behavior of the extracted factor with high accuracy. In this case the one-edge factors are fitted with genuinely high accuracy by the same coherent-state family. The fitted fidelities are $F_x = 0.997912, F_y = 0.998449$ and $F_z = 0.996353$ with the corresponding overlap moduli being $0.998956$, $0.999224$, and $0.998175$. The fitted parameters are
\begin{align}
\begin{split}
    (t_x,p_x,\theta_x)
    &=
    (0.0175434,\; 7.98990,\; 0), \\
    (t_y,p_y,\theta_y)
    &=
    (0.0156748,\; 5.11379,\; 0), \\
    (t_z,p_z,\theta_z)
    &=
    (0.0191079,\; 7.91532,\; 0).
\end{split}
\label{eq:structured-tcs-params}
\end{align}
Here the agreement extends well beyond the fidelities themselves. The fitted coherent states reproduce the mean spins and widths of the extracted factors with very high accuracy on every edge. In addition, the fitted phase slopes are essentially zero, in agreement with the extracted factors, so the corresponding local states are almost phase-flat over the region carrying significant support. In contrast with the MLP case, the structured fits are therefore quantitatively convincing at the level of both amplitude and phase, providing strong evidence for an emergent Thiemann coherent state organization of the factorized low-residual state. Repeating the same post-hoc coherent state fit for all 55 independently optimized structured states shown in Figure \ref{fig:vmc_residuals_scatter} gives an ensemble-averaged worst-edge fidelity of $\overline{F_\mathrm{Th}^\mathrm{worst}} = 0.9859 \pm 0.0365$, where the worst edge fidelity being defined as $F_\mathrm{Th}^\mathrm{worst} = \min_{a \in \{x, y, z\}} F_a$, and the quoted spread is the sample standard deviation across the independently optimised states rather than an uncertainty of the deterministic fit.

It is interesting to note that the fitted $t$-parameters in \eqref{eq:structured-tcs-params} fall in a similar range, and so do two of the three $p$-parameters. We observe a similar pattern in other solutions found with the  structured ansatz. It appears more consistently, the higher $j_\text{max}$. Above
$j_\text{max}=300$, the $t$-parameters are typically within 10\% - 20\% of each other, and the $p$-parameters of two edges are typically  within 0.5\% - 7\% of each other, whereas the third is much farther away.
We will see in section \ref{subsec:hartree_min} that the optimal \emph{product state} appears to have three coherent factors with similar parameters. The third factor  being different for the states considered here seems thus to be an artifact of imperfect optimization.

We note that nothing in the variational procedure instructs the state to resemble a Thiemann coherent state. The optimization has access only to the quadratic constraint, and the coherent-state family analysis is introduced only after the simulations are completed as an external diagnostic. Moreover, the extent to which such states provide an adequate semiclassical description has itself been the subject of discussion \cite{Thiemann:2000bw,Thiemann:2000ca,Thiemann:2000bx}. From this perspective, it is highly nontrivial that the factorized low-residual state produced by the structured ansatz is matched so accurately by the reduced Thiemann family. While this result is suggestive that the coherent state structure is not an artifact of the neural parametrisation, but a feature of the low-residual sector, we will study this question further in Section \ref{subsec:hartree_min}.

%
%

\section{\label{sec:ed_vs_vmc}Comparison of exact and variational low-residual states}

The exact and variational calculations reveal an apparent tension. Namely, the exact finite-matrix ground spaces are generally correlated, while the variational counterpart is factorized, with one neural architecture class producing states with one-edge factors being accurately described by reduced Thiemann coherent states. Given the different properties, the variational low-residual states at a given cutoff must contain contributions from more than just the ground space of the problem at that cut-off. Moreover, from the results so far, it appears that properties of both, the ground states and the variational states change with the cutoff. This raises a number of questions, in particular: 
\begin{itemize}
    \item How broad is the support of the VMC low-residual states over the spectrum of $\hat{\mathcal{Q}}_{j_\mathrm{max}}$? 
    \item Do the VMC low-residual states have physical interest, or should they be dismissed as un-physical?
    \item More generally, what can the ground states and variational low-residual states tell about the physical states of the theory without cutoff? 
    \item And finally: Is the appearance of Thiemann coherent states connected to the physical system under study, or is it an artifact of the ansätze for the variational states? 
\end{itemize}
We therefore compare the exact and variational states in three complementary manners in what follows. 
First, we compute the components of the structured variational state in the lowest eigenspaces of the finite constraint, for those cutoffs accessible to exact spectral methods. This clarifies where the variational states lie in the spectrum of the quadratic constraint. Next, we minimize once again the quadratic constraint at various cutoffs,  however now over general product states. We show that the resulting low residual states are Thiemann coherent states. Hence they seem to be linked to the constraint in a precise sense, and their appearance as variational low residual states is due to an apparent tendency of the variational procedure to find product states. Lastly, every obtained variational state and the respective exact ground states in the given range of cutoffs are embedded into the Hilbert spaces corresponding to successively higher cutoffs. Using this, we can evaluate the constraint residual of a state at a given cutoff $j_\mathrm{max}$ with respect to constraints at higher cutoffs. This tests whether their small residuals at $j_\mathrm{max}$ rely on cancellations created by the boundary of the truncated spin, i.e., cutoff effects, or whether they constitute near-kernel states of the full model.

%
%

\subsection{\label{subsec:vmc_state_weight}Low-residual spectral support of the structured variational branch}

The exact and variational results above establish that the obtained VMC states are not, in general, small deformations of the finite matrix ground state, but they do not determine how much of such states lie in the low-residual spectral sector of the finite constraint. At present, neither a reference state in the exact kernel of the full model nor a controlled cutoff limit state exists against which a distance from the kernel could be measured. The available quantitative notion of near-kernel support must therefore be based on the finite cutoff spectral information. In what follows, we focus the analysis below on the structured VMC solutions obtained and use such spectral information to answer two separate questions. First, we ask how much of the structured VMC state is carried by the low-lying eigenspaces of $\hat{\mathcal{Q}}_{j_\mathrm{max}}$. Second, we ask how far this spectral support window must be enlarged before the normalized projection recovers the coherent product structure of the full VMC state.

Let the distinct eigenvalues of the finite quadratic constraint be ordered such that $E_0 < E_1 < \cdots$, and let $\Pi_\alpha^{(j_\mathrm{max})}$ project onto the complete eigenspace associated with the $E_\alpha$ eigenspace. For a spectral threshold $\epsilon$, define
\begin{equation}
    P_\epsilon^{(j_\mathrm{max})} = \mathbf{1}_{[0, \epsilon]}(\hat{\mathcal{Q}}_{j_\mathrm{max}}) \quad,\quad W_\epsilon^{(j_\mathrm{max})} = \bra{\Psi_{\mathrm{VMC}}^{(\mathrm{str})}} P_\epsilon^{(j_\mathrm{max})} \ket{\Psi_{\mathrm{VMC}}^{(\mathrm{str})}},
\end{equation}
where $\ket{\Psi_{\mathrm{VMC}}^{(\mathrm{str})}}$ denotes the structured VMC state. Equivalently, 
\begin{equation}
    P_\epsilon^{(j_\mathrm{max})} = \sum_{E_\alpha \leq \epsilon} \Pi_\alpha \quad,\quad W_\epsilon^{(j_\mathrm{max})} = \left\|P_\epsilon^{(j_\mathrm{max})} \ket{\Psi_{\mathrm{VMC}}^{(\mathrm{str})}}\right\|^2.
\end{equation}
Thus, $W_\epsilon^{(j_\mathrm{max})} \in [0, 1]$ is the fraction of the squared VMC norm contained in the selected spectral band (e.g. $W_\epsilon = 0.9$ means the band contains 90\% of the state norm and that the remaining 10\% lies in eigenspaces above the threshold). Since every $\Pi_\alpha^{(j_\mathrm{max})}$ is a projector onto a complete eigenspace, singlets and doublets are always retained as whole multiples and consequently the result is independent of the basis returned by the eigensolver within a degenerate space.

This projector supplies the finite-matrix meaning of \emph{near-kernel} used form this point onward. Namely, a state has near-kernel support when a substantial part of its norm lies in a low-lying spectral band of $\hat{\mathcal{Q}}_{j_\mathrm{max}}$. This notions is explicitly cutoff dependent and distinct from membership in, or a controlled norm distance from, the exact kernel of the model after removal of the spin cutoff. It does, however, enable one to define a notion of nearness to the kernel controlled by the constraint residual, irrespective of the cutoff. Where we refer instead to a low expectation value of $\hat{\mathcal{Q}}_{j_\mathrm{max}}$, we retain the term \emph{low-residual state} to emphasize that it is a fixed cutoff variational statement.

As representative cutoffs, we once again consider $j_\mathrm{max} = 13, 23, 33$ and 43, in which case the exact ground eigenspace alone carries on average 0.5303, $7.64 \times 10^{-4}$, $7.76 \times 10^{-7}$ and $3 \times 10^{-4}$ of the structured VMC norm, respectively. Therefore it is evident that the VMC state is not, in general, a small deformation of the finite-matrix ground state. To examine broader low-lying bands, for a prescribed norm fraction $\eta \in (0, 1]$ we define the weight-selected threshold
\begin{equation}
    \epsilon_\mathrm{\eta}^{(W)}(j_\mathrm{max}) := \min \{E_\alpha : W_{E_\alpha}^{(j_\mathrm{max})} \geq \eta\}.
\end{equation}
Namely, for a fixed $\eta$, one starts from the ground eigenspace and successively includes complete eigenspaces in order of increasing $E_\alpha$ until their cumulative weight in the structured VMC state at the chosen cutoff first reaches $\eta$, i.e. $\epsilon_\mathrm{\eta}^{(W)}(j_\mathrm{max})$ is the smallest constraint residual boundary that supports at least the chosen fraction of the state. The associated projector contains the first $G^{(W)}$ complete eigenspaces, or $N^{(W)}$ eigenvectors when their multiplicities are included. While this is neither a cutoff independent measure nor a unique definition of the near-kernel sector, it provides a transparent finite-matrix based approach to assess how much of the low residual spectrum is required to support a prescribed fraction of the VMC state.

\begin{figure}[h]
    \centering
    \includegraphics[width=0.7\linewidth]{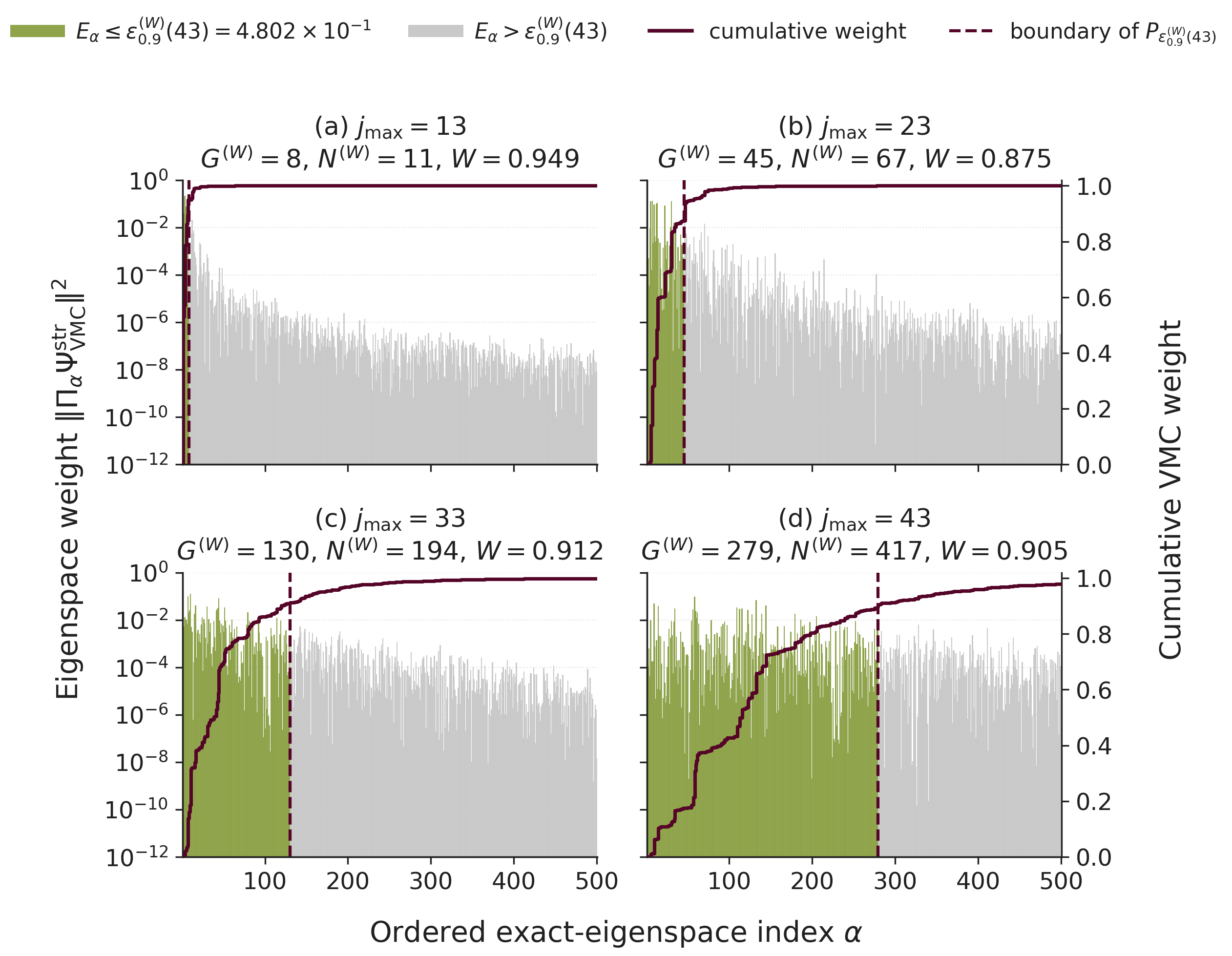}
    \caption{Spectral decomposition of the structured VMC state at $j_\mathrm{max} = 13, 23, 33$ and 43 over the first 500 complete eigenspaces. Bars show the VMC weight in each complete eigenspace of $\hat{\mathcal{Q}}_{j_\mathrm{max}}$. Green bars lie below the common weight selected threshold $\epsilon_{0.9}^{(W)}(43)$ while grey bars lie above it. The cumulative curve is read on the right axis. The quantities $G^{(W)}$ and $N^{(W)}$ count complete eigenspaces and eigenvectors including multiplicity respectively.}
    \label{fig:vmc_supp}
\end{figure}

For the comparison below, we choose $j_\mathrm{max} = 43$ and set $\eta = 0.9$, giving $\epsilon_{0.9}^{(W)}(43) = 4.802 \times 10^{-1}$. The same threshold is then applied without adjustment at the other three representative cutoffs. Their retained numbers of eigenspaces and captured VMC weights are allowed to differ, which permits a comparison at one common value of the constraint residual. Figure \ref{fig:vmc_supp} shows the weight carried by each of the first 500 complete eigenspaces and its cumulative sum.

\begin{table}[h]
    \centering
    \begin{tabular}{c|cccc}
        $j_\mathrm{max}$ & $G^{(W)}$ & $N^{(W)}$ & $100N^{(W)}/j_\mathrm{max}^3$ & $W_{\epsilon_{0.9}^{(W)}(43)}^{(j_\mathrm{max})}$ \\ \hline
        13 & 8   & 11   & 0.501\% & 0.949 \\
        23 & 45  & 67  & 0.551\% & 0.875 \\
        33 & 130  & 194  & 0.540\% & 0.912 \\
        43 & 279 & 417 & 0.524\% & 0.905
    \end{tabular}
    \caption{Retained spectral content at the common threshold $\epsilon_{0.9}^{(W)}(43)$. The quantities $G^{(W)}$ and $N^{(W)}$ are the number of complete eigenspaces and the rank of the corresponding projector, respectively. The fourth column gives this rank as a percentage of the full finite Hilbert space dimension, and the final column gives the captured structured VMC weight.}
    \label{tab:ret_ranks}
\end{table}

The retained ranks and weights are collected in Table \ref{tab:ret_ranks}. While the number of retained eigenspaces grows with the cutoff because the low-lying level density changes, the ranks of these projectors nevertheless represent only around 0.501\%, 0.551\%, 0.540\% and 0.524\%, of the full finite Hilbert space dimensions at each cutoff respectively. A large majority of the VMC norm is therefore concentrated in a small, low-lying fraction of the finite-matrix spectrum.

While the analysis above provides a measure of how near-kernel the representative structured VMC states are within each finite-matrix problem, it contains no information about the internal structure of the selected component. In particular, a large value of $W_\epsilon^{(j_\mathrm{max})}$ does not imply that the normalized projected state retains the Thiemann coherent state structure of the full VMC state. We therefore introduce a second criterion whose purpose is to determine the smallest complete low-residual spectral band for which the projected state recovers both the coherent one-edge profiles as well as the three-edge product structure of the original VMC state. Define the normalized projection
\begin{equation}
    \ket{\Psi_\epsilon} = \frac{P_\epsilon^{(j_\mathrm{max})} \ket{\Psi_\mathrm{VMC}^{(\mathrm{str})}}}{\sqrt{W_\epsilon^{(j_\mathrm{max})}}}.
\end{equation}
For each $\ket{\Psi_\epsilon}$, we determine its optimal approximation $\ket{\varphi_x \varphi_y \varphi_z}$ and fit each extracted factor to the reduced Thiemann coherent state in \eqref{eq:u1-thiemann-state} (note that here $\varphi_a$ denote the extracted one-edge factors of the projected state $\ket{\Psi_\epsilon}$ and not the full VMC state). To assess the quality of the Thiemann coherent state structure in the projected states, we compute both the worst one edge fidelity $F_\mathrm{Th}^\mathrm{worst}(\epsilon)$ and the full projected state fidelity $F_\mathrm{Th}(\epsilon)$, which we keep the $\epsilon$ dependence explicit in the notation. We now choose the threshold $\epsilon$ differently. For a prescribed full state fidelity $\gamma \in (0, 1]$, we define the cutoff dependent structure selected threshold
\begin{equation}
    \epsilon_\gamma^{(\mathrm{TCS})}(j_\mathrm{jmax}) := \min \{\epsilon \in \{E_\alpha\}_{j_\mathrm{max}} : F_\mathrm{Th}(\epsilon) \geq \gamma\}.
\end{equation}
Namely, for a fixed $\gamma$, one starts from the ground eigenspace and successively increases the threshold $\epsilon$ through the ordered elements of $\{E_\alpha\}_{j_\mathrm{max}}$. At each threshold, we form the normalized projected state $\ket{\Psi_\epsilon}$ and evaluate its full state TCS fidelity $F_\mathrm{Th}(\epsilon)$ and as such, $\epsilon_\gamma^{(\mathrm{TCS})}(j_\mathrm{jmax})$ is the smallest value of the spectral threshold $\epsilon$, at a given cutoff, for which the retained low-lying eigenspaces reconstruct the coherent product structure with fidelity at least $\gamma$. 

To now determine how much spectral information is needed to reach this threshold, we characterize the selected projector in four direct ways. First, we report the number of complete eigenspaces $G$ and the number of eigenvectors $N$ including multiplicity, as well as the retained rank $N/j_\mathrm{max}^3$ as a fraction of the full finite Hilbert space dimension as well as, lastly, the VMC weight $W_\epsilon^{(j_\mathrm{max})}$.

\begin{figure}[h]
    \centering
    \includegraphics[width=0.7\linewidth]{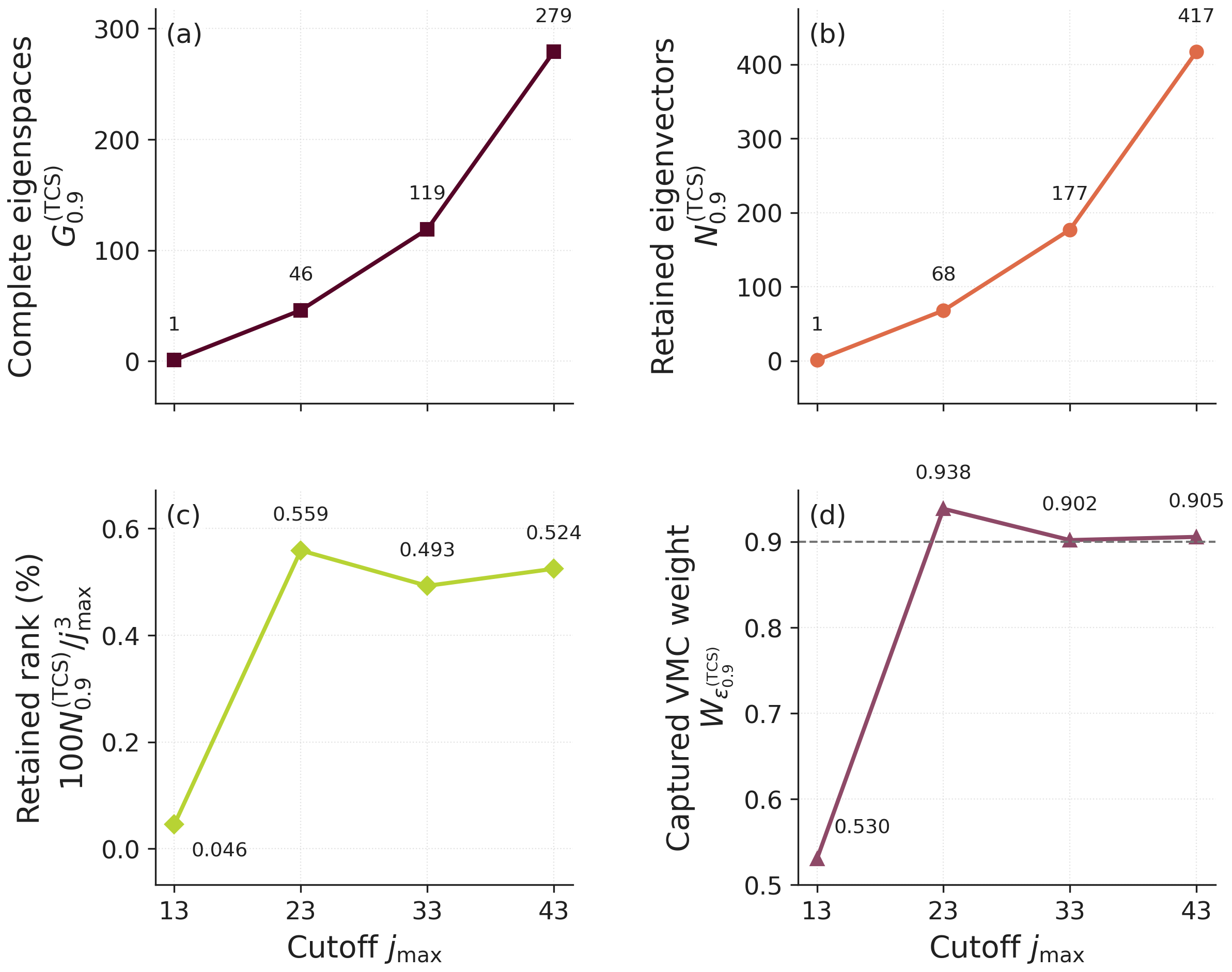}
    \caption{Spectral content required for the normalized structured VMC projection to first reach $F_\mathrm{Th} \geq 0.9$. Panels (a) and (b) show the number of complete eigenspaces and basis-eigenvectors, respectively while panel (c) gives the rank of the corresponding projector as a percentage of the full finite Hilbert space dimension. Panel (d) shows the VMC weight captured by the same projector.}
    \label{fig:vmc_tcs_summ}
\end{figure}

As shown in Figure \ref{fig:vmc_tcs_summ}, at the representative cutoffs $j_\mathrm{max} = 13, 23, 33, 43$ the first crossings of this threshold have fidelities 0.910, 0.917, 0.901 and 0.908. The corresponding projectors retain 1, 46, 119 and 279 complete eigenspaces, alternatively 1, 68, 177 and 417 eigenvectors when multiplicities are included. Despite this, their ranks however amount to only 0.046\%, 0.559\%, 0.493\% and 0.524\% of the respective finite Hilbert spaces. At the same time, they contain 0.530, 0.938, 0.902 and 0.905 of the structured VMC norm. The coherent product structure can be therefore reconstructed without retaining the entire VMC state, and can be reconstructed from a very small fraction of the finite Hilbert space whereby this fraction lies in the near-kernel sector of the finite-matrix constraint.

\begin{figure}[h]
    \centering
    \includegraphics[width=0.95\linewidth]{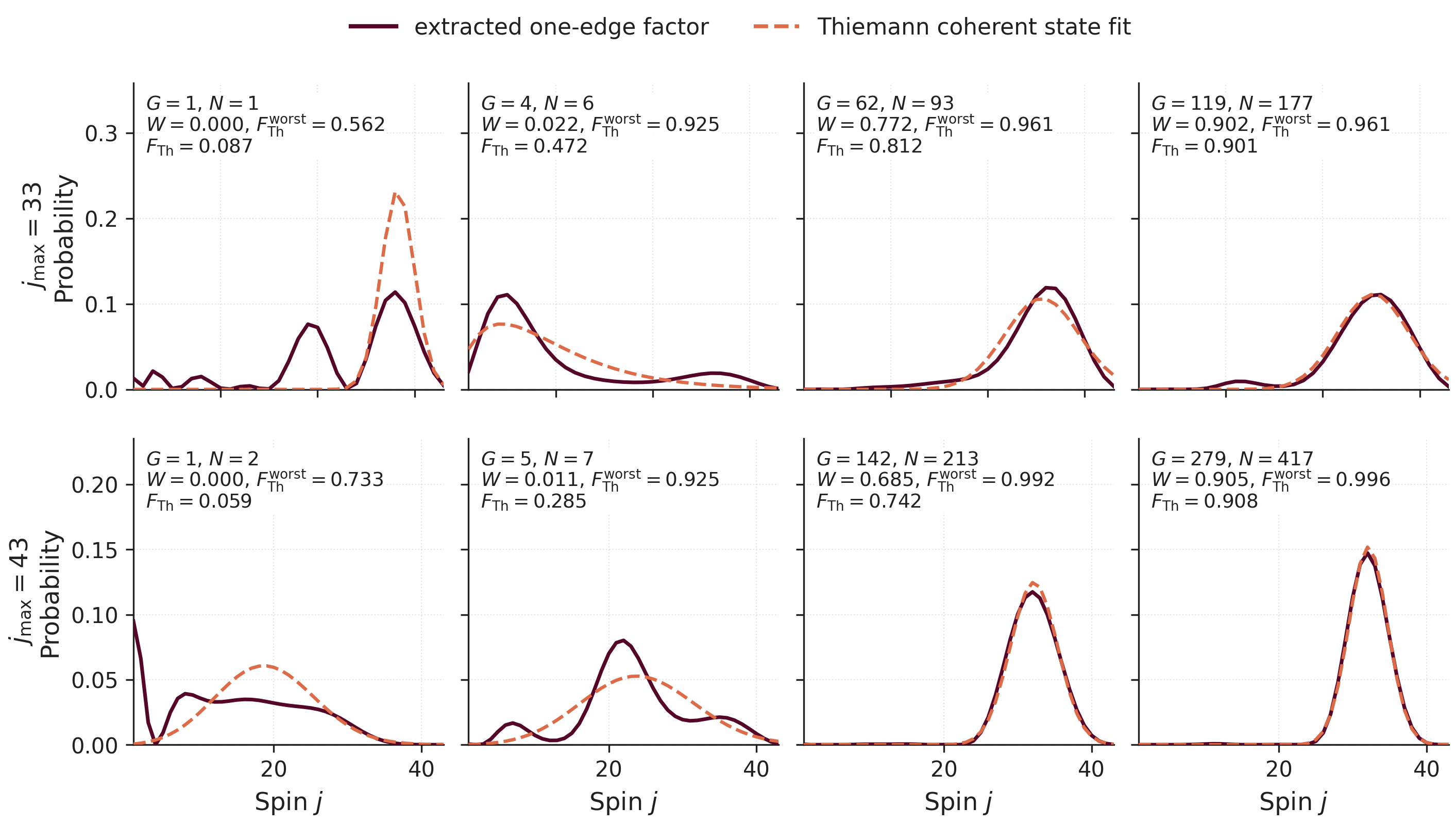}
    \caption{Evolution of the least accurately fitted one-edge factor in normalized spectral projections of the structured VMC state for $j_\mathrm{max}=33, 43$. The columns proceed from the exact ground space projection to the first locally coherent one-edge profiles, an intermediate number of complete eigenspaces and the first threshold with $F_\mathrm{Th} \geq 0.9$. Solid curves are factors extracted from the optimal rank-one approximation and dashed curves are their best reduced Thiemann coherent state fits.}
    \label{fig:tcs_evol}
\end{figure}

To further elucidate this, we also examine the worst one-edge factor of the $j_\mathrm{max}=33, 43$ under successively increasing projections reaching the obtained threshold in Figure \ref{fig:tcs_evol}. In each row, the first column shows the projection onto the exact ground space. The second shows the first spectral boundary at which all three extracted factors have one-edge coherent state fidelities above 0.9. The third is the midpoint in $G$ between this local threshold and the full state threshold obtained above and the last column shows the first boundary satisfying the $\epsilon$ threshold. Every column is closed at an exact eigenspace boundary and thus degenerate multiplets remain complete.

The evolution shown in Figure \ref{fig:tcs_evol} separates two stages. At the shown cutoffs, the ground space projections have very poor full state coherent product fidelities. All three one-edge factors first become coherent state-like after retaining $G = 4, 5$ complete eigenspaces for $j_\mathrm{max} = 33, 43$ respectively. At those boundaries, the worst one-edge fidelities are already above 0.9, while the full state fidelities are 0.472 and 0.285 respectively. Increasing the windows to $G = 62, 142$ raises the full state fidelities to 0.812 and 0.742, and the target value is finally crossed at $G = 119, 279$ respectively. A wider low-lying band is therefore required to reconstruct the global product organization and reduce the remaining inter-edge correlations.

\begin{figure}[H]
    \centering
    \includegraphics[width=0.7\linewidth]{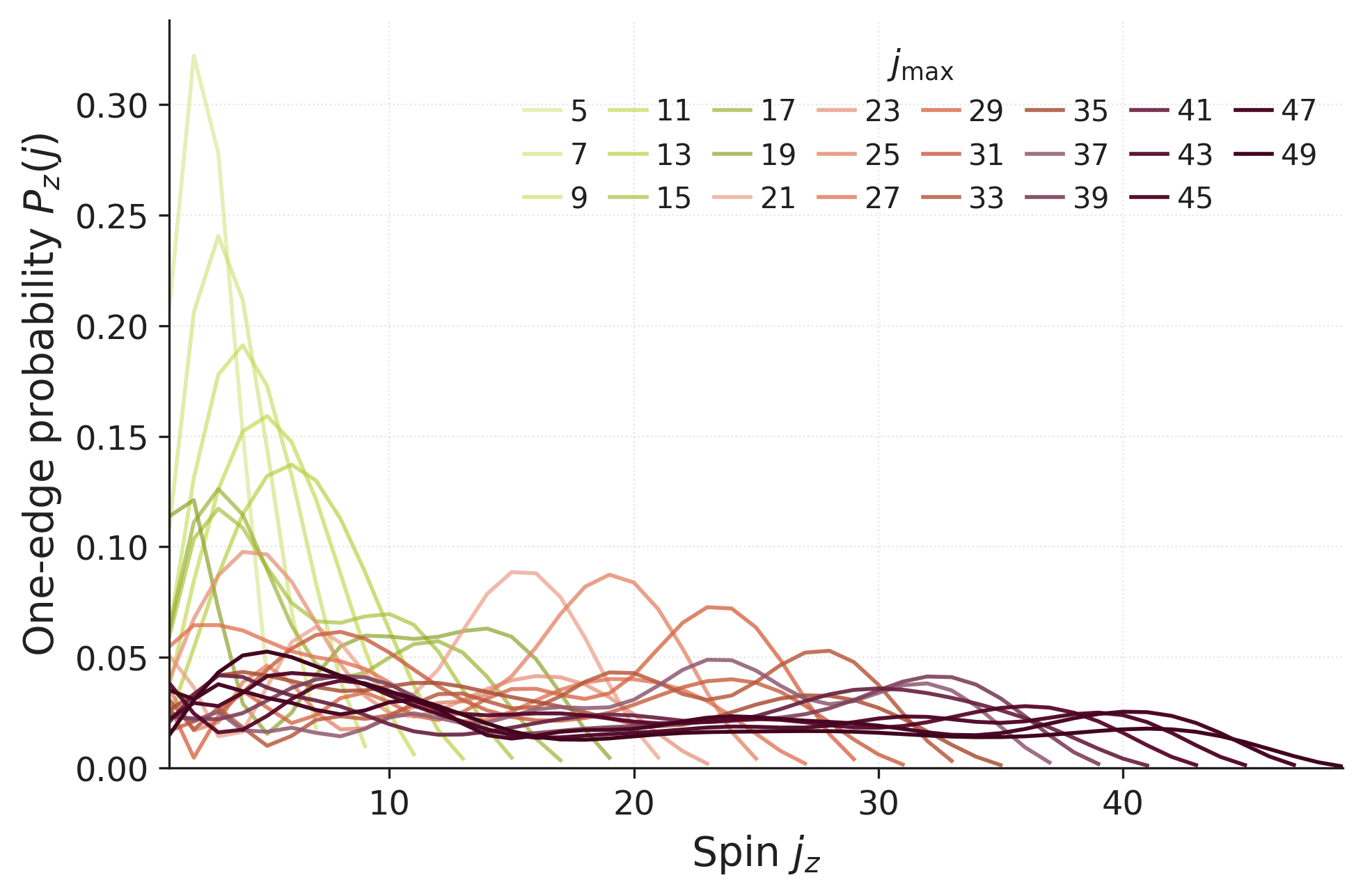}
    \caption{The $z$-edge spin marginal $P_z^{(j_\mathrm{max})}(j)$ of the representative exact finite matrix ground state $\ket{\Psi_0^\star}$ across the available cutoffs from $j_\mathrm{max}=5$ to 49. Each curve is normalized to unit total probability. The decreasing peak heights and broadening support therefore show that the probability becomes distributed over a larger range of spins as the cutoff grows. For a doublet ground space, $\ket{\Psi_0^\star}$ is the most product-like normalized vector in the complete ground multiplet.}
    \label{fig:exact_ground_edge_z}
\end{figure}

The poor coherent product fidelities of the ground space projections in Figure \ref{fig:tcs_evol}, as well as the changing structure of such projections between cutoffs, also motivate examining whether the one-edge spin content of the exact finite matrix ground state approaches a stable profile as the cutoff is increased. Denote by $\ket{\Psi_0^\star}$ the normalized eigenvector in a given ground space. For singlet spaces, this is trivial. For a doublet space, we choose the normalized vector in the complete ground space that realizes the maximal product fidelity in \eqref{eq:ed_fprod}. This prescription is independent of rotations of the numerical eigensolver basis within the doublet and selects the most product-like exact ground vector, thereby giving the exact ground space the most favourable representative for the present comparison. We then consider its z-edge spin marginal $P_z^{(j_\mathrm{max})}(j)$ and examine how this marginal changes with cutoff. Figure \ref{fig:exact_ground_edge_z} compares this probability distribution across the available exact cutoffs.

The absolute probabilities in Figure \ref{fig:exact_ground_edge_z} retain their common normalization and therefore display how concentrated the one-edge spin distribution is at each cutoff. As $j_\mathrm{max}$ grows, the support broadens while the maximum generally decreases. Probability is consequently redistributed over a larger set of spin values, so the marginal becomes flatter (i.e. more delocalized) on the discrete spin grid. The same decrease in peak height, however, makes secondary peaks and other shape information increasingly difficult to compare on the absolute probability scale. To separate this broadening from changes in the profile itself, we additionally define the peak-normalized quantity
\begin{equation}
    \tilde{P}_z^{(j_\mathrm{max})}(j) = \frac{P_z^{(j_\mathrm{max})}(j)}{\max_{j'}P_z^{(j_\mathrm{max})}(j')}.
    \label{eq:exact_ground_edge_peak_normalized}
\end{equation}

Thus $\max_{j'}\tilde{P}_z^{(j_\mathrm{max})}(j') = 1$ at every cutoff. Such a $\tilde{P}_z^{(j_\mathrm{max})}(j)$ allows one to examine the shape of the profile of the corresponding one-edge marginal as it retains the physical spin coordinate and the support of the original marginal while placing every curve on the same vertical scale.

\begin{figure}[H]
    \centering
    \includegraphics[width=0.72\linewidth]{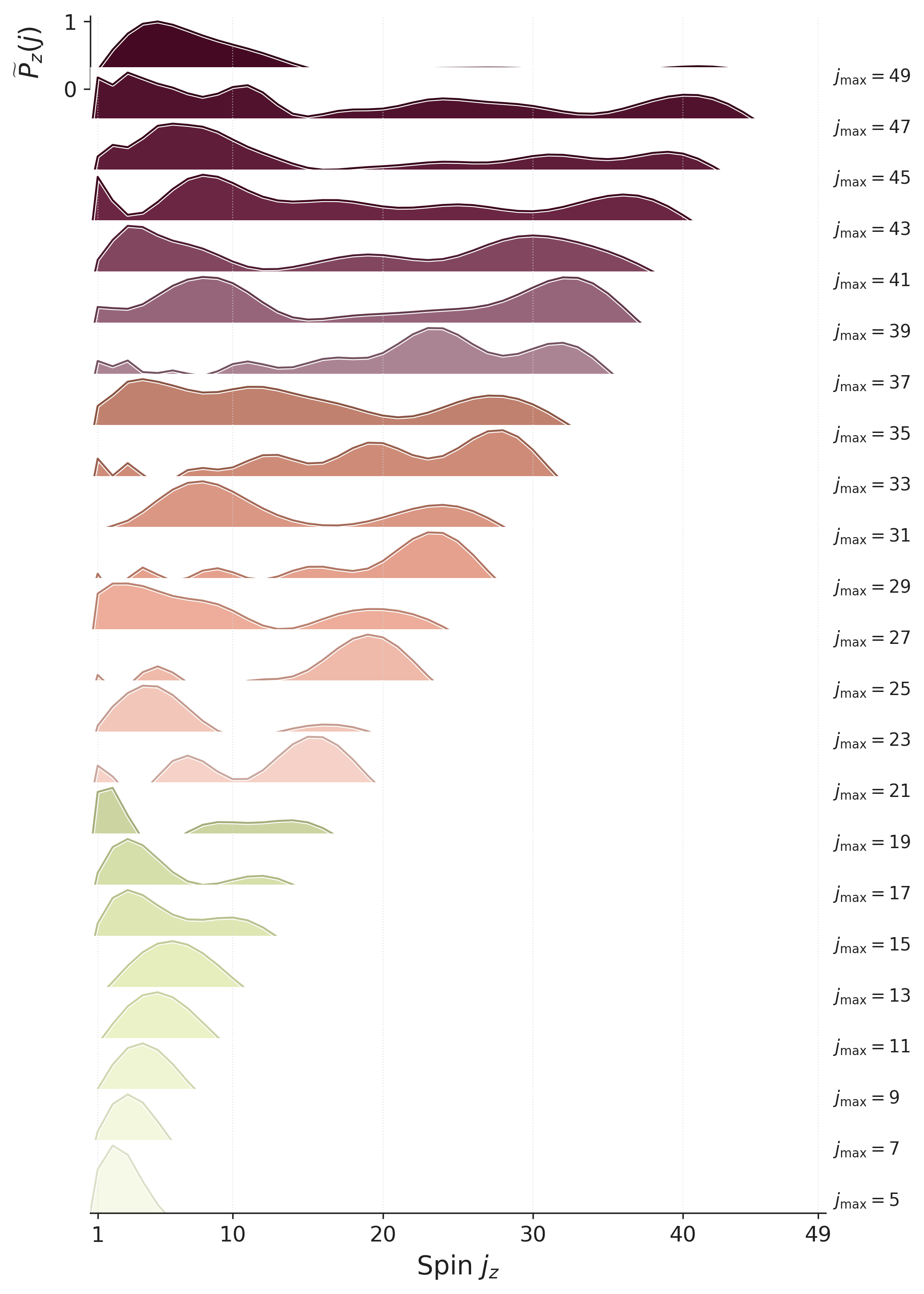}
    \caption{Exact ground one-edge marginals in Figure \ref{fig:exact_ground_edge_z}. Each row shows the peak-normalized profile $\tilde{P}_z^{(j_\mathrm{max})}(j)$ defined in \eqref{eq:exact_ground_edge_peak_normalized}.}
    \label{fig:exact_ground_edge_z_ridgeline}
\end{figure}

The peak-normalized view in Figure \ref{fig:exact_ground_edge_z_ridgeline} makes clear that the evolution is not a smooth displacement or broadening of one fixed packet. Rather, the exact ground marginal reorganizes non-monotonically as the cutoff changes and, at the larger cutoffs, generally contain several competing maxima rather than a single stable peak. This is already nontrivial structure in only one amplitude squared marginal. The corresponding marginals of the other two edges, the joint distributions that encode inter-edge correlations and the phases of the exact state contain further information which is not represented in either figure. Nevertheless, these figures show that even in this restricted one-edge probability content of the most product-like representative of the exact ground space, the picture remains strongly cutoff dependent and nontrivial. 

This analysis, however, must only be taken qualitatively. Namely, neither $\epsilon$-based projection defines a cutoff independent criterion and the numerical values chosen for such $\epsilon$ are also chosen arbitrarily. In the absence of an exact kernel state in the full model, the results discussed above cannot be converted into a quantitative statement regarding the distance from the true physical kernel of the full model. What they do show, however, is that most of the structured VMC norm and the coherent product organization of that state are both encoded in a small low-lying fraction of each finite matrix spectrum. This fixed cutoff result does not determine whether the coherent profile is selected by the constraint or induced by the structured neural architecture, which motivates the direct product state variational optimization considered next.

%
%

\subsection{\label{subsec:hartree_min}Unrestricted product state minimization}

The spectral analysis above locates the structured VMC state within the low lying finite matrix spectrum but it does not determine whether its coherent one edge profile is selected by the constraint or induced by the structured neural architecture. The near factorization found in both VMC states therefore motivates a direct variational calculation on the product manifold
\begin{equation}
    \mathcal{M}_\mathrm{prod} = \{\ket{a} \otimes \ket{b} \otimes \ket{c} \;|\; \|a\| = \|b\| = \|c\| = 1\}.
\end{equation}
The three one-edge factors are independent complex vectors with $j_\mathrm{max}$ components. We note that neither permutation symmetry is used, nor any assumptions on the shape of such one-edge factors are made. At each cutoff we then determine the product state variational level
\begin{equation}
    E_\mathrm{prod}(j_\mathrm{max}) = \min_{\ket{\Psi} \in \mathcal{M}_\mathrm{prod}} \bra{\Psi} \hat{\mathcal{Q}}_{j_\mathrm{max}} \ket{\Psi}.
    \label{eq:energy_prod_level}
\end{equation}
Since $\mathcal{M}_\mathrm{prod}$ is a strict subset of the full Hilbert space, a priori $E_\mathrm{prod}(j_\mathrm{prod})$ is an upper bound on the exact lowest eigenvalue $E_0(j_\mathrm{max})$. Hence, in this analysis it is not expected that the finite-matrix constraint ground states are obtained. We solve \eqref{eq:energy_prod_level} directly, without a Monte Carlo estimator, on a range of odd-valued cutoffs as done in the exact spectral analysis in Section \ref{sec:exact}. The real and imaginary parts of the components of the three factors provide $6 j_\mathrm{max}$ real variational parameters, subject to one normalization constraint for each factor. For every independent start, both real and imaginary components are drawn from a unit normal distribution and each factor is then normalized. At every step in the simulation, the product state is constructed on the complete $j_\mathrm{max}^3$ spin grid and the objective is evaluated as
\begin{equation}
    \bra{a \otimes b \otimes c} \hat{\mathcal{Q}}_{j_\mathrm{max}} \ket{a \otimes b \otimes c} = \|\hat{C}_{j_\mathrm{max}} \ket{a \otimes b \otimes c}\|^2.
\end{equation}
The sparse constraint action and its analytic derivatives with respect to all factor components are supplied to the sequential least squares programming (SLSQP) algorithm. We allow the optimizer to run to convergence, with a maximal iteration count limited 3000 optimizer steps and use an SLSQP stopping tolerance of $10^{-13}$. Thus, neither the objective nor its gradient are sampled stochastically and the only randomness in the process is the initialization of the state.

To monitor whether a reduced Thiemann coherent state profile develops during minimization, we fit each one-edge factor at approximately logarithmically spaced checkpoints along every trajectory. At a given checkpoint we evaluate the fidelity in \eqref{eq:one-edge-fidelity} exactly on the complete one-edge grid and maximized by bounded L-BFGS-B optimization with an analytic gradient. For the fitting parameters, we use $t \in [10^{-10}, 5]$, $p \in [-50, 50]$ and $\theta \in [- \pi, \pi]$. Further, five starts inferred from the moments and phase of the factor are supplemented by eight Latin-hypercube starts across the fitting domain, while the final checkpoint is tested with 24 Latin-hypercube starts, ensuring a thorough fitting. We note that all of this is done as a post-hoc diagnostic, i.e. only once the simulation is concluded, to avoid any interference with the simulations.

Although the factors are allowed to differ, all accepted minima reduce, within numerical precision, to the permutation invariant form
\begin{equation}
    \ket{\Psi_\mathrm{prod}} = \ket{\phi}^{\otimes 3},
    \label{eq:psi_h}
\end{equation}
up to an irrelevant overall phase. To make things precise, we will consider two representative cutoffs of $j_\mathrm{max} = 11, 41$ in the analysis below. For both cutoffs, the smallest value of $|\langle \phi_a | \phi_b \rangle|^2$ between any two optimized factors is 0.9999995, and thus after adjusting their independent phases the three factors represent the same one-edge state. We note that this is stronger than factorization alone, as a general point of $\mathcal{M}_\mathrm{prod}$ contains three different one-edge states and can distinguish the $x, y$ and $z$ directions. The form in \eqref{eq:psi_h} on the other hand assigns to all three edges the same probability distribution, phase profile and one-edge expectation values.

The permutation invariance of $\hat{\mathcal{Q}}_{j_\mathrm{max}}$  implies that the objective being optimized is unchanged under permutations of the three factors and makes the equal factor submanifold invariant under the exact variational dynamics. It does not, however, require an unrestricted minimum to lie on this submanifold. A symmetry broke minimum could instead occur together with its edge-permuted partners. In our multi-start calculations, the low residual states reached from independent, non-symmetric initializations consistently contain the same factor on every edge.

To test whether this branch is reproducible rather than tied to an isolated initialization, we performed ensembles of independently initialized optimizations at the two representative cutoffs. A separate preliminary multi-start calculation first provides the lowest product state constraint residual found for each cutoff, which we denote by $E_\mathrm{prod}^\mathrm{ref}$. We then generate independent starts until a total of 1000 runs were considered as admissible. Here, admissibility is based on several factors, first of which the successful termination of SLSQP. Further, aside from requiring the states to have a factor normalization error which is negligible, we also require that the final attained energy is \emph{at most} as high as the obtained reference energy $E_\mathrm{prod}^\mathrm{ref}$, with an absolute tolerance of $2 \times 10^{-10}$ and a relative tolerance of $2\times 10^{-7}$.

The last condition is needed because successful termination of a local optimizer only means that its stopping conditions have been met, but it does not by itself exclude termination at a higher constraint residual value. The residual criterion therefore prevents such endpoints from being mixed into the trajectory statistics for the branch being considered. We note, however, that it does \emph{not} constitute a proof that $E_\mathrm{prod}^\mathrm{ref}$ is the global minimum over $\mathcal{M}_\mathrm{prod}$. Lastly, we note that no admissibility criterion is based on the structure of the state, i.e. whether the state's one-edge factors resemble a reduced Thiemann coherent state is irrelevant in the admissibility process.

Given the above criteria, all 1000 starts at $j_\mathrm{max} = 11$ were admissible. At $j_\mathrm{max} = 41$, a total of 1033 starts were required to obtain 1000 admissible states, corresponding to 96.8\% of all starts. Of the remaining starts, 27 reached the iteration limit, and 3 terminated above the constraint residual threshold, and three encountered a rank-deficient equality constraint subproblem, i.e. the optimizer could not determine a unique update direction because the local linear system enforcing the factor normalization constraints became numerically singular. 

\begin{figure}
    \centering
    \includegraphics[width=0.75\linewidth]{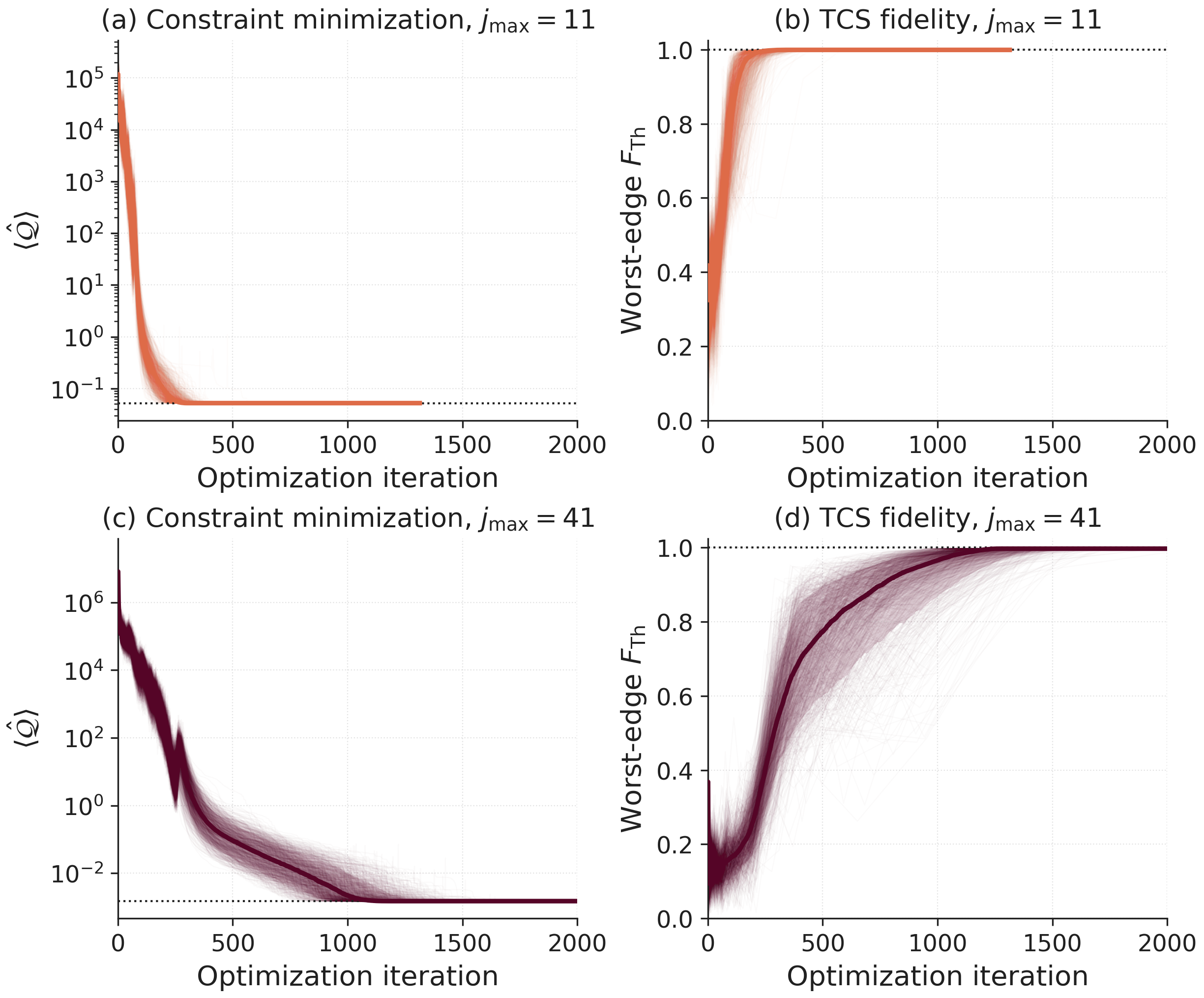}
    \caption{Minimization of the quadratic constraint over unrestricted complex product states at $j_\mathrm{max} = 11, 41$. The left panels show the exact full-grid constraint expectation value and the right panels show the lowest of the three post-hoc one-edge fidelities with the reduced Thiemann coherent state (TCS for brevity). Thin curves show 1000 independently initialized and accepted runs, thick curves show their pointwise medians and the shaded regions contain the 10\textsuperscript{th} and 90\textsuperscript{th} percentiles.}
    \label{fig:hartree_and_fth}
\end{figure}

Figure \ref{fig:hartree_and_fth} shows both the full state quadratic constraint expectation value and
\begin{equation}
    F_\mathrm{Th}^\mathrm{worst} = \min_{a \in \{x, y, z \}} \max_{t, p, \theta} |\langle \chi_{t, p, \theta} | \phi_a \rangle|^2,
\end{equation}
along the admissible optimization trajectories. The thin curves are the 1000 individual admissible simulations, the thick curve is their pointwise median and the shaded region shown spans the 10\textsuperscript{th} to 90\textsuperscript{th} percentiles. Note that we will use TCS (Thiemann coherent state) as shown in the figure for brevity. We note that the close overlap of many early trajectories does not arise from repeated initial conditions as the initial expectation values and complete optimizer histories are distinct for all admissible simulations. Their differences are compressed by the logarithmic scale in the left panels, while the common trend reflects rapid attraction towards the same low-residual product state branch. The fidelities are evaluated only at the fitting checkpoints and interpolated between them for visualization. 

At $j_\mathrm{max} = 11$, the ensemble reaches $E_\mathrm{prod} \approx 0.0520643$ with a run-to-run standard deviation of $6.7 \times 10^{-12}$ with a mean final worst edge fidelity of $F_\mathrm{Th}^\mathrm{worst} \approx 0.999229$ with a standard deviation of $5.5 \times 10^{-10}$. At $j_\mathrm{max} = 41$, the corresponding values are $E_\mathrm{prod} \approx 0.00148889$ with a standard deviation of $1.5 \times 10^{-11}$ and $F_\mathrm{Th}^\mathrm{worst} \approx 0.997195$ with a standard deviation of $8.8 \times 10^{-8}$. The spreads quantify reproducibility across independent optimizations and together with the acceptance fractions, they show that at least 100\% and 96.8\% of all independent starts at the two cutoffs, respectively, are explicitly verified to reach the same coherent product state branch.

Figure \ref{fig:hartree_and_fth} displays the central result of this analysis. Namely, the reduction of the quadratic constraint expectation value, with no Monte Carlo methods and over general product states, is accompanied, independently across the random starts, by an increase of the worse edge coherent state fidelity towards unity. This occurs despite the fact that every complex one-edge profile was available to the optimizer and the three one-edge factors were generally allowed to differ. Within the product state variational manifold, and specifically within the investigated cutoffs, this identifies the coherent profile as both a reproducible structure and an emergent structure which is chosen dynamically by the minimization of the constraint. While the analysis in the section therefore establishes the emergence of the coherent profile within the product state manifold, it does not determine how the resulting product states, the VMC states or the exact ground spaces respond when the cutoff is enlarged and consequently which, if any, of such states are better candidates to approximate near-kernel states to physical states in the full model.

%
%

\subsection{\label{subsec:cutoff_enlargement}Fixed states under cutoff enlargement}

The preceding section showed that minimization of the constraint on the product state manifold results in a Thiemann coherent state structure even though this was not imposed in any way on the variational family. The states obtained in this way are, of course, not the exact ground state of the finite matrix constraint problem since their constraint residual lies above $E_0(j_\mathrm{max})$ and the finite matrix constraint exact ground spaces are generally not product spaces. On the other hand, such exact ground spaces carry the optimal states \emph{within} each truncated Hilbert space only. Equivalently, this does not necessarily mean that they represent candidate physical states for the full model. The question is then what, if anything, either family of states tells about the solutions of the constraint in the full model.

A simple way in which a low-residual state in the theory at finite cutoff can be relevant to the full theory is if it stays low-residual when the cutoff is removed. We therefore compare how the exact ground spaces, the product states $\ket{\Psi_\mathrm{prod}}$ and the low-residual variational states, structured and MLP, respond when the spin cutoff is increased while each state is held fixed. If a low-residual state relies strongly on excitations at the boundary of the truncated spin grid, restoring the omitted configurations and corresponding matrix elements of the constraint can increase the constraint residual. If the state is supported away from that boundary, it stays low-residual under increase of $j_\mathrm{max}$ and hence also in the full theory.

Let $S < T$ denote two cutoff values $j_\mathrm{max}$ and define the isometry $\iota_{S \to T} : \mathscr{H}_S \to \mathscr{H}_T$ such that
\begin{equation}
    \iota_{S \to T} \ket{j_xj_y j_z}_S = \ket{j_xj_yj_z}_T,
\end{equation}
where every coefficient is placed in the entry carrying the same ordered triple $(j_x, j_y, j_z)$ in the larger basis and every newly available configuration is assigned zero amplitude, hence preserving the norm exactly. The state is not re-optimized after the embedding. For a normalized state $\Psi_S \in \mathscr{H}_S$, we consider
\begin{equation}
    R_{S \to T} [\Psi_S] = \bra{\iota_{S \to T} \Psi_S} \hat{\mathcal{Q}}_T \ket{\iota_{S \to T} \Psi_S} = \|\hat{C}_T \iota_{S \to T} \ket{\Psi_S}\|^2.
    \label{eq:emb_res}
\end{equation}
The native value is recovered when $T = S$. For the case of exact ground states, if the exact ground space $\mathcal{G}_S$ is degenerate, its individual eigenvectors are not uniquely defined since any unitary rotation within $\mathcal{G}_S$ gives an equally valid eigenbasis. We therefore work with the complete ground space projector, hereby denoted by $P_{\mathcal{G}_S}$, and restrict the embedded quadratic operator to this subspace,
\begin{equation}
    M_{S,T} = \left.P_{\mathcal{G}_S} \iota_{S \to T} \hat{C}^2_{T} \iota_{S \to T} P_{\mathcal{G}_S} \right|_{\mathcal{G}_S}.
\end{equation}
We characterize the multiplet by the basis independent average $\mathrm{Tr(M_{S,T})}/m_S$, where $m_S = \dim \mathcal{G}_S$. The smallest and largest eigenvalues of $M_{S, T}$ give, respectively, the minimum and maximum residual attainable by any normalized state in the same ground multiplet.\footnote{In the cases we considered, we find that the eigenvalues actually agree to to high precision, so no ground state vector is preferred regarding the residual under this embedding.} 

The dependence of the target cutoff follows directly from the finite shift range of the constraint. From \eqref{eq:C_E^sym}, each term in the Euclidean part contains two operators $\hat{s}_a^{(1)}$ and can change each of the two corresponding edge spins by one unit. Its action on a state supported in $\mathscr{H}_S$ can therefore produce upper boundary components in $\mathscr{H}_{S+1} \setminus \mathscr{H}_S$, but none beyond $S+1$. On the other hand, the two factors of $(\hat{s}_a^{(1/2)})^2$ in \eqref{eq:C_L^sym} allow the Lorentzian part to change the spin of one edge by as much as two units. Consequently, the image of the constraint on an embedded state from $\mathscr{H}_S$ is contained in $\mathscr{H}_{S+2}$. One therefore expects that, for the same fixed source state, no additional change happens when the target cutoff is increased beyond $S+2$, hence $R_{S \to T}$ should be independent of $T$ for $T \geq S+2$. 

\begin{figure}[h]
    \centering
    \includegraphics[width=0.8\linewidth]{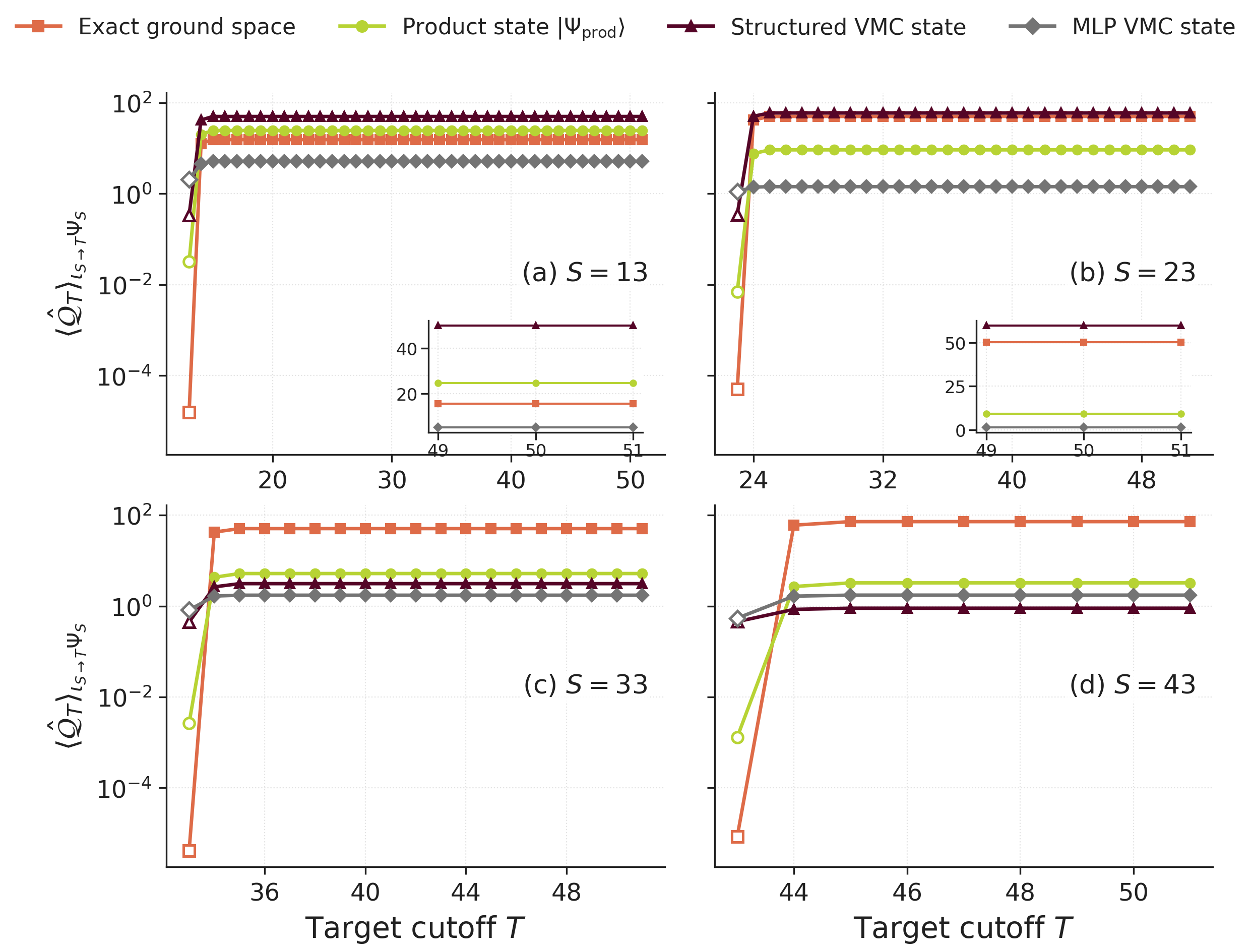}
    \caption{Quadratic constraint expectation value of fixed states under canonical cutoff enlargement for source cutoffs $S = 13, 23, 33, 43$. Open markers denote the native evaluation at $T = S$. Each subsequent point is obtained by embedding the same normalized state into $\mathscr{H}_T$ without re-optimization.}
    \label{fig:selected_s_to_t}
\end{figure}

\begin{figure}[h]
    \centering
    \includegraphics[width=0.8\linewidth]{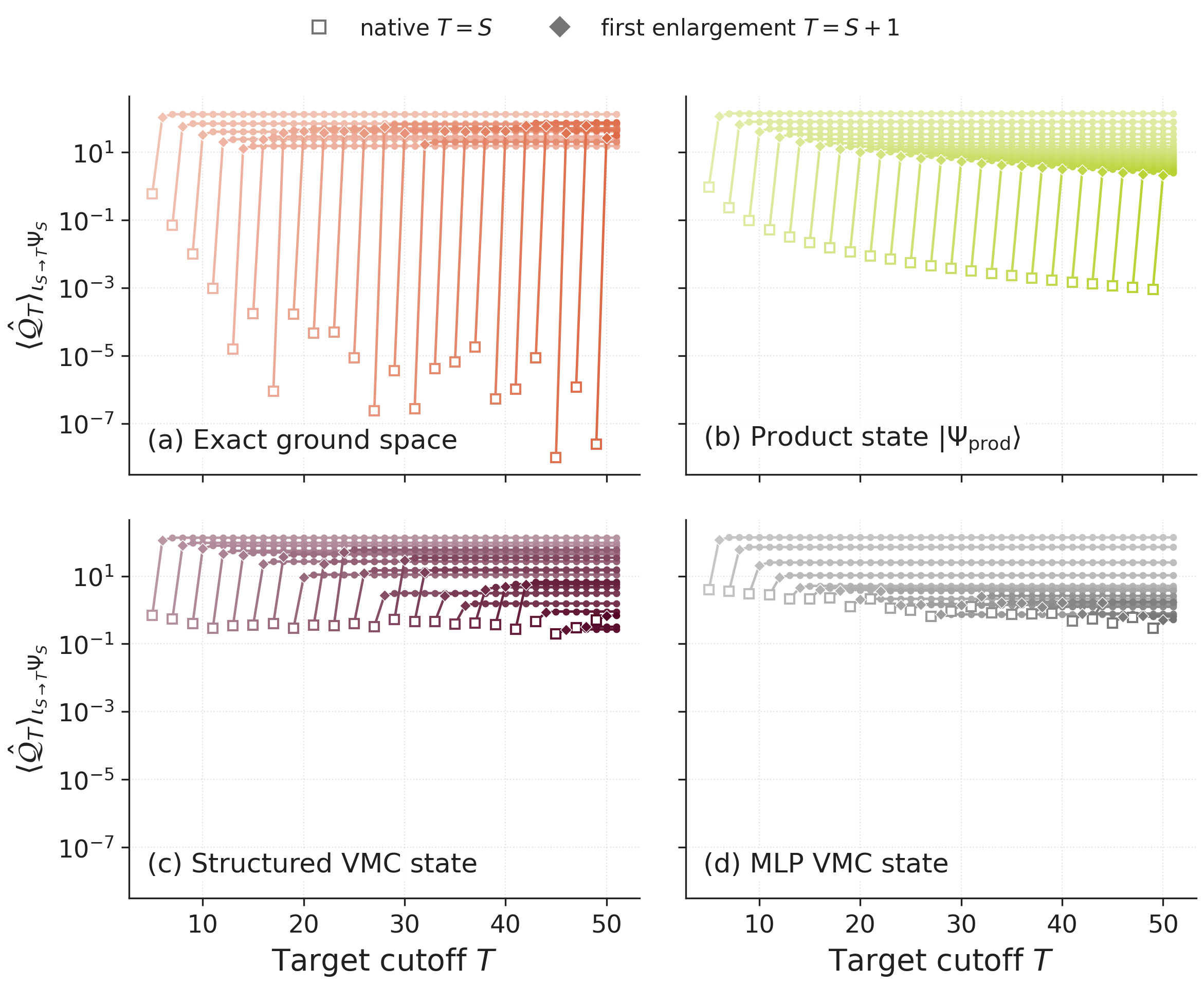}
    \caption{Fixed state cutoff enlargement test for all available source cutoffs. Each curve begins with the native value $T = S$ and follows the same state after canonical embedding into larger Hilbert spaces. Lighter curves correspond to smaller source cutoffs.}
    \label{fig:emb_flow}
\end{figure}

Figure \ref{fig:selected_s_to_t} compares the exact ground space, the unrestricted product state and the two VMC states for four representative source cutoffs. As expected from the shift structure described above, the fixed state residual can change at the first enlargement and then remains constant to numerical precision for $T \geq S + 2$. Figure \ref{fig:emb_flow} shows the same construction for every source cutoff at which all four state families are available.

The exact finite matrix ground space has the smallest possible native residual by construction, but this ordering is not preserved after the cutoff is enlarged. At $S = 33$, its trace averaged residual changes from $4.21 \times 10^{-6}$ to $5.06 \times 10^1$, an increase by a factor of approximately $1.20 \times 10^7$. At the same source cutoff, the unrestricted product, structured VMC and MLP residuals change from $2.67 \times 10^{-3}, 4.48 \times 10^{-1}$ and $8.19 \times 10^{-1}$ to 5.20, 3.11 and 1.74, respectively, corresponding to factors of $1.94 \times 10^3$, 6.95 and 2.12. The separation is remains pronounced for the exact ground space as the cutoff is increased. For example, at $S = 43$ the residual changes by a factor of roughly $10^6$, reaching an enlarged residual in the order of $10^1$ from a native residual of the order $10^{-6}$. The corresponding enlargement factors for the unrestricted product, structured VMC and MLP states are of the order $2.48 \times 10^{3}$, 1.98 and 3.22 respectively. 

The relative increase and the absolute residual carry different information. The exceptionally large ratios for the exact ground spaces partly reflect their extremely small native eigenvalues. More importantly, however, the same states acquire absolute residuals of order $10^1$ after enlargement. A state that appears to be an excellent solution of the truncated constraint can therefore cease to be low residual as soon as the transitions omitted at its native cutoff are restored. In this qualitative test, the exact finite matrix ground spaces are consequently the least compelling representatives of a cutoff-independent solution, despite being the optimal states of every native finite matrix problem.

The same qualification applies to the variational states. The unrestricted product states undergo increases of several orders of magnitude and retain enlarged residuals between a few units and several tens over the displayed range. They are therefore not approximate solutions of the full constraint merely because their native residuals are small. The structured VMC state also shows a large increase at the smaller source cutoffs, its enlarged residual is of order unity to a few units at the larger source cutoffs. Lastly, the MLP residual changes by a smaller factor although this is partly because its native residual is already comparatively large. At the largest source cutoffs, both VMC families have smaller absolute enlarged residuals than either the exact ground space or the unrestricted product state, making them more stable under this particular comparison. However, this does not by itself make either family a low-residual state of the constraint with the cutoff removed.

The difference between the native and enlarged residual cannot in general be interpreted entirely as strictly only the action into newly available configurations. In the finite construction used here, due to the boundary conditions imposed on the spins, $\hat{C}_S$ is not assumed to be exactly equal to the compression of $\hat{C}_{S+2}$ to $\mathscr{H}_S$. The enlarged residual can therefore also contain a change in the component that remains inside the original subspace. To isolate the genuinely outgoing part of one constraint action, we introduce the orthogonal projector
\begin{equation}
    P_S = \iota_{S \to S+2} \iota^\dagger_{S \to S+2},
\end{equation}
onto the embedded copy of $\mathscr{H}_S$ and define the outgoing residual
\begin{equation}
    \ell_S[\Psi_s] = \|(1-P_S)\hat{C}_{S+2} \iota_{S \to S+2} \ket{\Psi_S}\|^2,
\end{equation}
which is the absolute squared norm of the part of $\hat{C}_{S+2} \iota_{S \to S+2} \ket{\Psi_S}$ lying outside the old allowed spin hypercube. Since $P_S$ and $1 - P_S$ are orthogonal, $\ell_S[\Psi_S] \leq R_{S \to S+2}[\Psi_S]$.

\begin{figure}
    \centering
    \includegraphics[width=0.75\linewidth]{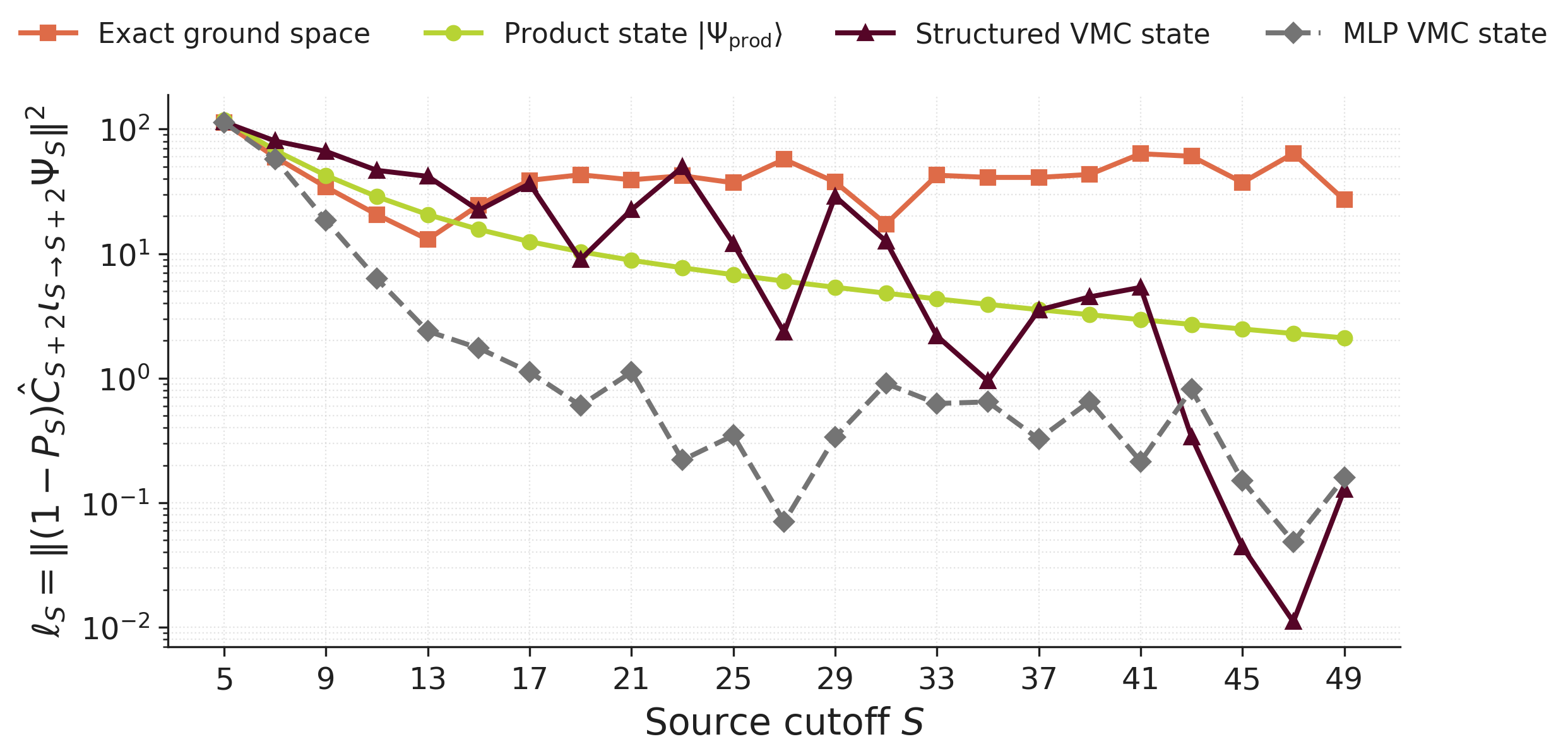}
    \caption{Absolute outgoing constraint residual $\ell_S$ for the exact ground spaces, unrestricted product states, structured VMC states and MLP states. Lower values mean that one action of the enlarged constraint produces a smaller component in the configurations absent from the source Hilbert space.}
    \label{fig:outgoing_residual}
\end{figure}

The outgoing comparison in Figure \ref{fig:outgoing_residual} leads to the same qualitative distinction at larger cutoffs. At $S = 43$, one finds $\ell_S = 60.57$ for the exact ground space, 2.705 for the unrestricted product state, 0.337 for the structured VMC state and 0.814 for the MLP. As can be seen from the figure, the VMC states have the smallest outgoing action in this range, while the unrestricted product state lies between them and the finite matrix ground space. The finite matrix ground spaces are the most sensitive to newly available spin configurations, which weakens their interpretation as successive approximations to one state of the constraint with the spin cutoff removed. The variational states are better behaved in this limited sense, especially at large cutoffs, but their enlarged residuals remain nonzero and the test supplies no quantitative distance from an exact kernel state.

Given the above, the three analyses provide complementary, albeit qualitative, evidence. The spectral projection places most of the structured VMC state and is coherent product organization in a small low-lying fraction of the finite matrix spectrum. The unrestricted product state minimization shows that the coherent profile is selected without imposing it on the variational manifold and lastly, the cutoff enlargement test shows that the VMC states avoid the strongest boundary sensitivity exhibited by the exact finite matrix ground spaces. In the absence of a reference exact full model physical state, however, none of these results establishes a quantitative distance from such an absent exact kernel state in the full model. Nevertheless, the combined evidence suggests that the VMC states capture nontrivial structural information about the near-kernel sector, in particular such coherent product structure may represent a genuine physical signature of the low residual constraint dynamics, although whether it survives in the exact kernel of the full model remains an open question.

%
%

\section{\label{sec:discussion}Discussion}

The results obtained in this work raise two related general questions, namely (i) the relation between the exact finite matrix ground spaces and the low-residual states selected by the variational process, and (ii) the possible interpretation of these states after removal of the spin cutoff. As for (i), in previous applications of neural quantum states to models in LQG, the variational states could be related to analytically known solution sectors or identified through their characteristic kinematical properties \cite{Sahlmann:2024kat,Sahlmann:2024pba,Sahlmann:2026qvs}. Once that structure had emerged, further reductions of the constraint residual did not alter the physical picture extracted from the states and a sufficiently small residual therefore provided a useful practical notion of near-kernel state. The present model is more subtle in the sense that states at different residual scales, although of small value, need not carry the same structure, as illustrated by the different factorization and spin profiles of the exact ground spaces, the two VMC ansätze and the product states. As for (ii), the exact ground spaces vary with the cutoff, in such a way that it is hard to extrapolate to the full theory. In particular, we have demonstrated that symmetry, spectral gap, correlation structure and one-edge spin probability distribution of the ground space change with $j_\mathrm{max}$. We also show that a small residual need not remain small when the expectation value is taken with respect to the constraint at higher cutoff. 

Altogether, we find that physical properties of the near-kernel states change with changing residual $\langle \hat{\mathcal{Q}}_{j_\mathrm{max}} \rangle$ and small, or even minimal, residual at a given $j_\mathrm{max}$ does not  provide a reliable measure of distance from the kernel of the full model. 

On the other hand, we note that the, variational states and the optimization procedure seem to capture the low-lying part of the spectrum of the constraint quite well: in the cases where we were able to obtain reliable data (i.e. up to $j_\mathrm{max}=43$), around 90\% of the weight in the states was contained within approximately the lowest 5 \textperthousand~ of the spectrum, as measured in terms of dimensions, see Table \ref{tab:ret_ranks}. For an extended system with more states close to the ground space that retain some of its physical properties, this may be enough to capture the ground space properties of the finite dimensional system. Also, in such systems, the cut-off dependence may be more benign. We have for example observed such behavior for U(1) BF theory \cite{Sahlmann:2024kat,Sahlmann:2024pba}. 

It is also an interesting outcome of the current investigation that the residual of a given state can depend sensitively on the cutoff at which the constraint is taken. In our case, the variational low-residual states at a given cutoff are closer to the kernel of the full theory than a ground state at the same cutoff, when measured in terms of constrained residual, in the range of cutoffs where we had reliable data, see Section \ref{subsec:cutoff_enlargement}.

Finally, it is interesting that the states obtained variationally have a Thiemann coherent state structure. To investigate whether this is a bias from the network architecture or a dynamically selected structure emergent due to the constraint, we conducted a minimization over otherwise unrestricted product states in Section \ref{subsec:hartree_min}. The calculation also did not use Monte Carlo sampling. Within the numerical resolution of the multi-start optimization, the state with a minimal residual is permutation invariant and its one-edge factor is a reduced Thiemann coherent state with near-unit fidelity. This observation is stronger than the fits of the structured VMC states to coherent states alone, as it does not rely on any (neural) parametrization. It identifies group coherent states as the factors for product states that are preferred by the constraint. 

We also note that the coherent state solutions we investigated in this respect are peaked on flat holonomies. In an interpretation in terms of (anisotropic) cosmology, this would mean a spatial slice with vanishing extrinsic curvature and hence a time-symmetric geometry. The interpretation in the present situation, however, is unclear because we have introduced no notion of time evolution, nor is this model derived as a symmetry-reduced sector of the full theory. 

It is surprising that the variational states converged to products of coherent states too, although the ansatz was much more general. We suspect that there was a bias towards separable states inherent in the optimization procedure. It might be interesting to investigate this issue more deeply in the future. 

Naturally, some of the questions raised and issues encountered here must also be tested against the modelling choices that define the present one-vertex problem. While it is the graph necessitated by the model by assumption, the one-vertex graph is much smaller than those considered in previous variational studies \cite{Sahlmann:2024kat,Sahlmann:2024pba,Sahlmann:2026qvs}, and the spin truncation introduces hard, non-cyclic boundaries in spin space rather than the cyclic boundary conditions used in previous work. The observed competition between the finite-matrix ground spaces and the low residual variational branches may therefore be affected by such choices. Further choices on which the properties of the model may depend include the regularization of the inverse flux, the symmetric factor ordering and the Barbero-Immirzi parameter, i.e. the relative strength of Euclidean and Lorentzian contributions to the constraint. Any of these ingredients could -- but need not -- affect the spectral organization and its sensitivity to the spin cutoff. Investigating possible dependencies of these choices defines a broader direction for future work. 

Altogether, this work shows that variational methods can be applied to QRLG models, but in this particular case it is hard to draw firm conclusions about physical states after removal of the cutoff.  States with a semiclassical structure seem to emerge easily in the low part of the constraint spectrum at least in the present one-vertex model. The emergence of high-accuracy Thiemann coherent states as states of low constraint residual when separability is imposed, is very surprising and interesting.

%
%

\section{\label{sec:ack}Acknowledgments}

The authors thank the two anonymous referees for an exceptionally careful reading of the manuscript and for many helpful recommendations. In particular their request for exact diagonalization results substantially improved the manuscript and the understanding of the system and of the variational states.

The authors gratefully acknowledge the scientific support and HPC resources provided by the Erlangen National High Performance Computing Center (NHR@FAU) of the Friedrich-Alexander-Universität Erlangen-Nürnberg (FAU). The hardware is funded by the German Research Foundation (DFG). For I.M.~this work was funded by National Science Centre, Poland through grant no.~2022\slash 44\slash C\slash ST2\slash 00023. H.S.~acknowledges the COST Action CA23130. 

Simulations for this work were performed using neuraLQX~\cite{neuralqx1:2026} and are available on GitHub \cite{qrlgvmc_github}. neuraLQX is a high-performance simulations toolkit for loop quantum gravity built on top of NetKet~\cite{netket3:2022,netket2:2019}. The implementation relies on JAX~\cite{jax2018github} and Flax~\cite{flax2020github}.

For the purpose of open access, the authors have applied a CC BY 4.0 public copyright license to any author accepted manuscript (AAM) version arising from this submission.

%
%

\section*{References}
\bibliographystyle{iopart-num-long}
\bibliography{references}

%
%

\appendix

\section{\label{app:networks}Structured network architecture}

The structured variational ansatz used in this work is defined on the three local edge variables of the one-vertex graph. After discretization, each edge label is represented by an index
\begin{equation}
    \iota_a \in \{0,\dots,d-1\},
    \qquad a \in \{x,y,z\},
\end{equation}
where \(d\) is the number of admissible local values in the chosen truncation. The network outputs the logarithm of the wavefunction in the form
\begin{equation}
    \log \psi(\iota_x,\iota_y,\iota_z)
    =
    A(\iota_x,\iota_y,\iota_z)
    +
    i\,\Theta(\iota_x,\iota_y,\iota_z),
\end{equation}
with separate amplitude and phase branches.

The logarithmic amplitude is decomposed into unary, pairwise, triple, and residual contributions,
\begin{equation}
    A = A^{(1)} + A^{(2)} + A^{(3)} + A^{(\mathrm{res})}.
\end{equation}
The unary part is a sum of one-edge tables,
\begin{equation}
    A^{(1)}(\iota_x,\iota_y,\iota_z)
    =
    b_x(\iota_x)+b_y(\iota_y)+b_z(\iota_z).
\end{equation}
This already contains the fully factorized family, since if only \(A^{(1)}\) is present then
\begin{equation}
    \psi(\iota_x,\iota_y,\iota_z)
    =
    \psi_x(\iota_x)\psi_y(\iota_y)\psi_z(\iota_z).
\end{equation}

The pairwise sector adds explicit low-rank couplings between the three pairs of edges,
\begin{align}
    A^{(2)}(\iota_x,\iota_y,\iota_z)
    &=
    \sum_{r=1}^{R_2}
    u^{L}_{xy}(\iota_x,r)\,u^{R}_{xy}(\iota_y,r)
    \notag\\
    &\quad
    +
    \sum_{r=1}^{R_2}
    u^{L}_{xz}(\iota_x,r)\,u^{R}_{xz}(\iota_z,r)
    \notag\\
    &\quad
    +
    \sum_{r=1}^{R_2}
    u^{L}_{yz}(\iota_y,r)\,u^{R}_{yz}(\iota_z,r),
\end{align}
where \(R_2\) is the pair rank. A genuine three-body contribution is then included through a low-rank canonical factorization,
\begin{equation}
    A^{(3)}(\iota_x,\iota_y,\iota_z)
    =
    \sum_{r=1}^{R_3}
    t_x(\iota_x,r)\,t_y(\iota_y,r)\,t_z(\iota_z,r),
\end{equation}
with triple rank \(R_3\).

To supplement these explicit low-order terms, each local index is mapped to an embedding vector
\begin{equation}
    e_a = E(\iota_a) \in \mathbb{R}^{m},
\end{equation}
and from these embeddings one constructs symmetric combinations
\begin{align}
    s_1 &= e_x + e_y + e_z, \\
    s_2 &= e_x \odot e_y + e_x \odot e_z + e_y \odot e_z, \\
    s_3 &= e_x \odot e_y \odot e_z,
\end{align}
where \(\odot\) denotes element-wise multiplication. These features are augmented by normalized polynomial invariants of the discrete labels and by Fourier features of the form
\begin{equation}
    \sin(\pi n \tilde{\iota}_a),
    \qquad
    \cos(\pi n \tilde{\iota}_a),
    \qquad
    n=1,\dots,N_f,
\end{equation}
with normalized indices \(\tilde{\iota}_a \in [0,1]\). The resulting feature vector \(h(\iota_x,\iota_y,\iota_z)\) is passed through a small feed-forward network, producing the residual correction
\begin{equation}
    A^{(\mathrm{res})}(\iota_x,\iota_y,\iota_z)
    =
    \mathcal{R}\!\left(h(\iota_x,\iota_y,\iota_z)\right).
\end{equation}

The phase is modelled by a separate branch built from the same residual feature vector,
\begin{equation}
    \Theta(\iota_x,\iota_y,\iota_z)
    =
    \lambda
    \tanh\!\left(
        \mathcal{G}\!\left(h(\iota_x,\iota_y,\iota_z)\right)
    \right),
\end{equation}
where \(\lambda\) fixes the overall phase scale.

Altogether, the ansatz combines an explicit low-order decomposition with a nonlinear residual correction,
\begin{equation}
    \log \psi
    =
    A^{(1)} + A^{(2)} + A^{(3)} + A^{(\mathrm{res})}
    +
    i\,\Theta.
\end{equation}
Its structure therefore distinguishes between product contributions, pair correlations, irreducible three-body couplings, and residual nonlinear effects. This is the sense in which the ansatz is biased towards factorized and weakly correlated states while retaining sufficient flexibility to represent more general solutions.

\section{\label{app:vmc_params}Variational simulation parameters}

For completeness, we summarise here the neural architecture and numerical parameters used for the VMC results in Section \ref{sec:results}. Both neural ansätze parametrise the logarithm of a complex wavefunction and are optimised against the Monte Carlo estimator for $\langle \hat{\mathcal{Q}} \rangle$. The structured architecture is described in \ref{app:networks}. For the MLP, let $h^{(0)} = (j_x, j_y, j_z)$. Its successive hidden layers are
\begin{equation}
    h^{(\ell)} = \log\cosh(W^{(\ell)} h^{(\ell - 1)} + b^{(\ell)}),
\end{equation}
with $\ell = 1, \ldots, L_\mathrm{MLP}$ and the logarithmic wavefunction is 
\begin{equation}
    \log \psi_\mathrm{MLP}(j_x, j_y, j_z) = \sum_r h_r^{(L_{\mathrm{MLP}})} + v_xj_x + v_yj_y + v_z j_z.
\end{equation}
Here, the $v_i j_i$ terms are the so-called visible bias terms and $r = 1, \ldots, 3\alpha$ indexes the neurons in the final hidden layer. The matrices, biases and visible coefficients are complex and every hidden layer has width $3\alpha$. They are initialised independently from a normal distribution with standard deviation $10^{-2}$. The cutoff-dependent depth and width multiplier are listed below, together with the depth $L_\mathrm{str}$ of the residual feed-forward component of the structured ansatz
\begin{center}
\begin{tabular}{c|cc|c}
    $j_{\mathrm{max}}$ & $L_{\mathrm{MLP}}$ & $\alpha$ & $L_{\mathrm{str}}$ \\ \hline
    21            & 2 & 2 & 10 \\
    51, 101       & 2 & 2 & 8  \\
    201, 301      & 2 & 3 & 10 \\
    401, 501      & 2 & 4 & 10 \\
    601, 701      & 4 & 5 & 10 \\
    801           & 4 & 5 & 12 \\
    901, 1001     & -- & -- & 12
\end{tabular}
\end{center}
The remaining structured ansatz parameters are common to the conducted runs, the embedding dimension is $m=16$, the pair and three-body ranks are $R_2 = R_3 = 8$, the residual hidden width is 32, the number of Fourier frequencies is $N_f = 6$ and the phase scale is $\pi$. For the independently initialised runs, the unary amplitudes, correlation tensors and unary phases are initialised from normal distributions with standard deviations 0.05, 0.01 and 0.15 respectively. The ensembles of structured ansatz states studied at each cutoff use real double precision parameters while the MLP uses complex parameters. 

All neural VMC simulations use the Adam optimiser with $\beta_1 = 0.9$, $\beta_2 = 0.999$ and $\epsilon_\mathrm{Adam} = 10^{-8}$ without stochastic reconfiguration. The learning rate is annealed linearly from $3\times 10{-3}$ to $7 \times 10^{-5}$ over 1000 optimiser steps. At every step, 2048 configurations are sampled using local Metropolis updates with 32 persistent chains and 12 sweeps between recorded samples. The chains are not reset between optimiser steps. After training, the reported residual and its Monte Carlo error are evaluated with 4096 samples. All simulations are carried out over the entire 1000 steps and therefore convergence here does not denote satisfaction of a gradient tolerance but operationally to the formation of a stable late time residual plateau.

The individual state plots used the seed 2472475101006. For the ensembles in Figure \ref{fig:vmc_residuals_scatter}, candidate seeds were drawn deterministically as $s = 2472475101006 + k$. The accepted seed suffixes $k$ are
\begin{center}
\begin{tabular}{c|c|c}
    $j_{\mathrm{max}}$ & structured ansatz & MLP \\ \hline
    51   & 1, 2, 3, 4, 5 & 1, 2, 3, 5, 6 \\
    101  & 1, 2, 3, 4, 5 & 1, 3, 9, 10, 11 \\
    201  & 1, 2, 3, 4, 5 & 1, 5, 6, 9, 10 \\
    301  & 1, 3, 4, 5, 6 & 1, 2, 3, 11, 13 \\
    401  & 1, 2, 3, 4, 5 & 1, 2, 3, 7, 11 \\
    501  & 1, 2, 3, 4, 5 & 1, 7, 11, 12, 13 \\
    601  & 1, 2, 3, 4, 5 & 1, 5, 7, 11, 13 \\
    701  & 1, 2, 4, 5, 6 & 1, 2, 7, 11, 13 \\
    801  & 1, 2, 3, 4, 5 & 2, 7, 11, 26, 28 \\
    901  & 2, 3, 4, 5, 6 & -- \\
    1001 & 1, 2, 3, 4, 5 & --
\end{tabular}
\end{center}

\end{document}